\documentclass[10pt,a4paper]{article}
\usepackage[ansinew]{inputenc}
\usepackage[hmargin=3cm, vmargin=1.5cm]{geometry}
\usepackage [T1]{fontenc} 
\usepackage{textcomp}
\usepackage{epsfig}
\usepackage{amsmath,amsfonts,latexsym, amssymb,amsthm}
\usepackage{hyperref}
\usepackage{array} 
\usepackage{makeidx}
\usepackage{graphics}
\usepackage{tikz} 
\usepackage{subfigure} 
\usepackage{xspace}
\usepackage[sort&compress]{natbib}
\usepackage{stmaryrd} 

\bibpunct{[}{]}{,}{n}{}{;} 

\newcommand {\EE}{{\ensuremath\mathbf{E}}}
\newcommand {\dd}{\ensuremath{\mathrm{d}}}
\newcommand{\ddh}{\ensuremath{\mathrm{d_H}}}
\newcommand {\sgn}{\mathrm{sgn}}
\newcommand {\PPr}{\mathrm{Prob}}
\newcommand {\PP}{\mathrm{P}}
\newcommand {\g}{\mathrm{g}}
\newcommand {\s}{\mathrm{S}}
\newcommand {\Q}{\mathrm{Q}}
\newcommand {\Tr}{\mathrm{Tr}}
\newcommand{\ket}[1]{\ensuremath{|#1\rangle}\xspace}
\newcommand{\bra}[1]{\ensuremath{\langle #1|}\xspace}

\newcommand{\ident}{\stackrel{\mbox{{\rm \tiny (def)}}}{=}}
\newcolumntype{C}{>{$}c<{$}}
\newtheorem* {summary*}{Summary}
\newtheorem* {definition*}{Definition}

\title{{\bf Classical description of quantum randomness using stochastic gauge systems }}
\author {Michel Feldmann
\thanks{Electronic adress: michel.feldmann@polytechnique.org}
}
\date {}

\begin{document}


\maketitle
\abstract{
We present a classical probability model appropriate to the description of quantum randomness. This tool, that we have called \emph{stochastic gauge system}, constitutes a contextual scheme in which the Kolmogorov probability space depends upon the experimental setup, in accordance with quantum mechanics. Therefore, the probability space behaves like a gauge parameter. We discuss the technical issues of this theory and apply the concept to classically emulate quantum entangled states and even `super-quantum' systems. We exhibit bipartite examples leading to maximum violation of Bell-CHSH inequalities like EPR pairs or exceeding the Tsirelson bound like PR-boxes, as well as tripartite cases simulating GHZ or W-states. We address also the question of partially correlated systems and multipartite entanglements. In this model, the classical equivalent of the entanglement entropy is identified with the  Kullback-Leibler divergence. Hence, we propose a natural generalisation of this function to multipartite systems, leading to a simple evaluation of the degree of entanglement and determining the bounds of maximum entanglement. Finally, we obtain a constructive necessary and sufficient condition of multipartite entanglement. 

} 

\tableofcontents

\section{Introduction}
\label{introduction}
While quantum mechanics is one of the pillars of modern physics, its very foundations remain controversial. A crucial point is that probability in quantum theory obeys different rules than do the classical probabilities, as emphasized by Richard Feynman in the 1950's. Later, E.~T.~Jaynes noted that quantum probabilities are striking similar to the Bayesian probabilities. One controversial issue concerns `local realism' highlighted by the well-known EPR paradox~\cite{einstein, bohm}: Nowadays, it is widely accepted that local realism does not hold in microphysics. This `spooky' property is derived from a famous theorem by Bell~\cite{bell,bell1} complemented by Kochen and Specker~\cite{kochen} and supported by a number of experimental verifications~\cite{genovese}. 
As a result, a huge literature has come to light, aiming to reconcile this apparent inconsistency, if not with  common sense, at least with a form of logic. Currently, a number of different approaches are competing, often at the interface between physics and philosophy (see for instance references~\cite{gudder,evans,goyal1,griffith,rovelli,caves1,fuchs}).
However, following Jaynes~{\cite{clearing} and among several authors,~\cite{kracklauer, hess, hess2, hess3, khrennikov, khrennikov2, cameleon, accardi, clover1,adenier} we have warned about a flaw in the proof of Bell's theorem~\cite{mf}. In order to falsify Bell's claims, we have exhibited a very simple classical counterexample with just three dice~\cite{mf1} based on the present model. In our opinion, this proves that the concept of quantum non-locality remains actually unfounded. Conversely, if quantum mechanics is really compatible with classical local realism, quantum phenomena can be classically emulated, at least to some extent. That is the purpose of the present paper.

We have constructed classical tools, namely, consistent parametric probability distributions, that we have called \emph{stochastic gauge systems}. This model constitutes a contextual probability scheme in which the Kolmogorov probability space depends upon the experimental setup, in accordance with quantum mechanics. Hence,  the recourse to non-locality is not necessary. By contrast, the flaw in Bell's theorem is to postulate the existence of an absolute probability space whatever the settings, in opposition to both theoretical and experimental evidences. Beyond the EPR paradox, we suggest that classical simulation of quantum phenomena may provide some insights in several fundamental questions. 

Our underlying picture is the following: We propose that the classical equivalent of a stationary quantum system is a spatially extended object in dynamic equilibrium. Therefore, in the absence of perturbation, its evolution is completely deterministic and there is no room for randomness. This analogy suggests that quantum collapse can be compared with a classical break of equilibrium. The stability of a classical equilibrium is similar to the resistance to decoherence of a quantum state. Based on Bell theorem, it is generally assumed that the notion of entanglement is strictly of quantum nature~\cite{plenio1,plenio}. By contrast, we argue that the classical analogue of quantum entanglement is the property of \emph{contextual dependence}, better underscored  when the break of equilibrium of a spatially extended object can be triggered from two distant locations separated by a boundary. We have found that this boundary behaves like an ignition front where the \emph{contextual gauge probability trial} is performed. Therefore, the \emph{contextual entropy} should be located on this boundary. This resembles the holographic hypothesis suggested by 't~Hooft~\cite{hooft} and Susskind~\cite{susskind} in black hole physics.

We will first deal with the technical issues of our gauge probability model before addressing the question of classical simulation of quantum entangled systems. We will then describe the classical Bell-type model and will exhibit a number of examples with several settings and total correlation between two regions. Specially, we will compute exactly some typical bipartite configurations leading to maximum violation of Bell or CHSH inequalities. We will next extend the model to general locally consistent systems, including non quantum devices like PR-boxes and multipartite systems like GHZ and W tripartite states. We also deal with partial correlation and define the concept of classical entanglement entropy. The entropy of a classical entangled system is not extensive contrary to the entropy of  a separable system. In thermodynamics, this supports the existence of long range interactions~\cite{dauxois, antoni}.  In this context, the classical entanglement entropy should be identified with the Kullback-Leibler divergence. We propose natural generalizations, including the definition of classical `multipartite entanglement entropies' and a concept of `entanglement scheme' to characterize complex multipartite entanglement. 

\section[Simulation of bipartite Bell-type systems]{Bipartite Bell type systems}
\label{classical}

In classical physics, suppose that a spatially extended object is in equilibrium and that this equilibrium can be broken from two distant locations. Equilibrium is a global concept: Break of equilibrium from one location signifies break of equilibrium for the whole object and consequently break of equilibrium at the second location as well. This means  \emph{instantaneous correlation} at a distance between the two locations. Nevertheless, this does not imply \emph{instantaneous physical effect} since the relaxation towards a new equilibrium will require a delay. A trivial picture is the evolution of a stationary huge soap bubble (possibly in dynamic equilibrium, for instance in uniform rotation). We compare this phenomenon with what happens in quantum EPR experiment. For this purpose, we identify the spatial extension of the quantum system with the locus of all possible regions of measurement. The quantum system in unitary evolution is considered as a stationary object in dynamic equilibrium. The measurement of one parameter in one location interrupts the unitary evolution or breaks the equilibrium. The final state is a new dynamic equilibrium. We claim that the behaviour of similar extended objects in equilibrium may be viewed as a form of \emph{classical entanglement}.

\subsection{Bell-type systems}

Consider a pair of quantum \emph{entangled} entities $\{{\mathfrak E}_0,{\mathfrak E}_1\}$ (e.g., particles).  The pair, or its global wave function, will be compared with a \emph{classical extended object} stretching out on two space regions $ {\mathcal R}_0$ and ${\mathcal R}_1$. We will first focus on totally correlated entities (as defined in Sec. \ref{totalcorrelation}) referred to as Bell-type systems.

In region ${\mathcal R}_0$, we suppose that an observer ${\mathcal O}_0$ selects freely a setting $u_0$ (e.g., an angle of polarization), element of a given set $\mathbf {\Theta_0}$, and measures on ${\mathfrak E}_0$  a random dichotomic observable $x_0$, element of a dyadic set $\mathbf {X} $ = \{0, 1\}, (or a spin $s_0 = 2x_0-1$). This measurement breaks the dynamic equilibrium of the whole extended object. Similarly, in region ${\mathcal R}_1$, a second observer ${\mathcal O}_1$ selects independently a setting $u_1 \in \mathbf {\Theta_1}$ and measures on ${\mathfrak E}_1$  a random dichotomic observable $x_1 \in \mathbf {X}$, (or a spin $s_1 = 1-2x_1$)%
\footnote{\label{footnote1}In Sec.~\ref{classical}, where we deal with only two regions, we use the convention $s_0 = 2x_0-1$ and $s_1 = 1-2x_1$  because the concept of total correlation is then simpler. We will drop this convention in the other sections.} %
For the sake of simplicity, we will assume that $\mathbf{\Theta_0 = \Theta_1= \Theta}$. 

The spins are well defined physical observables and thus the interchange of the numbering in the pair $\{{\mathfrak E}_0,{\mathfrak E}_1\}$ interchanges the binary digits `$0$' and `$1$'. More generally, we will call \emph{bit reversal} the exchange of `$0$' and `$1$' in $\mathbf {X} .$ Due to symmetry between ${\mathfrak E}_0$ and ${\mathfrak E}_1$, physical Bell-type systems are often invariant by bit reversal.

Note that for an ensemble of runs, the binary strings $\{ x_0\} $ and $\{ x_1\} $ can be viewed as digital signals emitted from two distant ports $ {\mathcal R}_0$ and ${\mathcal R}_1$.

\subsubsection{Contextuality and Bell's inequalities}
\label{contextuality}
As soon as the system $\{{\mathfrak E}_0,{\mathfrak E}_1\}$ interacts with the settings $u_0$ or/and $u_1$, \emph{instantaneously},  the initial equilibrium is broken. Therefore, there is some conditional probability $\PP(x_0;x_1| u_0;u_1)$ that ${\mathcal O}_0$  will observe (after a delay) the outcome $x_0$ and ${\mathcal O}_1$  the outcome $x_1$. In fact, the function $\PP(x_0;x_1| u_0;u_1)$ (or $\PP(\mathbf{x}|\mathbf{u})$ in short) describes the symmetry of the extended object. It will be convenient to have a special name for the entries $(\mathbf{x}|\mathbf{u})$.
\begin{definition*}[Target] We will name \emph{global target} or simply \emph{target} each entry $ (\mathbf{x}|\mathbf{u}) $. Let $\mathfrak{T}=\{ (\mathbf{x}|\mathbf{u})  \}$ be the set of all possible targets. 
\end{definition*}
For the purpose in hand, we can identify $\mathfrak{T}$ with a \emph{coarse-grained description} of the potential \emph{future} equilibrium of the extended object.
According to Bell's theorem~\cite{bell}, Bell's inequalities hold, but only when the theorem is valid, i.e., in \emph{non contextual} systems.
By contrast, if the conditional probability $\PP(x_0;x_1| u_0;u_1)$ depends upon the settings $u_0$ and $u_1$, Bell's inequalities can be violated.

Actually, there is a number of Bell's inequalities. We have shown~\cite{mf} that basically, these inequalities derive from the well known triangle inequalities with respect to a convenient metrics, namely the \emph{Hamming distance}%
\footnote{The Hamming distance~\cite{hamming} between two binary sequences is the number of places where the two sequences differ. Therefore, we have $ \ddh(\{ x_0 \} , \{ x_1 \} )=\sum (x_0\oplus x_1)$, where $\oplus$ denotes addition modulo $2$.}
  $ \ddh(\{ x_0 \} , \{ x_1 \}$ ) between the digital signals $\{ x_0\} $ and $\{x_1\} $ emitted from the two ports $ {\mathcal R}_0$ and ${\mathcal R}_1$ for an ensemble of runs. A similar metrics was proposed by Santos~\cite{santos1} using Boolean logic. 
Define the \emph{Hamming divergence} for one run as
\begin{equation}
\label{dhamming}
\dd(u_0, u_1)\ident \mathbf {E}[\ddh( x_0,  x_1)]= \sum_{x_0=0}^1 \sum_{x_1=0}^1 \ddh (x_0;x_1) \PP(x_0;x_1|u_0,u_1), 
\end{equation} 
where $\mathbf {E}[.]$ stands for the expectation value with respect to $ \PP(\mathbf{x}|\mathbf{u}) $. 
Clearly, we have
$$\dd(u_0, u_1)= \PP(1; 0 | u_0; u_1) + \PP(0; 1 |u_0; u_1)\le 1 .$$
The Hamming divergence is invariant by interchange of the two regions, but not, in general, by permutation of the settings%
\footnote{%
For this reason, we use the name `Hamming divergence' for `mean Hamming distance $\dd(k_0,k_1)$ between regions $\mathcal{R}_0$ and $\mathcal{R}_1$ for the settings $u_0=k_0$ and $u_1=k_1$ respectively'. 
}%
. Therefore, if $\theta_1, \theta_2 \in \mathbf{\Theta}$, in general $\dd(\theta_1, \theta_2 )\not= \dd(\theta_2, \theta_1)$. 
Let $ \theta_0, \theta_1, \theta_2 \in \mathbf{\Theta}$ be three different settings: Then, Bell's inequalities simply read: 
\begin{equation}
\label{triangle}
\dd(\theta_0, \theta_2) \le \dd(\theta_0, \theta_1) + \dd(\theta_1, \theta_2).
\end{equation}
In non contextual systems, we have $\PP(x_0;x_1|u_0;u_1) = \PPr (x_0|u_0)\PPr(x_1|u_1)$, and the proof can be directly derived from an elementary calculation.
The original Bell's inequalities make use of the expectation value $\s(u_0,u_1) = \EE[s_0 \cdot s_1]$. In non contextual systems we have~\cite{mf},
\begin{equation}
\label{substitution} 
\dd (u_0, u_1) = (1/2)[1+\s(u_0,u_1)].
\end{equation}  
Thus, starting from the following formulation of the triangle inequality, 
$$|\dd(\theta_0, \theta_1) -\dd(\theta_0, \theta_2)| \le \dd(\theta_1, \theta_2),$$ 
we obtain the well known original Bell's inequality,
\begin{equation}
\label{originalbell}
|\s(\theta_0, \theta_1) - \s(\theta_0, \theta_2)| \le 1+\s( \theta_1, \theta_2).
\end{equation}

Another popular formulation is the CHSH inequality~\cite{chsh}, that involves four settings, $A, A', B$ and $B'$, with $A, A', B, B' \in \mathbf{\Theta}$. In regions $\mathcal{R}_0$ (resp. $\mathcal{R}_1$), it is possible to select the settings $A$ or $A'$ (resp. $B$ or $B'$). For a particular situation, the outcomes are respectively $a$, $b$, $a'$ and $b'$ with $a, a', b, b' \in\{ 0,1\}$. We have,
$$
0\le \ddh (a, b) + \ddh (a', b) + \ddh (a, b') - \ddh (a', b')  \le 2.
$$
These inequalities are valid by inspection given that all outcomes can only take the values $0$ or $1$. Therefore,  they will hold true for any convex combination of these inequalities.  This will be the case if it is possible to prepare different experiments with different settings governed by the \emph{same probability space}. As  a result, accounting for Eq.~(\ref{dhamming}), we have
\begin{equation}
\label{distanceCHSH}
0\le \dd (A, B) + \dd (A', B) + \dd (A, B') - \dd (A', B')  \le 2.
\end{equation}
or using Eq.~(\ref{substitution}),
\begin{equation}
\label{originalCHSH}
|\s (A, B) + \s (A', B) + \s (A, B') - \s (A', B')|  \le 2.
\end{equation}

 In conclusion, Bell-CHSH inequalities are always valid provided that the target probabilities derive from the same underlying probability space, irrespective of the contextual parameters $u_0$ and $u_1$. In this classical framework, violation of Bell's inequalities is simply a criterion of \emph{contextual dependency} of the probability space and does not in the least imply  instantaneous effect at a distance nor violation of \emph{local realism}. 

\subsubsection[Partial measurement]{Partial measurement, nonsignaling correlations, local consistency and causal horizon}
\label{nonsignaling}
Quantum mechanical collapse may be viewed as a break of equilibrium of the wave function. When the particle ${\mathfrak E}_1$ interacts with the setting $u_1$ in region $ {\mathcal R}_1$,  the whole system decoheres and the pair of particles is no more entangled. Then, possibly after a delay, the wave function splits into two local wave functions. However, in region ${\mathcal R}_0$, this splitting is not observable and the observer ${\mathcal O}_0$ is not aware of what happens in $ {\mathcal R}_1$; he independently selects his own setting $u_0$ and proceeds to the measurement of $x_0$, while the local marginal probability $\PPr(x_0|u_0)$ of $x_0$ given $u_0$ is not affected, as long as both the setting $u_1$ and the outcome $x_1$ are ignored.  In other words, the local probability $\PPr(x_0|u_0)$ depends on the knowledge of the observer $\mathcal{O}_0$ and is thus an observer-dependent concept. This property characterizes quantum \emph{partial measurements} in two different locations, when the global Hilbert space is regarded as a  tensor product of two partial subspaces. This means that partial measurement is not a way to communicate and is thus compatible  with space-like separation of the two regions.  Adopting the point of view of the observers, we will call  \emph{local consistency} this \emph{nonsignaling} correlation~\cite{popescu, rohrlich,masanes}. The same property  is also encountered in a number of situations in relativity, cosmology and black hole physics, where quantum entanglement holds between two regions separated by a so called \emph{causal horizon}, and even in some classical laboratory analogues of black holes~\cite{rousseaux, philbin}. On the contrary, it is rarely met in usual classical physics, e.g., a soap bubble is not locally consistent. Nevertheless,  as we shall show in the following sections, it is quite easy to design \emph{ad hoc} classical devices, like dice games~\cite{mf1},  which fulfil this criterion.

Consider a system with $K$ discrete settings, $ \mathbf{\Theta}= $ $\{ \theta_0, $ $\theta_1,\dots,$ $\theta_{K-1} \}$. When local consistency holds, the partial  probability in one region, say ${\mathcal R}_0$,   does not depends on the setting $u_1$ in the second region ${\mathcal R}_1$. Let $u_0=\theta_k$ be selected in region ${\mathcal R}_0$  we have:
\begin{equation}
\label{localdiscret}
\forall u_1 \in \mathbf{\Theta}: \PPr(x_0|u_0=\theta_k) = \PP(x_0;0|\theta_k,u_1)+\PP(x_0;1|\theta_k;u_1) 
\end{equation}
and similarly for the second port. When the 2-region system mimics a quantum situation, Eq.~(\ref{localdiscret}) describes the partial state in region $\mathcal{R}_0$, obtained by  partial trace over the region $\mathcal{R}_1$.

Accounting for the normalization relation $$\PPr(x_i=0|u_i=\theta_k)+ \PPr(x_i=1|u_i=\theta_k)=1$$ for $i=0,1$, Eq.(\ref{localdiscret}) provides a number of $2(K-1)$ relations between the target probabilities $\PP(\mathbf{x}|\mathbf{u})$ for each setting $\theta_k$ and $2K(K-1)$ for all settings (Table~\ref{parameter}).

 For a continuous ensemble of settings, when $\PP(x_0;x_1|u_0;u_1)$ is continuous and differentiable, we have for $i=0,1$,
\begin{equation}
\label{localcontinu}
\frac{\partial \PPr(x_i|u_i)}{\partial u_{1-i}}=0.
\end{equation}

\subsubsection{Total correlation}
\label{totalcorrelation}
In Bell-type experiment the outcomes $x_0 = x_1$ are identical when the same setting $u_0 = u_1 = \theta \in \mathbf{\Theta}$ is selected in the two regions. Then:

\begin{equation}
\label{correlation}
\PP (1;0|\theta;\theta)=\PP(0;1|\theta;\theta)=0.
\end{equation}
This provides a number of $2$ additional relations between the target probabilities $\PP(\mathbf{x}|\mathbf{u})$  for each setting $\theta$ and $2K$ for all settings (Table~\ref{parameter}).
We will call this property \emph{total correlation}.

For \emph{totally correlated and locally consistent systems}, it follows from Eqs. (\ref{localdiscret}), (\ref{correlation}) that for $\xi \in \mathbf{X}$ and $\theta\in\mathbf{\Theta}$
\begin{equation} 
\label{localprob}
\PPr(x_0=\xi|u_0=\theta)=\PPr(x_1=\xi|u_1=\theta)=\PP(\xi;\xi|\theta;\theta).
\end{equation}

Note that `total correlation' is different from `maximum entanglement'.  We will discuss the concept of maximum entanglement in Sec. \ref{inequalities}. In general, totally correlated systems are not maximally entangled and we will meet later examples of bipartite maximally entangled systems which are not totally correlated (Table~\ref{PRbox}c and~\ref{superGHZ}d). When the system mimics a quantum situation, totally correlated systems describe pure states.
\subsection{Classical evolution of a Bell-type system}

As soon as the system interacts with the settings  $u_0$ and/or $u_1$, the initial equilibrium is broken and  the system relaxes towards a future equilibrium. Violation of Bell's inequalities proves that a simple random trial is unable to describe the contextual correlations $\PP (x_0;x_1| u_0;u_1)$.  Hence, we address the following issue: Is it nevertheless possible to implement the joint conditional probability $\PP(\mathbf{x}|\mathbf{u})$ in classical physics? In other words, is it possible to classically simulate a quantum collapse? 

We claim that the answer is definitively \emph{yes}: We are going to show that any \emph{locally consistent} and \emph{totally correlated} Bell-type system relaxation can be achieved by cascading (1) propagation with finite velocity towards a so-called \emph{ignition point}, (2) random trial at this ignition point and (3) backwards propagation, again with finite velocity. We will next identify the locus of all possible ignition points with a \emph{causal horizon}.

\subsubsection[Ignition states and projection function]{Ignition states, ignition set and projection function}

In this paragraph, we will construct a convenient outcome set, that we will call \emph{ignition set}, which meets the requirements of contextual systems. We will call its elements \emph{ignition states}. Let us first examine discrete systems.
\paragraph{Discrete settings}
\label{discretesettings}
Consider a finite ensemble of $K$ settings, $\theta_0,$ $ \theta_1,\ldots,$ $\theta_{K-1}$. In Sec. \ref{contextuality}, we have compared the set of targets $\mathfrak{T}=\{ (\mathbf{x}|\mathbf{u}) \} $ with a coarse-grained description of the potential future equilibrium of the extended object. Now, we will similarly construct a convenient \emph{fine-grained description}\footnote{This designation describes a method but does not fit very well, because the number of useful fine grains is often less than the number of coarse grains!} as follows: 
In region $\mathcal{R}_i$ define the \emph{local target} $(x_i|u_i)$. When $x_i=\xi$ and $u_i=\theta_k$, we write $(x_i|u_i)=(\xi|\theta_k)_i$. Each local target, e.g. $ (\xi|\theta_k)_0 $ in region $\mathcal{R}_0$, can be regarded as the union of all (global) targets compatible with $x_0=\xi$ and $u_0=\theta_k$.

$$(x_0|u_0)=(\xi|\theta_k)_0 = \bigcup_{x_1=0}^1 \bigcup_{u_1= 0}^{K-1} (x_0;x_1|u_0;u_1 ).$$
Conversely
$$(x_0;x_1|u_0;u_1)  = (x_0|u_0) \cap (x_1|u_1). $$
According to Eq.(\ref{localprob}), we have $ \PPr(\xi|\theta_k) = \PP(\xi;\xi|\theta_k;\theta_k) $. When total correlation holds, the marginal probabilities are equal in the two regions and the local targets $ (\xi|\theta_k)_0$ and $(\xi|\theta_k)_1$ are identical. 
Let $\mathfrak{T}_{\mathrm{loc}} =\{ (\xi|\theta_k)\} $ be the set of local targets (relevant for both regions). The cardinality of $\mathfrak{T}_{\mathrm{loc}}$ is $2K$. 

Let $ \mathfrak{I} = \{ j \} $ be a set of integers.
We aim to  construct a sequence of card$(\mathfrak{I} )$ fine grains. Define an application of $\mathbf{\Theta}\times\ \mathfrak{I} \rightarrow\mathbf{X}=\{0,1\}$ or rather an application $\Pi(k,j)$: $ \llbracket 0, K-1 \rrbracket\times\ \mathfrak{I} \rightarrow\mathbf{X}=\{0,1\}$ 
\begin{equation}
\label{xij}
\xi = \Pi(k, j) .
\end{equation} 
We will call \emph{projection function} the function $\Pi(k,j). $
Now, we construct the sequence of card$( \mathfrak{I} )$ fine grains $\hat\lambda_j$  by the following intersection of local targets: 
\begin{equation}
\label{lambdaj1}
j\in  \mathfrak{I} \mapsto\hat\lambda_j \ident \bigcap_{k=0}^{K-1}  \Big( \Pi(k,j) |\theta_ k\Big) .
\end{equation} 
Therefore, the maximum number of distinct fine grains is $2^K$ and then $ \mathfrak{I} = \llbracket 0,2^K-1\rrbracket$. 
Let $$j = \sum_{k=0}^{K-1} j_k 2^k$$ be the binary expansion of an integer $ j, (0 \le j \le 2^K -1)$ and $j_k$ the coefficient of $2^k$. A convenient definition of the projection function is
\begin{equation}
\label{lambdajbis}
\Pi(k,j)\ident j_k.
\end{equation} 

For reasons explained in Sec. \ref{collapse}, we will call \emph{ignition state} each fine grain $\hat\lambda_j$. 
Clearly, the ignition states included in a given global target $ (x_0;x_1|u_0;u_1)  $ are also included in the local targets $(x_0|u_0)$ and  $(x_1|u_1) $ and conversely. Therefore, we have
$$ \hat\lambda_j \in (x_0;x_1|u_0;u_1)  \Leftrightarrow \hat\lambda_j \in (x_0|u_0) \cap (x_1|u_1) .$$
It is also convenient to define the index subset,
\begin{equation}
\label{Ixj}
\mathbf{I}(\xi,k) \ident \{j\ |\ \Pi(k,j)=\xi\}.
\end{equation}
Clearly, we have $\mathbf{I}(0,k) +\mathbf{I}(1,k) = \mathfrak{I} .$
For simplicity, we will often identify $k$ with $\theta_k$, $j$ with $\hat\lambda_j$  and use the same symbol, 
$$\mathbf{I}(\xi,\theta_k) \ident \mathbf{I}(\xi,k) \ ; \ \Pi(\theta_k,\lambda_j) \ident \Pi(k,j).$$
In addition define the index subset,
\begin{equation}
\label{Jxu}
\mathbf{J} (\mathbf{x},\mathbf{u}) \ident \{ j \ / \ \hat\lambda_j \in (\mathbf{x},\mathbf{u}) \ \} = \mathbf{I}(x_0,u_0) \cap \mathbf{I}(x_1,u_1). 
\end{equation}
Finally, let $\mathbf{\Lambda}$ be the set of all ignition states, 
$$\mathbf{\Lambda} \ident \{ \hat\lambda_0, \hat\lambda_1,\ldots, \hat\lambda_{ \mathrm{card}( \mathfrak{I} ) -1}\} .$$
We will call  this ensemble \emph{ ignition set}. While the maximum value of $\mathrm{card}( \mathfrak{I} )$ is $2^K$, we shall see that only $2K$ ignition states are necessary and thus, this set is strongly redundant for large $K$. A more convenient set $ \mathfrak{I} $ of cardinality just $2K$ is the set of `double-plateau' integers $\mathbf{D}_K$ defined as the integers of $\llbracket 0,2^K-1 \rrbracket$ whose  binary expansion is a chain of identical bits except for a maximum of one jump. For instance,  $\mathbf{D}_3= \{ \overline{ 011 } , \overline{ 111 } , \overline{ 110 } , \overline{ 100 }, \overline{ 000 } , \overline{ 001 }  \} = \{ 3,7,6,4,0,1\}$ and  $\mathbf{D}_4 = \{ \overline{ 0011 } , \overline{ 0111 } , \overline{ 1111 } , \overline{ 1110 }$, $\overline{ 1100 } , \overline{ 1000 } , \overline{ 0000 } , \overline{ 0001 }  \} = \{  3 , 7, 15 ,14, 12, $ $8, 0, 1 \}$. For reasons explained in Sec. \ref{continuous} this particular order in $\mathbf{D}_K$, beginning with $K/2$ zeros  for $K$ even or $(K-1)/2$ zeros for $K$ odd, will be referred to as the \emph{trigonometric order}. For $r \in\llbracket 0,2K-1 \rrbracket$, this order defines a \emph{double-plateau} function $D_K(r)$,
\begin{equation}
\label{biplateau}
D_K \ : \
\llbracket 0,2K-1 \rrbracket\rightarrow\llbracket 0,2^K-1 \rrbracket \ : \
r\mapsto j = D_K(r)
\end{equation}
For example, we have $D_4(0)=3$ or $D_4(6)=0$.

Note that the ignition states $\hat\lambda_j$  have no definite probability for $K >1$. Therefore, in general, they cannot be regarded as hidden variables. Nevertheless, we will define in the next section (Sec. \ref{distributions}) a number of \emph{gauge} probability distributions $\g_k(\hat\lambda_j)$ on  $\mathbf{\Lambda}$, each compatible with $\PP(\mathbf{x},\mathbf{u})$. For instance, for $K=3$, only $6$ ignition states are necessary: The ignition states can be pictured by dice sides, while each probability distribution corresponds to a particular biased die~\cite{mf1}.
\begin{table}[htb]
\begin{center}
\begin{tabular}{||l||c||}
\hline
\hline
 Number of settings     &  $K$  \\  
\hline
\hline
Number of targets $(\mathbf{x}|\mathbf{u})$ & $4K^2$\\
\hline
Number of local targets card$( \mathfrak{T}_\mathrm{loc} )$  & $2K$\\
\hline
Constraints of normalization & $K^2$  \\
\hline
Constraints of local consistency&$2K(K-1)$\\
\hline
Constraints of total correlation & $2K$\\
\hline
Degrees of freedom & $K^2$\\
\hline
\hline
Number of gauge distributions  $\g_k(\lambda_j)$ & $K$\\
\hline
Rank of the gauge linear system & $2K$\\
\hline
\hline
\end{tabular}
\caption{\label {parameter} {\footnotesize Degrees of freedom in general bipartite Bell-type systems with $K$ possible settings. Due to total correlation, the $K^2$ degrees of freedom are shared between the two regions}}
\end{center}
\end{table}
\paragraph{Continuous settings}
The discrete projection function Eq. (\ref{xij}) is not appropriate for continuous settings. We have thus to guess a convenient \emph{projection function} $\Pi(\theta,\lambda) : \mathbf{\Theta}\times\mathbb{R}\rightarrow\mathbf{X}=\{0,1\}$. The fine-grained description for continuous settings is  next defined by: 
\begin{equation}
\label{lambdac}
\lambda\in\mathbb{R}\mapsto\hat\lambda = \bigcap_{\theta\in\mathbf{\Theta}}  \Big(   \Pi(\theta,\lambda) |\theta \Big).
\end{equation} 
We will use such a projection function in Sec. \ref{continuous}. More generally, the structure of the ignition set should reflect the system symmetry. 

\subsubsection{Gauge probability distributions}
\label{distributions}
For ease of exposition and simplicity, we will focus on finite systems with $K \in \mathbb{N}$ different settings.

We have emphasized that the ignition states $\hat\lambda_j$  have no definite probability. Now, we are going to define a number of $K$ parametric probability distributions $\g_k(\lambda_j)$ on the ignition set $\mathbf{\Lambda}$, labelled $k$ from 0 to $K-1$. Of course, each distribution has to be compatible with the target probabilities $\PP(\mathbf{x}|\mathbf{u})$. We regard each target as an union of ignition states. Therefore, the probability of each target $\PP(\mathbf{x}|\mathbf{u})$ will be identified with the sum of the probabilities of its ignition states. There are a number of consistent possibilities and two families of solutions. 

A first family is obtained from region ${\mathcal R}_0$: Let $u_0 = \theta_k$  and let $p_{kj}$ be the unknown probability of $\hat\lambda_j$ in the probability distribution labelled $k$.
For each setting $u_1 \in \mathbf{\Theta}$  and each outcome pair $x_0, x_1$, we write one compatibility condition: 
\begin{equation}
\label{defPkj}
\forall u_1, \forall x_0, \forall x_1:\sum_{j \in \mathbf{J}(\mathbf{x},\mathbf{u}) } p_{kj}= \PP(x_0;x_1| u_0=\theta_k;u_1)
\end{equation}
where $\mathbf{J}(\mathbf{x},\mathbf{u}) $ is defined by Eq.(\ref{Jxu}). From Eq.(\ref{Ixj}), we derive
$$\mathbf{J}(x_0,x_1, u_0,u_1)= \mathbf{I}(x_0,u_0) \ \cap \ \mathbf{I}(x_1,u_1).$$
We have clearly 
$$\forall u_1 \in \mathbf{\Theta} : \mathbf{I}(x_0, u_0)= \mathbf{I}(x_0,u_0) \  \cap \  \big[\  \mathbf{I}(0,u_1) \cup \mathbf{I}(1,u_1)\big]  ,$$ 
and thus, local consistency, Eq.(\ref{localdiscret}), is automatically encoded. Similarly, we have
$$\mathbf{J}(0,1, \theta_k,\theta_k)= \mathbf{I}(0,\theta_k) \ \cap \ \mathbf{I}(1,\theta_k)=\emptyset,$$ and thus, total correlation is also secured.

Accounting for local consistency and total correlation, each parametric probability distribution labelled $k$ is defined by a linear system of $4K-2(K-1)-2=2K$ independent equations and card$(\mathfrak{I})$ unknowns. There is generally a set of positive solutions. A second family of solutions could be obtained similarly from region ${\mathcal R}_1$, by interchanging the indices $0$ and $1$.
Select one solution $p_{kj}$ and define
 
\begin{equation}
\label{defraupkj} 
\g_k(\hat\lambda_j) \ident p_{kj}  
\end{equation}
Clearly, we have for any $u_1$:
$$\sum_{j\in\mathfrak{I}}\g_k(\hat\lambda_j) = \sum_{x_0=0}^1 \sum_{x_1=0}^1 \PP(x_0;x_1| \theta_k;u_1)=1.$$
We will call the function $\g_k(\hat\lambda_j)$ \emph{gauge probability distribution}.

\subsubsection{Classical collapse}
\label{collapse}
We aim to describe the random system relaxation by use of the ignition set $\mathbf{\Lambda}= \{\hat\lambda\}$ and the $K$ probability distributions $\g_k(\hat\lambda)$. The process is the following: In each region, the observer selects his own setting $u_0$ or $u_1$. These settings are transmitted with finite velocity towards a so-called \emph{ignition point} ${\mathfrak I}$. Only \emph{one setting} $u_0$ or $u_1$, labelled $\theta_{\mathrm{igni}} $ is kept, e.g.,  the first received setting. 
Now, at ${\mathfrak I}$, we perform a trial in the ignition set $\mathbf{\Lambda}$ using the only probability distribution $\g_{\mathrm{igni}} (\hat\lambda_j)$ to draw a single ignition state $\hat\lambda_{j_{\mathrm{igni}}} $. Let $ j_{u_0}= \Pi(u_0,\hat\lambda_{j_{\mathrm{igni}} })$ and  $j_{u_1}= \Pi(u_1,\hat\lambda_{j_{\mathrm{igni}} })$. These coefficients, $j_{u_0}$ and $j_{u_1}$, are transmitted with finite velocity towards the end regions $ {\mathcal R}_0$ and $ {\mathcal R}_1$ respectively, where the final outcomes are $x_0 = j_{u_0}=\Pi(u_0,\hat\lambda_{j_{\mathrm{igni}} })$ and $x_1 = j_{u_1}= \Pi(u_1,\hat\lambda_{j_{\mathrm{igni}} })$. 

Conversely, to compute the probability $\PP (\mathbf{x}|\mathbf{u}) $  we have to collect all ignition states $\hat\lambda_j$  within the target $(\mathbf{x}|\mathbf{u}) $:

\begin{equation}
\label{defPkj2}
\PP(x_0; x_1|u_0; u_1) = \sum_{\lambda_j \in (\mathbf{x}|\mathbf{u})} \g_{\mathrm{igni}} (\hat\lambda_j)
\end{equation}
where the label of $\g_{\mathrm{igni}}$ is such that $\theta_{\mathrm{igni}}$ is any one of the two settings, $u_0$ or $u_1$, and $\lambda_j \in (x_0;x_1| u_0;u_1)$ means  that $x_0=\Pi(u_0,\hat\lambda_j)$ and $x_1=\Pi(u_1,\hat\lambda_j)$.

Now the proof is a straightforward verification of the identity of Eq.(\ref{defPkj}) and Eq.(\ref{defPkj2}). By construction, Eq.(\ref{correlation}) is automatically satisfied and the choice of  $u_0$ or $u_1$ for $\theta_{\mathrm{igni}}$ to select the distribution $\g_{\mathrm{igni}}$ at the ignition point ${\mathfrak I}$ \emph{does not affect} the overall probability $\PP(x_0; x_1|u_0; u_1) $. 

\paragraph{Ignition front and causal horizon}
Note first that the process implies the existence of an ignition point located between the two end regions. Then, the locus of all possible ignition points draws a form of frontier between ${\mathcal R}_0$ and ${\mathcal R}_1$, whereas no transfrontier communication is possible between the two regions. Such an \emph{ignition front} is not recognized in quantum mechanical collapse since this process is generally assumed to be superluminal and non local~\cite{pearle}, but it could be \emph{light cones}, \emph{light-sheets}~\cite{bousso} or \emph{null surfaces}. In black holes physics, it looks like the \emph{event horizon}. We will thus identify this ignition front with a \emph{causal horizon}~\cite{jacobson}.

\paragraph{Gauge probability distributions}  
There are two possible probability distributions for the trial at ${\mathfrak I}$, defined by the settings of the two end regions. The real process, if any, is not observable and the choice between these two distributions is indifferent. Therefore,  the probability distributions $\g_k(\hat\lambda)$ behave like global \emph{gauge parameters}: The choice of one particular distribution correspond to a gauge selection at ${\mathfrak I}$ and this choice \emph{does not} affect the observable result. This \emph{gauge invariance} is fundamental. In our opinion, it is the core of the EPR paradox. We will call \emph{stochastic gauge system} the ensemble of the ignition set $\mathbf{\Lambda}=\{\hat\lambda\}$ and a number of parametric distributions $\g_k(\hat\lambda)$. 
\paragraph{Analogy with the holographic hypothesis}
The parametric trial is performed on the ignition front, i.e., on the causal horizon and not in regions ${\mathcal R}_0$ and ${\mathcal R}_1$. This meets a suggestion by `t Hooft and Susskind in black hole physics, namely, the \emph{holographic hypothesis}~\cite{hooft, susskind, bousso}, that space of quantum states in a region must be associated with the two-dimensional boundary rather than the volume of the system. Note that this assumption might be checked by experiments on entangled entities more easily than in black holes.

\paragraph{Consistency conditions} Due to gauge invariance, the parametric probability distributions $\g_k(\lambda_j)$ are not independent~\cite{mf}. Suppose that the observers ${\mathcal O}_0$ and ${\mathcal O}_1$  select respectively the settings $u_0$ and $u_1 \in \mathbf{\Theta}$. Let $\g_{u_0}$ and $\g_{u_1}$  be the gauge distributions associated with $u_0$ and $u_1$ respectively. For each run, the ignition state can be obtained by a parametric trial on $\mathbf{\Lambda}$ with respect to either $\g_{u_0}$ or $\g_{u_1}$. Then, Eq.(\ref{localdiscret}) implies that for $i=0,1$ and $x=0$ or $1$,

\begin{equation} 
\label{compatibility1}
\PPr(x_i|u_i) 
=\sum_{\hat\lambda_j\in  (x_i|u_0)}\g_{u_0}(\hat\lambda_j) 
=\sum_{\hat\lambda_j\in  (x_i|u_1)}\g_{u_1}(\hat\lambda_j)
\end{equation}
Clearly, these conditions mean that $\PPr(x_i|u_i)$ is well defined, i.e., that gauges distributions are compatible with local consistency.
Conversely, when the \emph{consistency conditions}, Eq.(\ref{compatibility1}), are fulfilled, using Eq.(\ref{defPkj2}), the probability distributions $\g_k(\lambda_j)$ define a single random system $\PP(x_0;x_1| u_0;u_1)$. An alternative formulation with expectation values reads,
\begin{equation} 
\label{compatibility2}
\EE(x_i|u_i) 
=\sum_{\hat\lambda_j\in (1|u_0)}\g_{u_0}(\hat\lambda_j) 
=\sum_{\hat\lambda_j\in (1|u_1)}\g_{u_1}(\hat\lambda_j).
\end{equation}

We will give some examples, but beforehand, let us summarize our results so far: 

\begin{summary*}[Emulation of Bell-type systems]
Any Bell-type system can be emulated by a \emph{stochastic gauge system} composed of an auxiliary {outcome set} $\mathbf{\Lambda}=\{ \hat\lambda_j \}$, called \emph{ignition set}, and a set of \emph{gauge distributions} $\g_k(\hat\lambda_j)$. One distribution $\g_k$ is associated with each setting $\theta_k$. A setting $u_i$ is freely selected in each end region $\mathcal{R}_i$ and transmitted with finite velocity towards a so-called \emph{ignition point} $\mathcal{I}$, where each setting is associated with its gauge distribution. Only one of the two distributions, $\g_{\mathrm{igni}}$, is used at $\mathcal{I}$. This choice is called \emph{gauge selection}.   A single trial is performed at $\mathcal{I}$ to draw a unique \emph{ignition state} $\hat\lambda_j$. This result is transmitted with finite velocity backwards the end regions. The final outcomes $x_i$ are next  computed by a projection function $\Pi( u_i, \hat\lambda_j)$. Gauge selection at $\mathcal{I}$ does not affect the probability of the final outcomes. The ignition point is located at the boundary of the two regions. Therefore, if the two regions are separated by a causal horizon, the gauge probability system is located on the horizon.
\end{summary*}

In classical systems, we will call this random process \emph{classical collapse}.

\subsubsection{General 2-setting totally correlated systems}
Consider first a Bell-type system with just $K=2$ settings, $\theta_0$ and $\theta_1$. When accounting for local consistency and total correlation, it is easily shown that the general system depends on 4 parameters, $q_i, i=1,2,3,4$ (Table \ref{bell2}), subjected to the following constraints: $0 \le q_i \le 1$; $q_3 \ge q_1,q_2$; $q_4 \ge q_1,q_2$; and $q_1+q_2 \ge q_3,q_4$. The linear system Eq.(\ref{defPkj}) admits $2K=4$ unknowns for $4$ independent equations. The gauge distributions $\g_{k}(\lambda_j)$  are given in Table \ref{jbell2}.  Suppose that the observers choose two different settings, e.g., $u_0=\theta_0$ and $u_1 =\theta_1$. Then, at the ignition point ${\mathfrak I}$, any distribution of the working set can be selected. The trial with respect to this particular distribution gives an ignition state $\lambda_j$. Finally, since $u_0=\theta_0$, we get $x_0 = 0$ or $1$ depending on whether $j$ is even or odd. Similarly, $x_1 = 0$ or $1$ according to $j < 2$ or $j \ge 2$.
Gauge invariance means that all statistical properties of $x_0$ and $x_1$ do not depend on the gauge selection, because they can be computed directly from $\PP(x_0;x_1|u_0;u_1)$.
The entanglement is detected by the CHSH inequality Eq.(\ref{originalCHSH}) only\footnote{In Sec. \ref{inequalities} we will define a more efficient witness derived from the entanglement entropy.}  when $q_3\not= q_4$. Suppose that $q_4>q_3$ and select $A = \theta_0$,  $A' = \theta_1$,  $B = \theta_1$,  and $B' = \theta_0$. Then we have, 
$$\s(\theta_{0}, \theta_{1}) + \s(\theta_{0}, \theta_{0}) + \s(\theta_{1}, \theta_{1}) - \s(\theta_{1}, \theta_{0}) = -2-4(q_4-q_3) < -2.$$
When $q_3=q_4$, the two distributions $\g_0(j)$ and $\g_1(j)$ are identical. The concept of  gauge distribution  degenerates and thus each ignition state $\hat\lambda_j$ has a definite probability, $\g(j)$, whatever the settings. Therefore, the ignition states can be regarded in this case as random \emph{hidden variables} located at the ignition point. 

\begin{table}[htb]
$$
\begin{array}{||c||c|c|c|c||}
\hline
\hline
(x_0, x_1) & (\theta_0,\theta_0 )& (\theta_1,\theta_1)&(\theta_0,\theta_1)&  (\theta_1,\theta_0)  \\  
\hline
\hline
(0, 0) & 1-q_1&1-q_2&1-q_3&1-q_4 \\
\hline
(0, 1)& 0&0&q_3-q_1&q_4-q_2 \\
\hline
(1, 0)& 0 & 0& q_3-q_2&q_4-q_1 \\
\hline
(1, 1)& q_1&q_2&q_1+q_2-q_3& q_1+q_2-q_4 \\

\hline
\hline
\dd(u_0,u_1)& 0&0&2q_3-q_1-q_2&2q_4-q_1-q_2\\
\hline
\s(u_0,u_1)& -1&-1&4q_3-2q_1-2q_2-1&4q_4-2q_1-2q_2-1\\
\hline
\hline
\end{array}
$$
\caption{\label {bell2} {\footnotesize General \emph{locally consistent} \emph{totally correlated} Bell-type system with two possible settings $\theta_0$ and $\theta_1$ depending on 4 parameters $q_1, q_2, q_3$ and $q_4$. The table gives the conditional probabilities $\PP(x_0; x_1|u_0; u_1)$ as well as the Hamming divergence $\dd(u_0, u_1)$ and the mean product spin $\s(u_0,u_1)$ between binary strings.}}
\end{table}

\begin{table}[htb]
{
$$
\begin{array}{||c||c|c||}
\hline
\hline
j &  \g_{ 0 }(j) &  \g_{ 1 }(j)\\
\hline
\hline
{ 0 } & 1-q_3 & 1-q_4\\
\hline
{ 1 } & q_3 - q_2 & q_4 - q_2\\
\hline
{ 2 } &  q_3 - q_1 & q_4 - q_1\\
\hline
{ 3 } & q_1 + q_2 - q_3 & q_1 + q_2 - q_4\\
\hline
\hline
\end{array}
$$
} 
\caption{\label {jbell2} {\footnotesize Gauge probability distributions corresponding to Table \ref{bell2}.}}
\end{table}

\subsection{Simulation of the EPR spin experiment}
\label{EPR}
We are going to classically simulate conventional EPR-B experiments, i.e., Bell-type setups for a pair of totally correlated spins with particular probabilities computed from quantum mechanics. 
In 2D-configuration, let $\mathbf{\Theta}$ be the unit circle $\mathbf{U}(1)=[0, 2\pi]$. From quantum mechanics we know that for any $\theta_a, \theta_b \in \mathbf{\Theta}$ the target probabilities are given by the following equations and Table (\ref{rhotheta}). 

\begin{equation}
\PP(0; 1|\theta_a; \theta_b) = \PP(1; 0|\theta_a; \theta_b) = (1/4)[1-\cos(\theta_a-\theta_b )]. 
\end{equation} 
\begin{equation}
\PP(0; 0|\theta_a; \theta_b) = \PP(1; 1|\theta_a; \theta_b) = (1/4)[1+\cos(\theta_a-\theta_b ) ].
\end{equation} 
Clearly, the system is locally consistent and totally correlated. In addition the two entities are symmetrical and therefore the two families of solutions are identical.

\begin{table}[htb]
$$
\begin{array}{||c||c|c||}
\hline
\hline
\quad \ (u_0, u_1)\rightarrow & (\theta_a,\theta_a)  & (\theta_a,\theta_b)  \\
(x_0, x_1) & (\theta_b,\theta_b)  & (\theta_b,\theta_a)   \\
\downarrow & &  \\

\hline
\hline
(0, 0) & 1/2 & (1/4)(1+\cos\theta) \\
\hline
(0, 1)& 0 & (1/4)(1-\cos\theta) \\
\hline
(1, 0)& 0 &  (1/4) (1-\cos\theta) \\
\hline
(1, 1)& 1/2 & (1/4)(1+\cos\theta) \\

\hline
\hline
\dd(u_0,u_1)& 0 & (1/2)  (1-\cos\theta) \\
\hline
\hline
\end{array}
$$
\caption{\label {rhotheta} {\footnotesize Conditional probability $\PP(x_0;x_1|u_0; u_1)$ of a typical EPR-B system for two different settings $\theta_a$ and $\theta_b$ with $\theta=\theta_a-\theta_b$.  }  }
\end{table}

\subsubsection[Two settings]{2-setting EPR-B experiment}
Consider first a 2-setting EPR-B experiment: Let $\theta_0$ and $\theta_1$ be the polarization angles and $\theta_{01}=\theta_1-\theta_0$. This is obtained with $q_1=q_2=1/2$ and $q_3=q_4= (1/4) (3-\cos \theta_{01})$ in Table \ref{bell2}. Then Table \ref{jbell2} reduces to Table \ref{T2symb}. The two gauge distributions are identical and the ignition states can be viewed as hidden variables with a definite probability. Such 2-setting EPR-pairs are referred to as `Bell's states' when $\theta_{01} = \pi/2$.

\begin{table}[htb]
{
$$
\begin{array}{||c||c||}
\hline
\hline
j & 4 \times \g_{ 0 } (j) = 4 \times \g_{ 1 }(j) \\
\hline
\hline
 { 0 } & 1 + \cos\theta_{01} \\
\hline
 { 1 } & 1 - \cos\theta_{01} \\
\hline
{ 2 } & 1 - \cos\theta_{01} \\
\hline
{ 3 } & 1 + \cos\theta_{01} \\
\hline
\hline
\end{array}
$$
} 
\caption{\label{T2symb}
{\footnotesize Gauge probability distributions for a symmetric EPRB system with 2 settings, $\theta_0$ and $\theta_1$. (We have $\theta_{01}=\theta_1-\theta_0)$. The two gauge distributions are identical}
}
\end{table}

\subsubsection[Three settings and Bell's inequality violation]{3-setting EPR-B experiment}
We now consider an EPR-B system with $K=3$ different settings, leading to a violation of Bell's inequalities Eq.(\ref{triangle}). 

The linear system Eq.(\ref{defPkj}) admits $2^3 =8$ unknowns and $2K = 6$ independent equations: Each probability distribution depends on 2 arbitrary parameters and therefore the general solution depends on 6 arbitrary parameters. We can take advantage of this degeneracy to cancel the gauge probability of two ignition states. It is possible to use the 6 double-plateau indices $\mathbf{D}_3$, 
$$\mathbf{D}_3=\{ j \} = \{ 3,7,6,4,0,1\}.$$ 
The linear system admits then only one solution provided that all gauge probabilities $p_{kj}$ are non negative. 

Let $\theta_0, \theta_1$ and $\theta_2$ be the three settings with $0\le\theta_0\le\theta_1\le\theta_2 <\pi$. Let $\theta_{ab} = \theta_b-\theta_a$. The solutions of Eq.(\ref{defPkj}) define a working set of effective gauge ignition states given by Table~\ref{T3symb}. The maximum violation of Bell's inequality,  Eq.(\ref{triangle}) or (\ref{originalbell}), is obtained, e.g.,  for $\theta_0 =0, \theta_1= \pi/5$ and $\theta_2= \pi/2$. This could be checked by a computer simulation using Table \ref{T3num}.

\begin{table}[htb]
{
$$
\begin{array}{||c||c|c|c||}
\hline
\hline
j & 4 \times \g_{ 0 } (j) & 4 \times \g_{ 1 } (j) & 4 \times \g_{ 2 }(j) \\
\hline
\hline
{ 0 } & 1+\cos\theta_{02} & \cos\theta_{01}+\cos\theta_{21} & 1+\cos\theta_{02}\\
\hline
{ 1 } & 1-\cos\theta_{01} & 1-\cos\theta_{01} & \cos\theta_{12}-\cos\theta_{02}\\
\hline
{ 3 } & \cos\theta_{01}-\cos\theta_{02} & 1-\cos\theta_{12} & 1-\cos\theta_{12}\\
\hline
{ 4 } & \cos\theta_{01}-\cos\theta_{02} & 1-\cos\theta_{12} & 1-\cos\theta_{12}\\
\hline
{ 6 } & 1-\cos\theta_{01} & 1-\cos\theta_{01} & \cos\theta_{12}-\cos\theta_{02}\\
\hline
{ 7 } & 1+\cos\theta_{02} & \cos\theta_{01}+\cos\theta_{21} & 1+\cos\theta_{02}\\
\hline
\hline
\end{array}
$$
} 
\caption{\label{T3symb}
{\footnotesize Working set of gauge probability distributions using only 6 ignition states (out of $8$), $j\in \mathbf{D}_3=\{ 0, 1, 3, 4, 6, 7\}$ for 3 settings. This working set is only valid when all entries are positive.}
}
\end{table}

\begin{table}[htb]
{
$$
\begin{array}{||c||c|c|c||}
\hline
\hline
j &  \g_{ 0 }(j)  &  \g_{ 1 }(j)  &  \g_{ 2 }(j) \\
\hline
\hline
{ 0 } &  0.250 &  0.349 &  0.250\\
\hline
{ 1 } &  0.048 &  0.048 &  0.147\\
\hline
{ 3 } &  0.202 &  0.103 &  0.103\\
\hline
{ 4 } &  0.202 &  0.103 &  0.103\\
\hline
{ 6 } &  0.048 &  0.048 &  0.147\\
\hline
{ 7 } &  0.250 &  0.349 &  0.250\\
\hline
\hline
\end{array}
$$
} 
\caption{\label{T3num}
{\footnotesize Gauge probability distribution with 6 ignition states derived from Table \ref{T3symb}, for the 3 settings: $\theta_ 0  =  0 ; \theta_ 1  =  \pi/5 ; \theta_ 2  =  \pi/2 $ leading to the maximum violation of Bell's inequality, Eq.(\ref{triangle}) or (\ref{originalbell}). }
}
\end{table}

\subsubsection[Four settings and CHSH inequality violation]{4-setting EPR-B experiment}

We now consider an EPR-B system with $K=4$ different settings, leading to a violation of CHSH inequalities Eq.( \ref{originalCHSH}). We proceed similarly as for 3 settings.

Let $\theta_0, \theta_1, \theta_2$ and $\theta_3$ be the four settings. A violation of CHSH inequality is obtained, e.g.,  for $\theta_ 0  =  0 , \theta_ 1  =  \pi/8 , \theta_ 2  =  \pi /4$ and $\theta_3= 3\pi/8$. The maximum violation is obtained, e.g., for $\theta_ 0  =  0 , \theta_ 1  =  \pi/4 , \theta_ 2  =  \pi /2$ and $\theta_3= 3\pi/4$. 

The linear system Eq.(\ref{defPkj}) admits $2^4 =16$ unknowns and $2K = 8$ independent equations: Each probability distribution depends on 8 arbitrary parameters and the general solution depends on 32 arbitrary parameters. We will take advantage of this degeneracy to force to zero the gauge probability of 8 ignition states. It is possible to keep the double-plateau indices,
$$\mathbf{D}_4=\{ j \} = \{  3,7, 15,14,12, 8,0, 1 \}.$$ 
Let again $\theta_{ab} = \theta_b-\theta_a$. The linear system admits then only one solution given in Table \ref{T4symb}, which turns out to be positive in the range of interest (Table \ref{T4num}).

\begin{table}[htb]
{
$$
\begin{array}{||c||c|c|c|c||}
\hline
\hline
j & 4 \times \g_{ 0 } (j) & 4 \times \g_{ 1 } (j) & 4 \times \g_{ 2 } (j) & 4 \times \g_{ 3 }(j) \\
\hline
\hline
 { 0 } & 1+\cos\theta_{30} & \cos\theta_{10} + \cos\theta_{13} & \cos\theta_{20} + \cos\theta_{23} & 1+\cos\theta_{30}\\
\hline
{ 1 } & 1-\cos\theta_{10} & 1-\cos\theta_{10} & \cos\theta_{21} - \cos\theta_{20} & \cos\theta_{31} - \cos\theta_{30}\\
\hline
{ 3 } & \cos\theta_{10} - \cos\theta_{20} & 1-\cos\theta_{12} & 1-\cos\theta_{12} & \cos\theta_{32} - \cos\theta_{31}\\
\hline
{ 7 } & \cos\theta_{20} - \cos\theta_{30} & \cos\theta_{21} - \cos\theta_{31} & 1-\cos\theta_{23} & 1-\cos\theta_{23}\\
\hline
{ 8 } & \cos\theta_{20} - \cos\theta_{30} & \cos\theta_{21} - \cos\theta_{31} & 1-\cos\theta_{23} & 1-\cos\theta_{23}\\
\hline
{ 12 } & \cos\theta_{10} - \cos\theta_{20} & 1-\cos\theta_{12} & 1-\cos\theta_{12} & \cos\theta_{32} - \cos\theta_{31}\\
\hline
{ 14 } & 1-\cos\theta_{10} & 1-\cos\theta_{10} & \cos\theta_{21} - \cos\theta_{20} & \cos\theta_{31} - \cos\theta_{30}\\
\hline
 { 15 } & 1+\cos\theta_{30} & \cos\theta_{10} + \cos\theta_{13} & \cos\theta_{20} + \cos\theta_{23} & 1+\cos\theta_{30}\\
\hline
\hline
\end{array}
$$
} 
\caption{\label{T4symb}
{\footnotesize Working set of gauge probability distributions using only  8 ignition states (out of $16$), 
$j\in  \mathbf{D}_4=\{ 0, 1, 3, 7, 8, 12, 14, 15\}$, for 4 settings. This working set is only valid when all entries are positive.}
}
\end{table}

\begin{table}[htb]
{
$$
\begin{array}{||c||c|c|c|c||}
\hline
\hline
j &  \g_{ 0 } (j) &  \g_{ 1 } (j) &  \g_{ 2 } (j) &  \g_{ 3 }(j) \\
\hline
\hline
{ 0 } &  0.073 &  0.177 &  0.177 &  0.073\\
\hline
{ 1 } &  0.073 &  0.073 &  0.177 &  0.177\\
\hline
{ 3 } &  0.177 &  0.073 &  0.073 &  0.177\\
\hline
{ 7 } &  0.177 &  0.177 &  0.073 &  0.073\\
\hline
{ 8 } &  0.177 &  0.177 &  0.073 &  0.073\\
\hline
{ 12 } &  0.177 &  0.073 &  0.073 &  0.177\\
\hline
{ 14 } &  0.073 &  0.073 &  0.177 &  0.177\\
\hline
{ 15 } &  0.073 &  0.177 &  0.177 &  0.073\\
\hline
\hline
\end{array}
$$
} 
\caption{\label{T4num}
{\footnotesize Gauge probability distributions derived from Table \ref{T4symb}, for the four settings: $\theta_ 0  =  0 ; \theta_ 1  =  \pi/4 ; \theta_ 2  =  \pi /2; \theta_ 3  =  3\pi/4.$ This setup leads to the maximum violation of the CHSH inequalities, Eq. (\ref{originalCHSH}).}
}
\end{table}

\subsubsection[Regular settings]{Discrete regular $K$-setting EPR-B experiment}
\label{regular}
Finally consider a discrete EPR-B system with $K$ regular settings 
$$k\in\llbracket 0, K-1\rrbracket \ : \ k\mapsto\theta_ k  =  \frac{k\pi}{K}.$$ 
We proceed similarly as above. The rank of the system is $2K$ and the relevant ignition states $\lambda_j$ are defined by the double-plateau function $j=D_K(r)$, Eq.(\ref{biplateau}), with $ r\in\llbracket 0, 2K-1\rrbracket $. If we use the same projection function $\Pi(k,j),$ Eq.(\ref{lambdaj}), we have,
$$\Pi[k,D_K(r)] = \Pi[0, D_K(r-k)] \quad ; \quad \g_k[D_K(r)]=\g_0[D_K(r-k)]$$ 
where $r-k \ge 0$ is computed modulo $2K$. 

An alternative formulation is to define a number of $2K$ ignition states, $$\lambda_{r}^D\ident \lambda_{D_K(r)},$$ a new projection function
$$ \Pi^D(k,r)\ident \Pi[k,D_K(r)]$$ and then
$$ \g_k^D(r)\ident \g_k[D_K(r)]$$
The functions, 
$$ \g(r)\ident \g_0^D(r),  \quad \mathrm{and} \quad \Pi(r)\ident \Pi^D(0,r) $$
with  $ r\in\llbracket 0, 2K-1\rrbracket$ are sufficient to describe all distributions since
\begin{equation}
\label{eqregular}
\g_k^D(r) = \g(r-k) \quad \mathrm{and} \quad \Pi^D(k,r)=\Pi(r-k)
\end{equation}
where $r-k\ge 0$ is computed modulo $2K$.  It is a simple exercise to derive the following expression, 

\begin{equation}
\g(r)= \frac{1}{2} \sin\alpha \times
\begin{cases}
|\cos (2r+1) \alpha\ | & K \equiv 0\pmod{2}\\
|\cos 2r\alpha\ | & K \equiv 1\pmod{2}
\end{cases}
\end{equation}
where $\alpha = \pi/(2K)$, and,
\begin{equation}
\Pi(r)= 
\begin{cases}
0 & \mathrm{if} \ 2r\le K \ \mathrm{(computed~modulo~} {2K} )\\
1&  \mathrm{if} \ 2r  > K \ \mathrm{(computed~modulo~} {2K} )\\

\end{cases}
\end{equation}
(where the residues modulo $2K$ are taken into $\llbracket 0, 2K-1\rrbracket$).
For instance, Table \ref{T5symb} gives the result for $K=5$.
For large values of $K$,  regular settings approach  continuous settings which deserve a special derivation.

\begin{table}[htb]
{
$$
\begin{array}{||c||c|c|c|c|c||}
\hline
\hline
r & \g_ 0^D  (r) : M  & \g_ 1^D (r) : M  & \g_ 2^D  (r) : M  & \g_ 3^D  (r) : M  & \g_ 4^D  (r) : M \\
\hline
\hline
{ 0 } &    1      & \cos 2\alpha & \cos 4\alpha & \cos 4\alpha & \cos 2\alpha\\
\hline
{ 1 } & \cos 2\alpha &    1      & \cos 2\alpha & \cos 4\alpha & \cos 4\alpha\\
\hline
{ 2 } & \cos 4\alpha & \cos 2\alpha &    1      & \cos 2\alpha & \cos 4\alpha\\
\hline
{ 3 } & \cos 4\alpha & \cos 4\alpha & \cos 2\alpha &    1      & \cos 2\alpha\\
\hline
{ 4 } & \cos 2\alpha & \cos 4\alpha & \cos 4\alpha & \cos 2\alpha &    1     \\
\hline
{ 5 } &    1      & \cos 2\alpha & \cos 4\alpha & \cos 4\alpha & \cos 2\alpha\\
\hline
{ 6 } & \cos 2\alpha &    1      & \cos 2\alpha & \cos 4\alpha & \cos 4\alpha\\
\hline
{ 7 } & \cos 4\alpha & \cos 2\alpha &    1      & \cos 2\alpha & \cos 4\alpha\\
\hline
{ 8 } & \cos 4\alpha & \cos 4\alpha & \cos 2\alpha &    1      & \cos 2\alpha\\
\hline
{ 9 } & \cos 2\alpha & \cos 4\alpha & \cos 4\alpha & \cos 2\alpha &    1     \\
\hline
\hline
\end{array}
$$
} 
\caption{\label{T5symb}
{\footnotesize Gauge distributions $\g_k^D (r)/M$ for 5 settings $M=(1/2)\sin\alpha \ ;\ \alpha =\pi / 10 $}
}
\end{table}

\subsubsection[Continuous settings]{Continuous 2D-EPR-B experiment}
\label{continuous}
For a continuous ensemble of settings, we have to construct a continuous ignition set (i.e., of cardinality $\aleph_1$) and thus to define a convenient projection function. Next we will derive a continuous set of gauge probability distributions.

\paragraph{Ignition set}
For any setting labelled $\theta$ let $(\xi|\theta)$ be the local target  $\xi \in \mathbf{X}$, given $ \theta \in \mathbf{\Theta}$. For any real $\lambda$, define  the ignition state $ \mathbf{\hat\lambda} $ by the  intersection of local targets:
\begin{equation}
\label{xi}
 \mathbf{\hat\lambda} = \bigcap_{\theta \in [0,\pi]} \{ (  \Pi(\theta,\lambda)  |\theta )\}  \in \mathbf{\Lambda},
\end{equation} 
where according to Eq.(\ref{Ixj}), the binary digit $\xi \in \mathbf{X}$  is computed by a projection function $\Pi(\theta, \lambda)$. It is convenient to define the following dyadic square wave function of $\theta$ and $\lambda$ guessed by analogy with Eq.(\ref{eqregular}):
\begin{equation}
\label{picontinue}
\Pi(\theta,\lambda)=\frac{1}{2}[1+\sgn\cos(\theta-\lambda)].
\end{equation}
Then, we will replace $\hat\lambda$ by $\lambda$ when necessary and we will use a subset of $\mathbf{\Lambda}$ defined by Eq.(\ref{xi}) and corresponding to $0\le\lambda\le 2\pi$. This leads to a single solution. When compared with the regular discrete settings, we find a close similarity. This justifies the name of \emph{trigonometric order} for the double-plateau functions in Sec. \ref{discretesettings}.

\paragraph{Gauge probability distributions} 
For each $\theta\in\mathbf{\Theta}$ we are going to derive a gauge distribution density, $\g_\theta(\lambda)\dd\lambda$. Define the interval 
$$\mathbf{I}(x, \theta)= \{ \lambda \  | \  \Pi(\theta,\lambda)=x\} $$
and let $\mathbf{J}(x_0,x_1, u_0,u_1)= \mathbf{I}(x_0, u_0)\bigcap \mathbf{I}(x_1, u_1)$.
To compute the solution, let $u_0=\theta$. For each setting $u_1 \in \mathbf{\Theta}$  and each outcome pair $x_0, x_1$,  we have the functional equation: 
\begin{equation}
\label{defPkjEPR}
\forall u_1, \forall x_0, \forall x_1:\int_{\mathbf{J}(x_0,x_1,u_0,u_1)} \g_\theta(\mathbf{\hat\lambda})\dd\lambda = \PP(x_0;x_1| u_0=\theta;u_1)
\end{equation}
where the unknown function is $\g_\theta(\mathbf{\hat\lambda})$ or simply, $\g_\theta(\mathbf{\lambda})$. The solution has to be computed for each interval. For instance, for $u_1>\theta$, we have $\mathbf{J}(1,1, \theta,u_1)= [u_1-\pi/2,\theta+\pi/2]$ and $\PP(1;1|\theta; u_1)= (1/4) [1+\cos(\theta-u_1)]$. Finally, the result is the following working set of gauge distributions~\cite{mf}: 
\begin{equation}
\label{gcontinue}
\g_ {\theta}(\lambda) = (1/4) |\cos(\theta-\lambda)|
\end{equation} 
Eqs. (\ref{picontinue}) and (\ref{gcontinue}) are closely related with Eq. (\ref{eqregular}).
Now the classical collapse is described by the process of Sec. \ref{collapse} with some straightforward changes. Clearly, this model exactly emulates the 2D-conventional EPR-B experiment and leads to a violation of Bell's inequalities.


\section{Simulation of multipartite entangled systems}
We will now extend the above results beyond the particular case of Bell-type system to general systems with any number of regions. For simplicity, we will only deal with dichotomic outcomes. This means that we only consider an ensemble of qubits when the system mimics a quantum mechanical situation. In any case, the concept of local consistency remains the key property of classical analogues of quantum systems. This condition allows only nonsignaling correlation in the formulation of Popescu and Rohrlich~\cite{popescu}.

\subsection {Classical analogues of quantum systems}
Consider an extended object in equilibrium composed of $n$ \emph{correlated} entities 
$\{{\mathfrak E}_0, {\mathfrak E}_1,\dots, {\mathfrak E}_{n-1}\} $,  located in $n$ space regions ${\mathcal R}_0$, ${\mathcal R}_1$,\dots, ${\mathcal R}_{n-1}$. The equilibrium can be broken and the future outcome $x_i$ will depend randomly upon both the symmetry of the object and the free choice $u_i$ of the observer. 
When $u_i=\theta_k$, it is convenient to give a name to the pair $(i,k)$ of one setting in one region:

\begin{definition*} [Configuration]
We will call \emph{configuration} a pair $(i,k)$ of one region $\mathcal{R}_i$ and one setting $\theta_k$. Each configuration is uniquely labelled by an index $\gamma=k+iK \ ,\ \gamma\in\llbracket 0,nK-1\rrbracket$.
\end{definition*}
Let $ \mathbf{x} $ be the outcome vector $(x_0;x_1;\dots;x_{n-1})$ and $\mathbf{u}$ the setting vector $(u_0;u_1,\dots;u_{n-1})$. We will name again \emph{local target} the entry $(x_i|u_i)$ and \emph{global target} or simply \emph{target} the pair $ (\mathbf{x}|\mathbf{u}) $. Let $ \mathfrak{T}= \{  (\mathbf{x}|\mathbf{u})  \}$ be the set of all targets. Depending upon the symmetry of the object, each target has a given conditional probability $\PP(\mathbf{x}|\mathbf{u})$. There are card$(\mathfrak{T}) = 2^n \times K^n =(2K)^n$ distinct targets, subjected to $K^n$ normalization constraints, namely:
$$\sum_{x_0=0}^{1 } \dots \sum_{x_{n-1}=0}^{1 } \PP(x_0;x_1;\dots ;x_{n-1}|u_0;u_1;\dots ;u_{n-1}) =1$$

When the system mimics a quantum situation, each setting-vector $\mathbf{u}$ can be regarded as an orthonormal basis of a Hilbert space $\mathcal{H}$  whose unit vectors are labelled by the  $2^n$ outcome-vectors $\mathbf{x}$. Therefore, each target $ (\mathbf{x}|\mathbf{u}) $ may be identified with a unit ket  $ \ket{\mathbf{x},\mathbf{u} } $ of the Hilbert space. When the system is in the reduced state $\rho$, where $\rho$ is a density operator, the target probability is computed as 
$$\PP (\mathbf{x},\mathbf{u})=\Tr (\rho \ket{\mathbf{x},\mathbf{u}} \bra{\mathbf{x},\mathbf{u}}) = \bra{\mathbf{x},\mathbf{u}} \rho \ket{\mathbf{x},\mathbf{u}} $$ 

More generally, given an extended object in equilibrium defined by a probability distribution $\PP(\mathbf{x}|\mathbf{u})$, we aim to describe classically a break of equilibrium of the system. Beforehand, we will clarify the concepts of local consistency in this context.

\subsubsection[Local consistency]{Nonsignaling correlations, complete local consistency and degrees of entanglement}
We no more necessarily require total correlation, i.e., equal outcomes for identical settings in different regions, and accept any degree of entanglement.  By contrast, we do conserve \emph{local consistency} in order to allow independent partial measurements. This means that a definite local probability holds in each region, irrespective of the settings in the other regions.  More generally, we will define the concept of \emph{complete local consistency} when it is possible to derive consistent subsystems in  $n-r$ regions simply by ignoring $r$ regions. Again, this condition only allows nonsignaling correlations between any pair of regions. A limit case concerns independent regions, when the probability of the whole system is simply the product of the probabilities of each region. Then, the system is separable and the entities are no more correlated.

\begin{definition*} [Complete local consistency]
A  set of $n$ \emph{entangled} classical entities 
$\{{\mathfrak E}_0, {\mathfrak E}_1$, $\dots$, ${\mathfrak E}_{n-1}\} $,   is \emph{completely local-consistent} when the measurement in one or several regions does not affect the marginal probabilities in the other regions.
\end{definition*}
For any partition of $\{ 0, 1,\dots , n-1 \} $ into two subsets, say (after reordering the indices if necessary), $\{ 0, 1,\dots , r-1 \} $ and $\{ r, r+1,\dots , n-1 \} $ with $1\le r \le n-1$, irrespective of $u_r,u_{r+1},\dots,u_{n-1}$, 
we have 
\begin{equation}
\label{completeConsis}
\PPr(x_0;\dots ;x_{r-1}|u_0;\dots ;u_{r-1}) =\\ 
\sum_{x_r=0}^{1 } \dots \sum_{x_{n-1}=0}^{1 } \PP(x_0;\dots ;x_{n-1}|u_0;\dots ; u_{n-1}). 
\end{equation}
In words, $\PPr(x_0;\dots ;x_{r-1}|u_0;\dots ;u_{r-1})$ only depends on the local settings, $(u_0;\dots ;u_{r-1})$. Thus,  the free choices of $u_r,u_{r+1},\dots,u_{n-1}$ and the subsequent measurements of $x_r,\dots,x_{n-1}$ do not affect the probabilities in regions $ \mathcal{R}_0$,  $\mathcal{R}_1, \dots, \mathcal{R}_{r-1} $, as far as these choices and the subsequent outcomes are ignored. It will be convenient to name \emph{partial target} the entry $(x_0;\dots ;x_{r-1}|u_0;\dots ;u_{r-1})$. Therefore, \emph{when complete local consistency holds, any \emph{partial target} has a definite probability. } Note that  when the $n$-region system mimics a quantum situation, Eq.~(\ref{completeConsis}) describes the partial state in the subsystem $\{ \mathcal{R}_0,\dots, \mathcal{R}_{r-1}\} $, obtained by  partial tracing over the subsystem $\{ \mathcal{R}_r,\dots, \mathcal{R}_{n-1}\} $,

Conversely, it is possible to increase the  number of regions simply by adding  successively new regions. This is similar to the process of quantum purification. Any subset of regions is consistent and the global and local viewpoints are compatible. Then, a multipartite system can be decomposed into its subsets without local perturbation.
Note that computation of the partial probabilities $\PPr(x_0;\dots ;x_{r-1}$ $|u_0;\dots ;u_{r-1}) $ implies that the regions $\mathcal{R}_r, \mathcal{R}_{r+1}, \dots$ are ignored. As stressed in Sec.(\ref{nonsignaling}),  partial probabilities are subjective by definition.

\subsubsection{Degrees of freedom}
Besides the $K^n$ normalization constraints, complete local consistency provides a number of additional constraints between the target probabilities $\PP(\mathbf{x}|\mathbf{u}) $. Since any subset of regions is consistent, it is possible to enumerate the degrees of freedom in each $m$-partite subsystem in isolation.
In each single region, we have $K$ local independent parameters, e.g. $\PPr(0|u_i)$, because $\PPr(1|u_i)$ is derived by normalization, and thus $nK$ degrees of freedom for $n$ regions and $2K$ for $2$ regions. In each pair of regions, we have seen in Sec. \ref{distributions}, that we have $K^2+2K$ degrees of freedom. Since $2K$ local parameters are already counted we have $K^2$ \emph{joint bipartite} degrees of freedom, e.g., $\PPr(0;0|u_i;u_j)$. As a result, we have $ \binom{n}{2} K^2 $  bipartite degrees of freedom for  the general $n$-region system. Similarly, it is easily shown by induction that there are $K^m$   \emph{joint $m$-partite} degrees of freedom for a general $m$-region system and hence $ \binom{n}{m} K^m $ $m$-partite degrees of freedom for the general $n$-region system. Finally,  in the general multipartite system, the number of degrees of freedom is
$$\binom{n}{1} K + \binom{n}{2} K^2 +\binom{n}{3} K^3 +\dots +\binom{n}{n} K^n =(K+1)^n -1.$$
Because only $nK$  degrees of freedom are local, in general,  $(K+1)^n - nK -1$ degrees of freedom are \emph{shared}. These results are summarized in Table \ref{parametern}.  Note that only the total number of constraints is meaningful, and for $n>1$ one may prefer to consider that there is only $1$ constraint of normalization and $2^n K^n -(K+1)^n$ constraints of local consistency. One or several degrees of freedom can be suppressed by additional constraints. For instance, in order to force all degrees of freedom to be shared, besides the constraints of normalization and total consistency, we have to add a number of $nK$ new constraints. For $n=2$, these $2K$ new constraints may be $ \PP(0;1|\theta_k;\theta_k) = \PP(1;0|\theta_k;\theta_k) =0$ forcing total correlation. 
Similarly, for $n>2$, we can suppress both only-local and only-bipartite correlations with a number of $nK +\binom{n}{2} K^2$ additional constraints. 
%
\begin{table}[htb]
\begin{center}
\begin{tabular}{||l||c||}
\hline
\hline
  Number of regions     &  $n$  \\  
\hline
Number of settings     &  $K$  \\  
\hline
\hline
Number of targets $(\mathbf{x}|\mathbf{u})$ & $(2K)^n$\\
\hline
Total number of local targets & $2nK$\\
\hline
Constraints of normalization & $K^n$  \\
\hline
Constraints of local consistency& $K^n(2^n-1) -(K+1)^n+1$\\
\hline
$m$-partite degrees of freedom  & $ \binom{n}{m} K^m $ \\
\hline
Total degrees of freedom  & $(K+1)^n -1$ \\
\hline
Shared degrees of freedom  & $(K+1)^n -Kn-1$ \\
\hline
\hline
Number of gauge distributions $\g_\gamma(\lambda_j)$ & $nK$\\
\hline
Rank of the gauge linear system & $2(K+1)^{n-1}$\\
\hline
\hline
\end{tabular}
\caption{\label {parametern} {\footnotesize Degrees of freedom in general multipartite locally consistent systems.
Gauge parameters refer to one-step collapse.
}}
\end{center}
\end{table}

Specially, when the system mimics a quantum set of $n$ spins, 
the density operator $\rho$ is defined by $2^{2n}-1$ real coefficients. On the other hand,  the number of degrees of freedom of  $\PP(\mathbf{x}|\mathbf{u})$ is $(K+1)^n -1$ (Table \ref{parametern}). Therefore, a given reduced state $\rho$ may be uniquely described by a locally consistent system of $n$ regions and $K$ different settings where $K=2^2-1=3$. This number of independent settings, $K=3$, is clearly related to the fact that spinors belong to a representation of the rotation group in $\mathbb{R}^3$. As a result, we may \emph{conjecture} that a number of properties in quantum systems may be proved simply in a classical model with just three independent settings.

However, in the present model, we will accept any number of independent settings.  Thus, if a locally consistent system of $n$ regions is defined by a set of independent target probabilities for $K>3$ settings, and if the system has to mimic a quantum situation, the settings are subjected to a number of compatibility conditions  (in addition with  other constraints in form of inequalities, e.g., Eq.~\ref{tsirelsonbound} below).

The degrees of freedom are only relevant for general systems. Indeed, a particular system is completely definite and has thus no free parameter. It is useful to define instead a number of \emph{coefficients of entanglement}, characterizing the  behaviour of the system subject to splitting. The main issue, arguably surprising, is the emergence of \emph{randomness}. 

\subsection{Randomness and entropy}

We regard the classical equivalent of a stationary quantum system as a spatially extended object in equilibrium. As long as the dynamic equilibrium holds, the system evolution is completely deterministic. The potential probabilities $  \mathbf{\PP}(\mathbf{x}|\mathbf{u}) $ describe only the symmetry of the object, exactly as a spinning roulette can be characterized by the prior probabilities of its wheel pockets, but before any break of equilibrium, by definition, the entropy remains zero. Similarly, in quantum physics the von Neumann entropy of a pure state in unitary evolution is zero%
\footnote{%
Actually,  the von Neumann entropy is equal to zero when the reduced state of the system is pure, i.e., not mixed (See Sec.\ref{entmesure}). This implies that the evolution is deterministic.
}
. 

Consider for a start the case of bipartite systems. Randomness can arise from two causes, measurement and partition of the system.

\subsubsection[Entropy of measurement]{Entropy of measurement in bipartite systems}
\label{entmesure}
Firstly, randomness arises by measurement, because the initial equilibrium is broken. This is a well known situation in classical physics: After a break of symmetry, several final outcomes are possible at random. The randomness is then described by the conventional Shannon entropy~\cite{shannon}, accounting for all constraints  at the very best. Since the break of equilibrium follows a free choice of the settings, there are as many entropy coefficients as possibilities. For a finite number of $K$ settings in the 2-region context, the entropy is described by a square $K$-matrix, $\s_1(u_0,u_1)$. 

\paragraph{Global measurement}
For a global measurement, we have, 

\begin{equation}
\label{completemeasure}
\s_1(u_0,u_1)= \sum_{x_0=0}^1 \sum_{x_1=0}^1 - \PP(x_0;x_1| u_0;u_1) \log_2 \PP(x_0;x_1| u_0; u_1)
\end{equation}
We will name $\s_1$ \emph{global measurement entropy}. The point `global'  is a matter of context.

Note that this function $\s_1$ characterizes the classical randomness following a conventional break of equilibrium. When the system mimics a quantum situation in a reduced state $\rho$, $\s_1$ should not be mistaken for the von Neumann entropy, $-\Tr (\rho \log \rho)$, which characterizes the `purity' of the state and is not related with its the amount of information%
\footnote%
{In the present model, if we are concerned by the degree of `purity', i.e., the correlation of the outcomes for identical settings in the two regions, a witness function could be
$$ c(\theta) = \sum_{x=0}^1 - \PP(x;x| \theta;\theta) \log_2 \PP(x;x| \theta; \theta)$$
With the current convention (see footenote \ref{footnote1}) we have $c(\theta) = 0$ for pure states and 
this function is closely related to the von Neumann entropy, $-\Tr (\rho \log \rho)$. With the opposite convention, used in Sec.~\ref{classical} we have $c(\theta) = 1$~bit for totally correlated systems.
}
.
\paragraph{Partial measurement}
It is possible also to consider only a partial a measurement, e.g., in region $\mathcal{R}_0$, and we have
\begin{equation}
\label{partialmeasure}
\s_1(u_0)= \sum_{x_0=0}^1  - \PPr(x_0| u_0) \log_2 \PPr(x_0| u_0) 
\end{equation}
where $\PPr(x_0|u_0)$ is computed from Eq.(\ref{localdiscret}, \ref{completeConsis}) or Eq.(\ref{compatibility1}). 
We will name $\s_1$ \emph{partial measurement entropy}. Due to local consistency, the point `partial'  is again a matter of context.

When the system mimics a quantum situation in a \emph{pure state} $\rho$, this function $\s_1$ is closely related to the so-called `entanglement entropy' $E_1$ in quantum information~\cite{bennett1}, namely e.g., in region $\mathcal{R}_0$,
\begin{equation}
\label{S1quantic} 
E_1 (\mathcal{R}_0) = -\Tr_0 (\rho_0 \log \rho_0) 
\end{equation}
where $\Tr_i(.)$ stands for partial tracing over the region $ \mathcal{R}_i$ and $\rho_i = \Tr_{1-i}(\rho)$, ($i\in\{ 0, 1\})$.
Still for pure states, we have~\cite{araki},
\begin{equation}
\label{S1quantic01}  
E_1 (\mathcal{R}_0) = E_1(\mathcal{R}_1).
\end{equation} 
In the present model, owing to Eq.~\ref{localprob}, the same equality $ \s_1 (u_0) = \s_1 (u_1) $ applies to  Bell's-type systems when $u_0=u_1$. 
However the equality does not hold in general. For separable systems, $E_1 (\mathcal{R}_0)$ and  $E_1(\mathcal{R}_1)$, or $ \s_1 (u_0) $ and $ \s_1 (u_1) $ in the present model, are even independent. 
 What is most important, neither the function $E_1$ in Eq.(\ref{S1quantic}) nor $\s_1$ in Eq.(\ref{partialmeasure}) do grasp the level of entanglement. Therefore, we regard the partial measurement entropy, Eq.(\ref{partialmeasure}), as completely distinct from the entanglement entropy, even if the two values coincide for bipartite totally correlated systems, and we will use a different definition for the `entanglement entropy' in the next section.

\subsubsection[Entropy of entanglement ]{Entropy of entanglement in bipartite systems}

Less expected, a second source of randomness arises in entangled systems without break of equilibrium.  Indeed, it is possible to split the whole system into its two parts and consider each region separately. Then two cases are possible: either the two subsystems are independent, or the two regions are entangled. In the first case,  the two regions are separable, all information is strictly local and no randomness appears. The entropy of the whole system is the sum of the entropy of each subsystem, i.e, \emph{the entropy is extensive}~\cite{gibbsparadox}. 

By contrast, in the second case, local information is insufficient to completely describes the whole system and additional information is needed. By splitting,  we decide to ignore all non local information  and our description of the system becomes incomplete. In other words, the system is no more completely definite. Therefore, without any break of equilibrium, the entropy of the two separated subsystems is now positive, clearly greater than the zero entropy of the full system. As a result, \emph{entropy is no more extensive}. 

In quantum theory, this is generally viewed as an astonishing aspect of entanglement. Actually, this situation is  well known in classical thermodynamics in the face of \emph{long range interactions}~\cite{dauxois, antoni}. In the present case, the only possible paradox lies perhaps in the fact that the splitting is nonsignaling, and thus each region remains locally consistent. 

Quantitatively, we aim to evaluate non-local correlations between a pair of local targets, $(x_0|u_0)$ and $(x_1|u_1)$, where $x_i$ is regarded as a random variable while $u_i$ is viewed as a parameter. A similar issue holds in the conventional Shannon transmission theory. The concept of \emph{mutual information} was indeed devised to evaluate the correlation between the data at the two ends of a communication link. This concept was extended to multipartite systems by R.~Fano~\cite{fano}. In this background, the entropy of measurement $\s_1$ can also be called \emph{self information}.
For ease of exposition, it is convenient to use a shorthand $X_i$ for the local target, $(x_i|u_i)$. Then $(X_0;X_1)$ describes the global target  $(x_0;x_1| u_0;u_1)$. 
Mutual information is defined as~\cite{yeung},
$$ I(X_0 ; X_1) \ident \EE [\log_2 \frac{ \PPr (X_0;X_1) }{\PPr (X_0)\PPr (X_1)}]$$
where $\EE[.]$ stands for the expectation value with respect to $\PPr (X_0;X_1) =\PP (x_0;x_1| u_0;u_1)$. Mutual information does not depend on the order of the two targets,
\begin{equation}
\label{mutualinformation}
 I(X_0 ; X_1) = I(X_1 ; X_0) 
\end{equation}
Define, $$\s_2(u_0,u_1)\ident I[(x_0|u_0) ; (x_1|u_1)] =I[(x_1|u_1) ; (x_0|u_0)] $$
We compute easily,
\begin{equation}
\label{entangentropy}
\s_2(u_0,u_1)= \sum_{x_0=0}^1 \sum_{x_1=0}^1 
 \PP(x_0;x_1| u_0;u_1) \log_2 \Big( \frac{\PP(x_0;x_1| u_0;u_1)}{\PPr(x_0|u_0)\PPr(x_1|u_1)}\Big)
\end{equation}
In the present context, it is convenient to call this coefficient $\s_2$ \emph{bipartite entanglement entropy}.
Depending upon the background,  \emph{mutual information} is also named (apart from the sign) \emph{relative entropy}~\cite{caticha} or \emph{Kullback-Leibler divergence}%
~\cite{kullback}. 

Now, rename the local probability distributions $\PPr(x_0|u_0) = \Q_0(x_0|u_0)$ and $\PPr(x_1|u_1) = \Q_1(x_1|u_1) $ and define a  global distribution $\Q(\mathbf{x}|\mathbf{u}) $, as
\begin{equation}
\label{definitionQ}
\Q(\mathbf{x}|\mathbf{u}) \ident\Q_0(x_0|u_0) \times \Q_1(x_1|u_1)
\end{equation}
$\Q= \Q(\mathbf{x}|\mathbf{u}) $ describes a separable system, derived from the original system $\PP= \PP(\mathbf{x}|\mathbf{u}) $ by removing all non-local correlation.
Using the \emph{relative entropy} $\s(\PP||\Q)$, 
$$\s(\PP||\Q) = \sum_{x_0=0}^1 \sum_{x_1=0}^1
 \PP(x_0;x_1|u_0,u_1) \log \PP(x_0;x_1|u_0,u_1)- \PP(x_0;x_1|u_0,u_1) \log \Q(x_0;x_1|u_0,u_1) $$
we have identically
\begin{equation}
\label{relative2}
\s_2(\mathbf{u}) \ident \s (\PP || \Q). 
\end{equation} 
In words, $\s_2(\mathbf{u}) $, is the relative entropy of the entangled distribution $\PP(\mathbf{x}|\mathbf{u})$ with respect to the  separable distribution  $\Q(\mathbf{x}|\mathbf{u}) $, Eq.(\ref{definitionQ}). Therefore, $\s_2(\mathbf{u})$ is always positive and is zero only for a pair of separable regions. It  is invariant by interchange of the \emph{regions} but not, in general, by permutation of the \emph{settings}. Note that the relative entropy is non symmetrical by interchange of the relevant \emph{distributions}, namely, the  joint distribution $\PP(\mathbf{x}|\mathbf{u})$  and the separable distribution $\Q (\mathbf{x}|\mathbf{u})$.

Let $\Q'_0(x_0|u_0)$ and  $\Q'_1(x_1|u_1)$ be two arbitrary local distributions in $\mathcal{R}_0$ and $\mathcal{R}_1$ respectively.  The distribution $\Q'$,
$$\Q'(\mathbf{x}|\mathbf{u}) \ident \Q'_0(x_0|u_0) \times \Q'_1(x_1|u_1)$$
is separable. Accounting for the definition of $\Q_0$ and $\Q_1$ we obtain,
$$\s(\PP||\Q)-\s(\PP||\Q')=- \s(\Q_0||\Q'_0) - \s(\Q_1|| \Q'_1) \le 0.$$ 
Therefore,
$$
\s_2(\mathbf{u}) = \s(\PP ||\Q) = \min_{\Q'\in\mathcal{D}} \s(\PP || \Q'), 
$$
where $\mathcal{D}$ is the set of all separable distributions. The minimum is clearly obtained for $\Q'=\Q$. 

Furthermore, if the two regions are totally correlated, the whole information is shared between the two parties and the entanglement entropy for a given setting, $\s_2$, is equal to the local measurement entropy, $\s_{1\mathrm{loc}}$, of any region.

When the system mimics a quantum situation in a reduced state $\rho$, the separable distribution $\Q$ mimics a separable system defined by a density operator $\sigma$,
\begin{equation}
\sigma = \sigma_0 \otimes \sigma_1 \ \mathrm{with} \ \sigma_0 = \Tr_1 (\rho) \ \mathrm{and} \  \sigma_1= \Tr_0 (\rho),
\end{equation}
where the partial trace $\Tr_i(.)$ stands for $\Tr_{\mathcal{R}_i}(.)$  ($i\in\{ 0,1\}$). The \emph{quantum relative entropy} of $\rho$ with respect to $\sigma$ is~\cite{umegaki},
$$ \s(\rho || \sigma) = \Tr(\rho\log\rho) - \Tr(\rho\log\sigma),$$
while the quantum entanglement entropy, ${E}_2(\rho)$ was defined by Vedral \emph{et al}~\cite{vedral1} as,
\begin{equation}
\label{relative2Q}
{E}_2(\rho) \ident \min_{\sigma'\in\mathcal{D}} \s(\rho || \sigma'), 
\end{equation}
where $\mathcal{D}$ is the set of all disentangled%
\footnote{
We take always the word `disentangled' as synonymous of `separable', while in reference~\cite{vedral1} the signification is complex and more subtle. However, the difference seems not very clear.
}
 states. Clearly, if the system is separable, $\rho=\sigma$ and $E_2=0$ and conversely. As a result, $E_2=0$ if and only if the system is disentangled. More generally, with the same proof as in the classical case, it can be shown that  
$\min\s(\rho || \sigma')$ is obtained for $\sigma' = \sigma$. Then ${E}_2(\rho) = \s(\rho || \sigma) $.
Therefore, in bipartite systems, the present definition, Eq. (\ref{entangentropy}, \ref{relative2}),  is closely related to the definition%
\footnote{%
\label{constructive}
The definition Eq.(\ref{entangentropy}) is constructive and provide a necessary and sufficient condition of classical bipartite entanglement.
We can obtain similarly a constructive definition of quantum bipartite entanglement by droping `$\min$' and  substituting $\sigma$ for $\sigma'$ in Eq.(\ref{relative2Q}).
}
of the quantum entanglement entropy Eq. (\ref{relative2Q})~\cite{vedral, horodecki, belavkin}.
We will revisit this concept for $n$-partite systems.

For \emph{pure state}, $\s(\rho || \sigma)$ coincide with the von Neumann entropy of the partial density $\rho_i $ in any region $\mathcal{R}_i$.
$$ \s(\rho || \sigma)=- \Tr_0(\rho_0\log\rho_0) =  - \Tr_1(\rho_1\log\rho_1),  $$
but in general, this relation does not hold.

In the present model, for a discrete classical system with $K$ possible settings, the entanglement entropy $\s_2(\mathbf{u})$ is a square matrix of order $K$, in general non symmetrical. This result is not surprising because we know that we have $K^2$ bipartite degrees of freedom (Tables~\ref{parameter} and \ref{parametern}). Therefore we can identify these $K^2$ parameters with the $K^2$ coefficients of bipartite entanglement. As a result, a bipartite system is separable if and only if its entanglement entropy matrix is zero. 

The bipartite entanglement entropy $\s_2$ measures the amount of information required to account for non-local bipartite correlation. For instance, in a previous paper~\cite{mf1}, we have described a dice game between two parties, Alice and Bob, to illustrate the concept of classical entanglement. In this example, it is easy to compute an improbable `entanglement entropy' between the two players as a 3-matrix, given in Table~\ref{eentropy613}! 

For a continuous ensemble of settings, the entanglement entropy is a function $\s_2(u_0,u_1)$. Specially, for a pair of EPR totally entangled entities, the entanglement entropy with respect to a pair of settings $\theta_a$ and $\theta_b$ is
$$\s_2(\theta_a,\theta_b)=(1/2)[(1+\cos\theta)\log_2(1+\cos\theta)+(1-\cos\theta)\log_2(1-\cos\theta)]$$
where $\theta=\theta_a-\theta_b$. Clearly, $\s_2 \le 1$ and the maximum is equal to 1 bit, obtained for $\theta=0$. This is just the amount of information required to transmit classically the non-local correlation between the two entities of the pair~\cite{toner}.

\begin{table}[htb]
{
$$
\begin{array}{||c||c|c|c||}
\hline
\hline
\s_2(\Delta_a,\Delta_b) &{ \Delta_1 } & { \Delta_2 } & { \Delta_3 } \\
\hline
\hline
{ \Delta_1 } &  1 &  0.35 &  0\\
\hline
{ \Delta_2 } &  0.35 &  1 &  0.35\\
\hline
{ \Delta_3 } &  0 &  0.35 &  1\\
\hline
\hline
\end{array}
$$
} 
\caption{\label{eentropy613}
{\footnotesize Example of classical entanglement entropy (in bits) between two players, Alice and Bob, in a totally correlated classical dice game described in Ref.~\cite{mf1}. The game is implemented with three dice labelled $\Delta_k$ ($k=1, 2$ and $3$). At each run, each player selects freely one die, $\Delta_a$ and $\Delta_b$ respectively. The coefficients measure the amount of information required to account for non-local correlation.
}
}
\end{table}


\subsubsection{Multipartite entanglements}
\label{multipartiteEE}
When considering multipartite \emph{measurements}, there is no new topic. 
By contrast,  \emph{splitting} of the system raises different questions because there is a number of different ways to group the regions together. Again, the same issue is encountered in communication networks: In transmission theory, the natural extension of the bipartite mutual information is the multivariate mutual information~\cite{fano}. Similarly, we will define multivariate mutual information coefficients   between local targets, computed recursively as~\cite{yeung},
\begin{multline}
\label{iFano}
I(X_0 ; X_1 ; \dots ; X_{n-1}) \ident 
I(X_0 ; X_1 ; \dots; X_{n-3}  ; X_{n-2}) \\
-I[(X_0 ; X_1 ; \dots ; X_{n-3} ; X_{n-2})|X_{n-1})] 
\end{multline}
In Eq.(\ref{iFano}), $X_i$ is still a local target, shorthand for $(x_i|u_i)$, where $x_i$ is a random variable and $u_i$ a parameter. 
It can be shown that multivariate mutual information is completely symmetrical with respect to $X_0,X_1,\dots,X_{n-1}$ even if this is not manifest in the definition.
Contrary to bivariate mutual information which is always positive, multivariate mutual information can be either positive or negative.

Shannon's information theory has a nice set-theoretic structure~\cite{hu, yeung1,yeung}  recalled in Table~\ref{mesure}. Thanks to this structure,  computation of entropic expressions is simplified by a graphical representation, named \emph{information diagram} and similar to the conventional Venn diagram. Entropy or mutual information $I$ are associated with a signed measure $\mu$. For simplicity, and without loss of generality, we will use the same symbol $I$ to describe both entropy and information (while entropy is usually symbolized by $H$).
%
\begin{table}[htb]
{
$$
\begin{array}{||c|c||}
\hline
\hline
\mathrm{Information} \ \mathrm{theory} &\mathrm{Set} \ \mathrm{theory}\\
\hline
\hline
{ I(X) } &  \mu(X) \\
\hline
{ (X_1; X_2)} &  X_1 \cap X_2\\
\hline
{ (X_1,X_2) } &  X_1\cup X_2\\
\hline
{ (X_1| X_2)} &  X_1- X_2\\
\hline
\hline
\end{array}
$$
} 
\caption{\label{mesure}
{\footnotesize Set-theoretic structure of Shannon's information theory. Each variable $X$ is represented by a set $X$. Entropy or mutual information $I$ are associated with a signed measure $\mu$. Information expressions, like $I(X_1,X_2)$, $I(X_1 ; X_2)$ or $I[(X_1|X_2)]$ are respectively associated with $\mu(X_1 \cup X_2$,  $\mu(X_1 \cap X_2$ and $\mu(X_1 - X_2) $ (where $X_1-X_2= X_1 \cap X_2^c)$. 
}
}
\end{table}
%
%
Multivariate mutual information, Eq.~(\ref{iFano}), is directly associated with 
\begin{equation}
\label{iFanoBis}
I(X_0;\dots;X_{n-1}) = \mu ( X_0 \cap \dots \cap X_{n-1})
\end{equation}
and  thus $I(X_0;\dots; X_{n-1})$ is clearly completely symmetrical with respect to $X_0,\dots,X_{n-1}$. 
For instance, when $n=3$, Eq.(\ref{iFano}) is represented by the information diagram shown in ~Fig.\ref{diagram}. Given a setting vector $\mathbf{u}$, each local distribution $X_i$ is represented by an area and $I(X_i)$  is the partial measurement entropy, or the self-information of this variable. We consider only binary outcomes and thus $I(X_i)\le 1$~bit.  For instance, Eq.(\ref{partialmeasure}), describes the self-information $I(X_0)$. The $n$ regions delimit a number of $2^n-1$ atoms of information, $\mu_1,\dots,\mu_{2^n-1}$. For $n=3$, we have $I(X_0;X_1;X_2) =  \mu_7$. 

In practice, it is convenient to derive first the measurement entropy $I(X_{ i_0},\dots,X_{i_{r-1}} )$ of each partial subsystem $( \mathcal{R}_{i_0},\dots,\mathcal{R}_{i_r})$ for $r=1,\dots,n$, because each probability distribution is easily computed. For instance, if $\{ i_0,\dots,i_{r-1} \} = \{ 0,\dots,r-1\}$,  the relevant partial probability in the $r$-region subsystem is given by Eq.(\ref{completeConsis}). We have  $2^n -1$ such partial subsystems. Next, we express the measure $\mu_i$ of each atom of the information diagram (Fig. \ref{diagram}) in terms of these measurement entropies. We have $2^n -1$ atoms and therefore we obtain the mutual information $I(X_0;X_1;\dots;X_{n-1})=\mu_{2^n-1}$ by solving a linear system of $2^n -1$ equations for $2^n -1$ unknowns. 

For example, when $n=3$, we derive easily the following equation by inspection of Fig. \ref{diagram},
%
%
%
$$
\begin{bmatrix}
I(X_0) \\I(X_1) \\I(X_1,X_0) \\I(X_ 2)\\I(X_2,X_0)\\I(X_2,X_1) \\I(X_2,X_1,X_0) \\
\end{bmatrix}
=
\begin{bmatrix}
1&0&1&0&1&0&1\\
0&1&1&0&0&1&1\\
1&1&1&0&1&1&1\\
0&0&0&1&1&1&1\\
1&0&1&1&1&1&1\\
0&1&1&1&1&1&1\\
1&1&1&1&1&1&1\\
\end{bmatrix}
\begin{bmatrix}
\mu_1\\ \mu_2\\  \mu_3\\ \mu_4\\ \mu_5\\ \mu_6\\ \mu_7\\ 
\end{bmatrix}
,
$$
and we obtain
$$
\begin{bmatrix}
\mu_1\\ \mu_2\\  \mu_3\\ \mu_4\\ \mu_5\\ \mu_6\\ \mu_7\\ 
\end{bmatrix}
=
\begin{bmatrix}
0 & 0& 0& 0& 0&-1& 1\\
0 & 0& 0& 0&-1& 0& 1\\
0 & 0&0& -1& 1& 1&-1\\
0 & 0& -1&0& 0& 0& 1\\
0 &-1& 1& 0& 0& 1&-1\\
-1& 0& 1& 0& 1& 0&-1\\
 1& 1& -1&1&1&-1& 1\\
\end{bmatrix}
\begin{bmatrix}
I(X_0) \\I(X_1) \\I(X_1,X_0) \\I(X_ 2)\\I(X_2,X_0)\\I(X_2,X_1) \\I(X_2,X_1,X_0) \\
\end{bmatrix}
$$
%
The last row reads
\begin{multline}
\mu_7= I(X_0;X_1;X_2)=  I(X_0) +I(X_1) + I(X_ 2)\\
-I(X_0,X_1) -I(X_0,X_2)-I(X_1,X_2) +I(X_0,X_1,X_2) 
\end{multline}
For instance, it is possible to mimic the 3-region GHZ-state with the setting $(Z;Z;Z)$, where $Z$ is  the axis of common entanglement. Similarly, we can mimic the W-state with the same setting.  The measures $\mu_i$ of all atoms are given in Table ~\ref{muatomeWG}. 
\begin{table}[htb] 
{
$$
\begin{array}{||c||c|c|c|c|c|c|c|c||}
\hline
\hline
\mathrm{states}  & \mu_1 & \mu_2 & \mu_3 & \mu_4 & \mu_5 &\mu_6& \mu_7  & \mathcal{E}\\
\hline
\mathrm{GHZ}  (ZZZ) &  0 &  0 & 0 & 0 & 0 & 0 & 1 & 2\\
\hline
\mathrm{W}  (ZZZ) &  0 &  0 & 0.667 & 0 & 0.667 & 0.667 & -0.415 & 1.170 \\
\hline
\hline
\end{array}
$$
} 
\caption{\label{muatomeWG}
{\footnotesize  
Signed measures $\mu_i$ and total entanglement $\mathcal{E}=2\mu_7+\mu_3+\mu_5+\mu_6$ (in bits) for the tripartite GHZ and W states with the setting $(Z;Z;Z)$.
}
}
\end{table} 

More generally we have,
\begin{multline}
I(X_0;X_1;\dots;X_{n-1})= \sum_i I(X_i) - \sum_{i,j} I(X_i,X_j) +\\ \sum_{i,j,k}  I(X_i,X_j,X_k)-\dots +(1)^{n-1}I(X_0,\dots,X_{n-1})
\end{multline}
where the summation is taken over all possible combinations of subscripts and,
\begin{multline*}
 I(X_i) = \s_1(u_i)= \sum_{x_i=0}^1 - \PPr(x_i|u_i) \log_2  \PPr(x_i|u_i)\\
=\sum_{x_0=0}^1 \dots \sum_{x_{n-1}=0}^1 -\PP(\mathbf{x},\mathbf{u}) \log_2  \PPr(x_i|u_i),
\end{multline*}
\begin{multline*}
 I(X_i,X_j) = \s_1(u_i,u_j)=\sum_{x_j=0}^1 \sum_{x_i=0}^1 - \PPr(x_i;x_j|u_i;u_j) \log_2  \PPr(x_i;x_j|u_i;u_j)\\
=\sum_{x_0=0}^1 \dots \sum_{x_{n-1}=0}^1 -\PP(\mathbf{x},\mathbf{u}) \log_2 \PPr(x_i;x_j|u_i;u_j),
\end{multline*}
etc. In the present context, we will use the following terminology:
\begin{definition*}[$n$-partite entanglement entropy, degree and coefficients of entanglement]
We will name \emph{$n$-partite entanglement entropy} the  expression
\begin{equation}
\label{entangentropyn}
\s_n(\mathbf{u}) = I(X_0;X_1;\dots; X_{n-1}).
\end{equation}
When different from zero, $\s_n(\mathbf{u})$ defines one \emph{degree of $n$-partite entanglement}, $e_n=1$. We will call \emph{entanglement coefficients} its $K^n$ components. 
\end{definition*}
The $K^n$ entanglement coefficients can be identified with the $K^n$ $n$-partite degrees of freedom of a general $n$-region system. As a result, if $M_n \not= 0$, we will regard  the $n$-region system as $n$-partite entangled.
In general, $\s_n(\mathbf{u})$ is not invariant by permutation of the settings but the object describes the $n$-partite entanglement as a whole and does not depends on the order of the $n$ regions. For $n=3$, we have
%
\begin{multline}
\label{entangentropy3}
\s_3(\mathbf{u})= \sum_{x_0=0}^1 \sum_{x_1=0}^1 \sum_{x_{2}=0}^1 
\PP(\mathbf{x}| \mathbf{u}) \ \times \\
\log_2 \Big( \frac{ \PPr(x_0;x_1|u_0;u_1) \PPr(x_1;x_2|u_1;u_2) \PPr(x_2;x_0|u_2;u_0) }{\PP(x_0;x_1;x_2|u_0;u_1;u_2) \PPr(x_0|u_0)\PPr(x_1|u_1)\PPr(x_{2}|u_{2})}\Big)
\end{multline}
For instance, from Table~\ref{muatomeWG}, we have $\s_3(Z;Z;Z)= 1$ bit for the GHZ state and $\s_3(Z;Z;Z)= -0.415$ bit for the W-state.
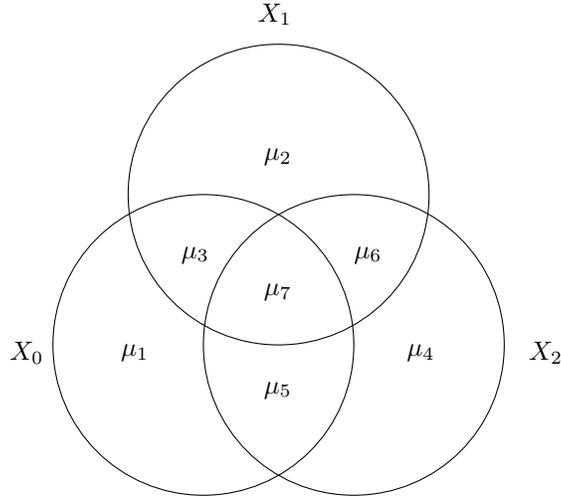
\begin{figure}[htb]
\begin{center}
\begin{tikzpicture}
\coordinate [label=left:$X_0$] (X0) at (-2,-0.1);
\coordinate [label=left:$X_1$] (X1) at (1.3, 4.4);
\coordinate [label=left:$X_2$] (X2) at (4.9,-0.1);
\coordinate [label=left:$\mu_7$] (0) at (1.3,.7);
\coordinate [label=left:$\mu_3$] (1) at (0.2,1.2);
\coordinate [label=left:$\mu_6$] (2) at (2.5,1.2);
\coordinate [label=left:$\mu_5$] (3) at (1.3,-0.6);
\coordinate [label=left:$\mu_1$] (4) at (-0.6,-0.1);
\coordinate [label=left:$\mu_2$] (5) at (1.3,2.5);
\coordinate [label=left:$\mu_4$] (6) at (3.2,-0.1);

\draw (0,0) circle (2cm);
\draw (2,0) circle (2cm);
\draw (1,2) circle (2cm);
\end{tikzpicture}.
\end{center}
\caption{\label{diagram}
{\footnotesize 
Information diagram of three parties, $X_0$, $X_1$ and $X_2$. Using Table \ref{mesure}, we have for instance $I(X_0)= \mu_1+\mu_3+ \mu_5+\mu_7$, $I(X_0;X_1)= \mu_3+\mu_7$, $I(X_0;X_1;X_2)= \mu_7$.
}
}
\end{figure}
%

\subsubsection[Entanglement scheme]{Characterizing entanglement}

\begin{definition*} [Degree of entanglement]
When a \emph{n}-region system is \emph{n}-partite entangled, we will say that its degree of entanglement is \emph{n} and set $e_n$=1. Otherwise,  its degree of entanglement is the larger number of regions, \emph{m}, which constitutes a \emph{m}-partite entangled subsystem.
\end{definition*}

Finally, any given system has a particular structure described by the {scheme} $( e_2, e_3,\dots,e_n)$ of  its  bipartite, tripartite,\dots, $n$-partite degrees of entanglement. The number of $m$-partite coefficients of entanglement non identically vanishing is thus $e_m K^m$, with  $e_m\le\binom{n}{m} $. 

\begin{definition*} [Entanglement scheme]
Given a particular $n$-partite system with $K$ settings, we will call \emph{entanglement scheme}  the sequence $( e_2, e_3,\dots,e_n)$ of  its  bipartite, tripartite,\dots, $n$-partite degrees of entanglements, where $e_m\le\binom{n}{m}$. The overall number of \emph{entanglement coefficients} is
$$N= \sum_{m=2}^n e_m K^m.$$
\end{definition*}

In addition, we will define later $K^n$ \emph{total entanglement coefficients} by Eq. (\ref{Tcorrelation}
) in Sec. (\ref{inequalities}).

Given the probability distribution $\PP(\mathbf{x}|\mathbf{u})$ the coefficients of entanglement are easily computed. For instance, Bell-type systems are bipartite, totally correlated systems, characterized by the scheme $e_2=1$ (with $\binom{2}{2}K^2 = K^2$ coefficients of entanglement). Among tripartite systems (Sec.~\ref{tripartite} below), GHZ-states (considered with only two settings) are characterized by the scheme $(e_2,e_3)=(0,1)$ and $K=2$, with $ \binom{3}{3}K^3= 8  $ coefficients of entanglement, while  general W-states (with two settings) are characterized by the scheme $(e_2,e_3)=(3,1)$ and $K=2$, with $\binom{3}{2}K^2 +K^3  = 20$ coefficients of entanglement, but only $2$ tripartite coefficients (up to $8$) are non zero. Separable systems are trivially characterized by $e_2=e_3=\dots=e_n=0$ and thus zero coefficient of entanglement.

The current results are rather paradoxical. We have elaborated upon  a close analogy with signaling networks except that the present system is \dots nonsignaling! However that may be, information theory has established a number of helpful inequalities between entropy functions, which are of course currently valid.

\subsubsection{Information inequalities}
\label{inequalities}
In quantum information theory,  entanglement in multipartite systems is not yet fully understood. By contrast, in the present model, the information of all atoms  is easily computed and we have a number of tools at disposal stemmed from information theory. We recall first the polymatroidal axioms. Next we define the notions of `total entanglement' and maximally entangled systems.
\paragraph{Polymatroidal axioms}
Let
 $ \alpha, \beta$ be two subsets of $\llbracket 0,n-1\rrbracket$. If $\alpha=\{ i_1,i_2,\dots , i_r\}$, define
$$I(\alpha)\ident I(X_{i_1},\dots,X_{i_r})$$ (for simplicity, we conserve the same symbol $I$).

The following \emph {polymatroidal axioms}~\cite{yeung} describe the Shannon information inequalities.
\begin{enumerate}
\item $I(\emptyset) = 0$
\item If $\alpha \subset \beta$ then $I(\alpha)  \le I(\beta)$
\item $I(\alpha)+I(\beta) \ge I(\alpha \cap \beta)+I(\alpha \cup \beta)$
\end{enumerate}
This tool is powerful to derive important entropic relations. 
Note that this is not the complete story since there are  inequalities not derivable from this set of axioms, called \emph{non-Shannon inequalities}.

We will just give one example, and define the concept of \emph{total entanglement}.
\paragraph{Total entanglement}
In transmission theory, Watanabe~\cite{watanabe} has defined a global coefficient $\mathcal{E}$, named \emph{total correlation} as,
$$
\mathcal{E}(X_0;X_1;\dots ; X_{n-1}) =
\sum_{i=0}^{n-1} I(X_i) - I(X_0,X_1, \dots,X_{n-1}) 
$$
For instance, in tripartite systems (Fig.~\ref{diagram}), we have 
$$\mathcal{E}(X_0;X_1;X_2) = 2 \mu_7 +\mu_3+\mu_5 +\mu_6$$
It is possible to express $\mathcal{E}(X_0;X_1;\dots ; X_{n-1})$ in terms of relative entropy as follows:
Given a setting vector $\mathbf{u}$, construct a  separable probability distribution $\Q = \Q(\mathbf{x}|\mathbf{u}) $, as
\begin{equation}
\label{definitionQn}
\Q(\mathbf{x}|\mathbf{u}) \ident\ \prod_{i=0}^{n-1} \PPr(x_i|u_i) 
\end{equation}
Now we have, with $\PP= \PP(\mathbf{x}|\mathbf{u})$,
$$
\mathcal{E}(\mathbf{u})= \s(\PP || \Q) 
$$
%
As a result, total correlation is always positive and is zero \emph{if and only if} all regions are independent.  
In the present context, it is relevant  to use a special term, 
\begin{definition*} [Total entanglement]
Given a particular $n$-partite system with $K$ settings, we will call \emph{total entanglement}  the function
\begin{equation}
\label {Tcorrelation}
\mathcal{E}(\mathbf{u})\ident \s[\PP(\mathbf{x}|\mathbf{u}) || \Q(\mathbf{x}|\mathbf{u})]. 
\end{equation}
\end{definition*}

Note that the concept of total entanglement can be  easily translated into quantum formalism. Consider a quantum $n$-region system in the reduced state $\rho$ and let $\sigma_i$ be the reduced state in region $\mathcal{R}_i$, obtained by tracing over the $n-1$ other regions.
Let $\sigma$ be the the separable state
$$\sigma =  \bigotimes_{i=0}^{n-1} \sigma_i.$$
Define the `quantum total entanglement' as

\begin{equation}
\label{quantumtotal}
\mathcal{E}(\rho)= \s(\rho||\sigma)
\end{equation}
Clearly, when all regions are separable,  $\mathcal{E}(\rho)= 0$ . 
%
We have already noted (Eq.~\ref{relative2Q}) that the quantum entanglement entropy, ${E}(\rho)$ was defined by Vedral \emph{et al}~\cite{vedral1} as,
$$
{E}(\rho) \ident \min_{\sigma'\in\mathcal{D}} \s(\rho || \sigma'). 
$$
where $\mathcal{D}$ is the set of all disentangled states. It can be shown again that the minimum is obtained for $\sigma' = \sigma$, then ${E}(\rho) = \s(\rho || \sigma) = \mathcal{E}(\rho)$.  As for bipartite systems (footnote~\ref{constructive} above), this provides a necessary and sufficient witness of entanglement.

%
In the present model, since we only consider binary outcomes, we have 
$$ I(X_i)  \le 1 \  \mathrm{bit}$$
As a result, we have the inequality
$0\le \mathcal{E}(\mathbf{u}) \le n \  \mathrm{bits}.$
Actually this maximum can be improved to $n-1$ bits. The proof is trivial for $n=1$. Now, thanks to the polymatroidal axioms, the addition of one region $X_m$ can only  increase the total entanglement by $I(X_m)\le 1 $ bit. Therefore, we have for a $n$-region system, 
\begin{equation}
\label{totalentang}
0\le \mathcal{E}(\mathbf{u}) \le n-1 \  \mathrm{bits}.
\end{equation}
The upper bound can be saturated, e.g., when the measures $\mu_i$ of all atoms of information are zero, except  for $I(X_0;X_1;\dots;X_{n-1})=1$ bit. This correspond to the general $n$-region GHZ-states, when the same setting, corresponding to the axis of entanglement, is used in each region. 
\begin{definition*}[Maximally entangled system]
We will say that a $n$-region system is maximally entangled when its total entanglement for at least one setting-vector is equal to $n-1$ bits.
\end{definition*}
 For instance, among bipartite systems, EPR-pairs and specially the singlet state are maximally entangled (Table \ref{singlet}c below) as well as the so-called PR-box (Table \ref{PRbox}c below), with $\mathcal{E} = 1$ bit. Among tripartite systems,  the GHZ-state with the setting $(Z;Z;Z)$ is maximally entangled with $\mathcal{E}(Z;Z;Z) = 2$~bits (Table~\ref{muatomeWG}). Note that with only two settings $X$ and $Y$, the GHZ-state is not maximally entangled (Table~\ref {toutGHZ}c below). The W-state is not maximally entangled with a maximum entanglement of $1.170 < 2$~bits (Tables~\ref{muatomeWG} and \ref{toutW}c below). 

This framework is likely to be helpful to discuss a number of issues like `entanglement monogamy'~\cite{coffman} or collapse mecanism (Sec. \ref{mechanism} below). These discussions are beyond the scope of this paper.

To sum up, we have a \emph{necessary and sufficient} condition of entanglement, namely,  that total entanglement $\mathcal{E}(\mathbf{u}) $, (Eq.~\ref{Tcorrelation}), is non zero for at least one  setting-vector $\mathbf{u}$ and a  bound of maximal entanglement, namely,   $\mathcal{E}(\mathbf{u}) \le n-1$ bits for a $n$-region system.

\paragraph{Classification of locally consistent systems}
In quantum information theory, an important challenge is to characterize entangled and separable systems. A number of entanglement witnesses are known.
Bell-CHSH inequalities were first defined for only two regions. In 1990, David Mermin~\cite{mermin3} constructed a set of multipartite inequalities similar to the CHSH expressions. Since then, a large range of witness operators have been described~\cite{horodecki}. In the present model, all these operators are valid, but \emph{the concept of \emph{total entanglement} provides an easy necessary and sufficient condition of entanglement}.

Beyond the separation between entangled and separable systems, a second dividing line may be drawn between quantum and `super-quantum' systems, i.e., systems that can or not mimic a quantum mechanical situation~\cite{popescu}. According to Tsirelson, the characterization of quantum state was first raised by Anatoly Vershik, but the very question was addressed quantitatively by Boris Tsirelson himself~\cite{tsirelson,tsirelson1}, who showed that the CHSH inequality, Eq.(\ref{originalCHSH}), became for quantum systems, 
\begin{equation}
\label{tsirelsonbound}
|\s (A, B) + \s (A', B) + \s (A, B') - \s (A', B')|  \le 2\sqrt{2}.
\end{equation}
In addition, Tsirelson identified this bound, $2\sqrt{2}$, as a special value for two regions of the Grothendieck constant\footnote{The algebraic nature of the Grothendiek constant is unknown and so, the constant is generally qualified of `enigmatic', possibly because Alexandre Grothendiek is viewed as an enigmatic mathematician.}, defined in general topological tensor product spaces~\cite{grothendieck}.
In terms of \emph{Hamming divergence}, Eq.(\ref{dhamming}), we have also,
$$ 1-\sqrt{2} \le \dd (A, B) + \dd (A', B) + \dd (A, B') - \dd (A', B') \le  1+\sqrt{2}$$

The generalization of Eq. (\ref{originalCHSH}, \ref{tsirelsonbound}) to multipartite systems was addressed by Mermin~\cite{mermin3}  but the surprising result was that the same bound is valid for both quantum and `super quantum' systems.
We will give in Sec. \ref{revisiting} a criterion based on the evaluation of the Tsirelson bound following a process of classical multipartite collapse.

\subsection{Classical multipartite collapse}
Consider a $n$-region classical system in equilibrium. The equilibrium can be broken as follows:
In one or several regions ${\mathcal R}_i$,  an observer, ${\mathcal {O}}_i,$ (1) selects freely a setting $u_i$, element of a given set $\mathbf {\Theta}$ of $K$ different settings, (2) breaks the local equilibrium in a way reliant on $u_i$, and (3) measures in ${\mathcal R}_i$  an observable $x_i$, element of a set $\mathbf {X}  = \{0, 1\}. $ 

\subsubsection{Collapse mechanism}
\label{mechanism}
We need first to discuss the `collapse mechanism', in the sense of  `reaction mechanism' in chemical kinetics. Indeed, the  global process defined for Bell-type systems turns out to be insufficient to describe all multipartite collapses, even for independent regions.

The EPR paradox was cleared up by a direct gauge implementation of the probability distributions in the two regions. Actually, a gauge selection  defined a \emph{leading region}, say $\mathcal{R}_0$, but afterward the two regions entered into the model on just the same footing. We have computed a  probability distribution, $g_{k_0}(\hat\lambda_j)$, located on the boundary of the two regions, accounting for the free choice $u_0=k_0$ in the leading region and governing \emph{both} $(x_0|k_0)$ and $(x_1|k_0;u_1)$. 

However, local realism does not  require that the same distribution governs the two processes. One can construct an \emph{indirect collapse} as follows. (1)~The settings $(k_0,k_1)$ are sent to an ignition point $\mathcal{I}$. (2)~A gauge selection defines the {leading region}, e.g.,  $\mathcal{R}_0$. (3)~At the boundary of the two regions, the outcome $x_0=\xi_0$ of the leading region is drawn at once, using    $ \PPr(x_0|k_0) $. (4)~A gauge distribution $\PPr(x_1|\xi_0;k_0;u_1)$ is used in the second region $\mathcal{R}_1$. Of course, thanks to  the well known product rule~\cite{jaynes}, accounting for local consistency, the two processes give exactly the same final result, 
$$ \PP(\xi_0;x_1|k_0;u_1)= \PPr(\xi_0|k_0) \PPr(x_1|\xi_0;k_0;u_1) $$ 
Clearly, the indirect process requires two successive trials. 
As a result, the number of gauge distributions is doubled. On the other hand, each distribution $\PPr(x_1|\xi_0;k_0;u_1)$ is simpler,  and even trivial in bipartite system, because we have only to account for compatibility conditions in one region instead of two. Therefore, in the computation of the gauge distributions, the direct route can fail, while the two-step process turns out to be successful.  As in chemical kinetics, this may be physically meaningful because the two-step route may require a delay, while the direct process can be viewed as quasi-instantaneous at the ignition point. 

In multipartite systems, we will proceed similarly. Thanks again to the product rule and local consistency, it is possible to define  multi-step collapses, for instance,
\begin{multline}
\label{product}
\PP(x_0;x_1;x_2;x_3|u_0;u_1;u_2;u_3)= \PPr(x_0|u_0)
\times \PPr( x_1|x_0;u_0;u_1) \\
\times  \PPr(x_2|x_0;x_1;u_0;u_1;u_2) 
\times  \PPr(x_3|x_0;x_1;x_2;u_0;u_1;u_2;u_3) 
\end{multline}
At each step, we have to select a new gauge region. For $m$ steps, we have $m$ gauge trials per run. 

Suppose, for instance, that several observers decide to proceed to a measurement and select freely a setting $\theta_{k_i}$.  They send their setting towards an ignition point, located at the boundary of the relevant regions. A first gauge selection at the ignition point defines a leading configuration, say $\gamma_0= k_0+K i_0$, i.e.,  a gauge region $\mathcal{R}_{i_0}$ and a gauge setting $\theta_{k_0}$. The first outcome $\xi_0$ is drawn using the local probability $\PPr(x_{i_0}|k_0)$. Now, if e.g., $i_0=0$, the probability distribution in the remaining $n-1$ regions is a gauge function
\begin{equation} 
\label{firstep}
\PPr(x_1;\dots;x_{n-1}|\xi_0;k_0;u_1;\dots;u_{n-1})= M \times \PP (\xi_0;x_1;\dots;x_{n-1}|k_0;u_1;\dots;u_{n-1})
\end{equation}
where $M = M( \xi_0,k_0,u_1,\dots,u_{n-1})$ is a normalization factor. Note that Eq.(\ref{firstep}) only describes the updated probability  of $(x_1,\dots,x_{n-1})$  for the observer $\mathcal{O}_0$, because he has  got alone a new information, namely $u_0=k_0$ and $x_0=\xi_0$. By contrast, the $n-1$ other observers are not aware of what happens in region $\mathcal{R}_0$, and their view of the $n-1$ regions is still given by Eq.(\ref{completeConsis}),
\begin{equation}
\label{others}
\PPr(x_1;\dots ;x_{n-1}|u_1;\dots ;u_{n-1}) = 
\sum_{x_0=0}^{1 }  \PP(x_0;\dots ;x_{n-1}|u_0;u_1;\dots ;u_{n-1}). 
\end{equation}
Due to local consistency, the right hand-side of Eq.(\ref{others}) does not depend on $u_0$ and we can plug $k_0$ in $u_0$. In general,   Eq.(\ref{firstep}) and Eq.(\ref{others}) are different, because they describe different level of knowledge.

At least formally, this process can describe a real quantum collapse. Therefore, if the original system mimics a quantum situation, the probability distribution Eq.(\ref{firstep}) will also mimic a quantum system. More generally, it can be seen that Eq.(\ref{firstep})  is still locally consistent. Therefore we can define a second-step probability distribution $ \PP' $ as
\begin{equation}
\label{newprob}
\PP'(x_1;\dots ;x_{n-1}|u_1;\dots;u_{n-1})\ident  \PPr(x_1;\dots;x_{n-1}|\xi_0;k_0;u_1;\dots;u_{n-1}) .
\end{equation}
Owing to Eq.(\ref{product}),  this distribution $\PP'$ describes now the $n-1$ remaining regions (instead of $\PP(\mathbf{x}|\mathbf{u})$). It is possible to iterate when wanted. Let $m$ be the number of iterations. If $m=n$, the overall collapse is completed. Each observer gets his own outcome, $x_i=\xi_i$ but ignores the other outcomes. If $m<n$, the last gauge trial has to be performed in a subsystem of $n-m$ regions. For this purpose, we will construct in the next section a convenient stochastic gauge system on the model of the Bell-type systems. If we fail to obtain a set of positive gauge probability distributions, this means that there is no locally realist collapse in $m$ steps and we have to try with $m+1$ steps. The last stage in just one region is trivial, and therefore, it is always possible to find a locally realist route whatever the number of regions and the degree of entanglement. In other words, given any locally consistent system, the global collapse can always be implemented by an arbitrary chain of $n$ one-region collapses. 

\begin{definition*}[m-step collapse]
We will call \emph{m-step collapse} a collapse composed of \emph{m-1} cascaded one-region trials and one final  trial into a subsystem of $n-m+1$ regions.
\end{definition*}
In bipartite systems, we have always found by computer simulation that the direct collapse in just one step is feasible. This is not always the case in multipartite systems. We will see that a one-step process fails is some tripartite `super quantum' systems. When more than three regions are involved, the one-step collapse may fail even when all regions are separable.

Note that a $n$-step collapse, requiring a minimum of computational resources, can be helpful to simulate quantum algorithms.

\subsubsection{Using the Tsirelson criterion for more than two regions}
\label{revisiting}
We have seen that any locally consistent system  of $n$ regions can always collapse by an arbitrary chain of $n$ one-region collapses. Since a quantum system can formally follow this process,  each step in the chain defines a quantum gauge subsystem of $m$ regions (with $m\in\llbracket 1,n\rrbracket$). Therefore, only a `super quantum' system can lead to a `super quantum' gauge subsystem.  
In other words, if one gauge subsystem is `super quantum'  we can conclude with certainty that the global system was `super quantum'. 

Specially, evaluation of the Tsirelson bound allows the $2$-region `super quantum' subsystems to be detected. Beyond, multi-step collapses provide a criterion of non-quantum behaviour for more than two regions. For instance, we will use this criterion in Sec.~\ref{superQGHZ} below to characterize a 3-region super-quantum system.

\subsubsection{One-step collapse}
In order to describe a classical one-step collapse in multipartite systems, we have to generalize the  concepts defined in Bell-type systems, namely, ignition states, gauge probability distributions, and measurement process. 

\paragraph{Ignition states}
The ignition states will define the potential outcomes in the $n$ regions. In Sec (\ref{discretesettings}) we have constructed a fine-grained description of Bell-type systems. Now, we aim to construct a fine-grained description of  multipartite systems by the same token. 

We have defined the \emph{local target} $(x_{i}|u_{i})= (\xi|\theta_k)_{i} $ in a region $\mathcal{R}_{i}$ by $x_{i}=\xi$ given $u_{i}=\theta_k$.  When the system is completely local-consistent, the probability  $ \PPr(\xi|\theta_k)_{i} $ in a single region is uniquely defined. The number of distinct local targets is thus $2K$ for one region and $2nK$ for the whole. Next, we construct the subset $\mathfrak{T}_{ (\xi|\theta_k)_{i} }$ of $\mathfrak{T}$ composed of all global targets compatible with one particular local target $ (\xi|\theta_k)_{i} $. 
$$ \mathfrak{T}_{ (\xi|\theta_k)_{i} }=\{ (\mathbf{x}|\mathbf{u}) \ /\  x_{i} =\xi \ ; \ u_{i} =\theta_k \} $$
Any local target $ (\xi|\theta_k)_{i}$ in region $\mathcal{R}_{i}$ can be  regarded as an union of global targets
$$ (\xi|\theta_k)_{i} = \bigcup_{(\mathbf{x}|\mathbf{u}) \in  \mathfrak{T}_{ (\xi|\theta_k)_{i} }}  \{ (\mathbf{x}|\mathbf{u})\} .$$
Conversely, any global target $ (\mathbf{x}|\mathbf{u}) $ can be viewed as the intersection of  $n$ local targets in regions $\mathcal{R}_0,\dots,\mathcal{R}_{n-1}$
$$ (\mathbf{x}|\mathbf{u}) = (x_0|u_0) \cap (x_1|u_1)\cap \dots \cap(x_ {n-1} |u_{n-1} ) $$
Let $\mathfrak{I} =\{ j \}$ be a set of integers. 
For $i\in \llbracket 0, n-1 \rrbracket$, define $n$ applications $\Pi_i(k,{j})$ of $ \llbracket 0, K-1 \rrbracket\times\ \mathfrak{I} \rightarrow\mathbf{X}=\{0,1\}$ 
\begin{equation}
\label{xijbis}
\xi_i = \Pi_i(k, {j}) .
\end{equation} 
We will call \emph{projection function} in region $\mathcal{R}_i$ the function $\Pi_i(k,{j}). $
More concisely, it is possible to define a global projection function $\Pi(\gamma, j) $ in terms of  configuration as
\begin{equation}
\label {projecgene}
\Pi(\gamma, j) \ident \Pi_i(k,j) \  \mathrm{with} \ \gamma=k+iK
\end{equation}

Now, we construct a sequence of card$( \mathfrak{I} )$ fine grains $\hat\lambda_{j}$  by the following intersection of local targets:
$${j}\in  \mathfrak{I} \mapsto\hat\lambda_{{j}} $$
\begin{equation}
\label{lambdaj1bis}
\hat\lambda_{{j}} \ident \bigcap_{i=0}^{n-1}  \bigcap_{k_i=0}^{K-1} \Big( \Pi_i(k_i,{j}) |\theta_{k_i}\Big)_i.
\end{equation} 
There are $K^n$ different setting vectors, labelled  $(k_0,k_1,\dots,k_{n-1})$.  Therefore, the maximum number of {distinct} fine grains is $2^{K^n}$. On the other hand, we have to define the outcome $x_i$ for each configuration $(i,k)$ given $j$. There are only $nK$ configurations $\gamma=k+iK$ and the maximum cardinality of $\mathfrak{I}$ is thus $2^{nK}$. We need a selection rule to select $2^{nK}$ ignition states among $2^{n^K}$ potential fine grains.
Let 
\begin{equation}
\label{jmu}
j = \sum_{\mu=0}^{K^n-1} j_{\mu} 2^\mu
\end{equation} 
be the binary expansion of an integer $ j, (0 \le j \le 2^{K^n} -1).$
For any $\mu\in \llbracket 0, K^n-1 \rrbracket$,  there is a unique base-$K$ expansion of $\mu$ as
$$\mu=\sum_{i=0}^{n-1} k_i K^i \ (0 \le k_i \le K-1 ) $$
Define a selection rule $\sigma$,
$$\sigma : \llbracket 0, Kn-1 \rrbracket\mapsto\llbracket 0, K^n-1 \rrbracket $$
$$(k,i) \rightarrow \mu =\sigma(k,i)$$
\emph{A priori}, any function $\sigma$ can be used$\dots$ provided that the linear system, Eq.(\ref{defPLJgene}) below, leads to feasible solutions. The simplest selection function is $\sigma_1(k,i) =k+iK=\gamma$. We have found by simulation that this simple function is sufficient, and more sophisticated selection rules only give a permutation of the label of the ignition states. Therefore, we will use this selection function throughout the rest of the paper, unless stated otherwise explicitly.

Let $j_{\mu}$ be the coefficient of $2^{\mu}$ in the binary expansion of $j$ in Eq.(\ref{jmu}). 
A convenient definition of the projection function is
\begin{equation}
\label{lambdaj}
\Pi_i(k,j)\ident j_{\sigma(k,i)} = j_{k+iK}=j_\gamma.
\end{equation} 
For ease of exposition, when no confusion is possible, we will write alternatively  
$$  \Pi_i(\theta_k, \hat\lambda_j)\ident \Pi_i(k, j).$$ 

Now, we are going to show that this problem is roughly similar to a bipartite Bell-type system with $Kn$ settings. 
\paragraph{Gauge probability distributions}
In the $n$-region-$K$-setting context, we will need $nK$ gauge probability distributions $\g_{\gamma}(\hat\lambda_j)$ (or  simply, $\g_{\gamma}(j)$) for $\gamma= k+iK \in\llbracket 0,nK-1\rrbracket$ with $ i\in\llbracket 0,n-1 \rrbracket $ and $ k\in \llbracket 0,K-1 \rrbracket $. Suppose that the observer in region $\mathcal{R}_{i_0}$ selects the setting $k_0$. Let us derive the gauge distribution $\g_{\gamma_0}(j)$  for $\gamma_0= k_0 +i_0 K$. Let $p_{\gamma_0  j}$ be the unknown probability of $\hat\lambda_j$ for $j\in\mathfrak{I}$. Thus, we have to solve a linear system. 

For each setting vector $\mathbf{u}$ with $u_{i_0}=\theta_{k_0}$ and each outcome vector $\mathbf{x}$, we will write one equation: 
\begin{equation}
\label{defPLJgene}
u_{i_0}=\theta_{k_0} : \forall u_i (i\not= i_0), \forall x_i   : \sum_{j \in \mathbf{J}(\mathbf{x},\mathbf{u})} p_{\gamma_0 j} = \PP(\mathbf{x}| \mathbf{u})
\end{equation}
$j \in \mathbf{J}(\mathbf{x},\mathbf{u})$ means that $\forall i : x_i = \Pi_i(u_i, \hat\lambda_j)$, i.e., 
$\hat\lambda_j \in (\mathbf{x}|\mathbf{u})$. 
For each gauge distribution $\g_{\gamma_0}(j)$ we have card$(\mathfrak{I})$ unknowns $p_{\gamma_0 j}$ and $2^n K^{n-1}$ equations among which it can be shown that only $2(K+1)^{n-1}$ are independent. This system is generally degenerate and provides a set of solutions, $\g_{\gamma_0}(j)= p_{\gamma_0 j} $. The computation fails if it is impossible to find non-negative solutions. Then, we have to try an indirect collapse, with more than one step.

\emph{Consistency conditions}
The gauge distributions are not independent. As in Bell-type  system, Eq. (\ref{compatibility1}), we have a number of consistency conditions, translating complete local consistency, Eq.(\ref{completeConsis}), into gauge formulation. Whatever the gauge distribution $\g_\gamma$ with $\gamma=k+Ki$, provided that $u_i=k$, we have:
\begin{equation} 
\label{compatibilityn}
\PPr (x_0;\dots ; x_{r-1}|u_0;\dots ; u_{r-1}) = \sum_{\lambda_j\in  (x_0;\dots ;x_{r-1}|u_0;\dots ;u_{r-1}) } \g_\gamma(j)
\end{equation}
%
\paragraph{Measurement}
We proceed similarly as for totally correlated Bell-type pairs: In each region $i$, each observer selects his own setting $u_i=\theta_k$. Each configuration $\gamma_i= k+iK$ is transmitted with finite velocity towards the ignition point  ${\mathfrak I}$. Only \emph{one configuration}, say $ \gamma_{\mathrm{igni}} $, is kept at random. 
Now, at ${\mathfrak I}$, we perform a trial in the ignition set $\mathbf{\Lambda}$ using the only probability distribution $\g_{ \gamma_{\mathrm{igni}} } (\hat\lambda_j)$ to draw a single ignition state $\hat\lambda_{j_{\mathrm{igni}} }$. Let $j_{\gamma_i}$  be the coefficients of $2^{\gamma_i}$  in the binary expansion of $j_{\mathrm{igni}}  $. These coefficients, $j_{\gamma_i}=\Pi(\gamma_i ,j_{\mathrm{igni}} ) $ are transmitted with finite velocity towards the end regions $ {\mathcal R}_i$, where the final outcomes are thus $x_i = \Pi(\gamma_i ,j_{\mathrm{igni}} ) $.  

Conversely, let  $\g_{\gamma_{\mathrm{igni}}} (\hat\lambda_j)$ be any gauge distribution,  with $\gamma_{\mathrm{igni}} = k_{\mathrm{igni}}  +i_{\mathrm{igni}}  K$. This gauge selection implies that the setting in region $i_{\mathrm{igni}} $ is $\theta_{k_{\mathrm{igni}}}$. It is possible to compute the probability $\PP(\mathbf{x}|\mathbf{u}) $. We have to collect all ignition states $\hat\lambda_j$ within the target ($\mathbf{x} ,\mathbf{u}$), given that $u_{i_{\mathrm{igni}}} =\theta_{k_{\mathrm{igni}}} $:
Then, we have,
\begin{equation}
\label{defPkj2multi}
\PP(\mathbf{x}|\mathbf{u}) = \sum_{\hat\lambda_j \in (\mathbf{x}|\mathbf{u})} \g_{\gamma _{\mathrm{igni}}} (\hat\lambda_j)
\end{equation}
where  $\hat\lambda_j \in (\mathbf{x}|\mathbf{u})$ means  that $\forall i, x_i=\Pi_i(u_i,\hat\lambda_j)$. 
\subsubsection{Multi-step collapse}
If we fail in computing a feasible set of gauge distributions, $\g_\gamma(\lambda_j)$, we try a two-step collapse. This means a gauge selection of one region and a local trial  to check again  a new one-step collapse, but now in  only $n-1$ regions. If we fail again, we iterate for a three-step collapse, etc. 
Finally, any $n$-region system admits at least a classical collapse in $n$ steps.

\section{Typical examples of multipartite systems}
We are now going to give some examples\footnote{In the previous versions of this paper, theses examples were widely erroneous.}. In bipartite systems, we will consider the cases of partially entangled regions as well as `super-quantum' states. In tripartite system, we will deal with the GHZ and the W states. For simplicity, we will only compute these tripartite systems with two settings, namely, $X$ and $Y$. We have already discussed of their entanglement behaviour with respect to the axis $Z$ in Sec.~\ref{multipartiteEE} (Table~\ref{muatomeWG}). In each example, we will derive the multipartite entanglement entropies.

\subsection{One-region systems}
It may be convenient to have a general gauge formalism, valid in one region systems. Only the direct collapse is relevant. A one-region system with $K$ settings (Table~\ref{proba5num}) has $K$ configurations and $K$ gauges distributions. It requires a working set of $2$ ignition states (Table~\ref{parametern}), as described in Table~\ref{gauge5num2}.

\begin{table}[htb]
\centering
\subtable[%
\label{proba5num}%
{\footnotesize  Probability distribution P(x|u).}
]%
{%
{
\begin{tabular}{||C||C|C|C|C|C||}
\hline
\hline
x_ 0 &\theta_{ 0 }&\theta_{ 1 }&\theta_{ 2 }&\theta_{ 3 }&\theta_{ 4 }\\
\hline
\hline
 0 & p_0 &  p_1 &  p_2 &  p_3 &  p_4\\
\hline
 1 & 1- p_0 & 1- p_1 &  1- p_2 &  1- p_3 &  1- p_4 \\
\hline
\hline
\end{tabular}
} 
}
\\
\subtable[%
\label{gauge5num2}%
{\footnotesize  Gauge distributions $\g_k(j)$ with 2 ignition states, $j = 0$ and $j =2^K-1=31$.}
]%
{
\begin{tabular}{||C||C|C|C|C|C||}
\hline
\hline
j &\g_{ 0 } (j) &\g_{ 1 }(j) &\g_{ 2 }(j) &\g_{ 3 }(j) &\g_{ 4 }(j) \\
\hline
\hline
 0 & p_0 &  p_1 &  p_2 &  p_3 &  p_4\\
\hline
 31 & 1- p_0 & 1- p_1 &  1- p_2 &  1- p_3 &  1- p_4 \\
\hline
\hline
\end{tabular}
} 
\caption{%
\label{oneregion}
{\footnotesize Typical one region system with K=5 settings.}
} 
\end{table}
Such a system is trivially not entangled, and therefore the Bell-CHSH inequalities are irrelevant. Similarly, it is impossible to characterize a super-quantum behaviour, but a given probability distribution $\PP(x|u)$ may force a number of constraints on the settings. 
Actually, when the system mimics a quantum situation, the targets $ (0 |\theta_k)$  and  $(1 |\theta_k)$  describe two qubits $\ket{\xi |\theta_k}$ with $\xi\in\{ 0,1\}$, or alternatively two opposite unit vector $\pm \mathbf{e}_k$ on the surface of the Bloch sphere in $\mathbb{R}^3$, while the density operator of the system, $\rho$,  is described by a vector $ \mathbf{r} $ in the volume of the Bloch sphere still in $\mathbb{R}^3$. For instance, if $p_0=p_1=1$ in Table~\ref{oneregion}, then the system is in the pure state $\ket{0|\theta_0}$ \emph{and} $\ket{0|\theta_1}$ and the constraint is that two settings $\theta_0$ and $\theta_1$ have to be identical, i.e., $\mathbf{e}_0 = \mathbf{e}_1$. 
More generally, the target probability is,
$$
\PP(0 |\theta_k) = (1/2)(1 + \mathbf{r} \cdot \mathbf{e}_k)
\ ; \ 
\PP(1 |\theta_k) = (1/2)(1 - \mathbf{r} \cdot \mathbf{e}_k)
$$ 
where the dot stands for the ordinary scalar vector in $\mathbb{R}^3$. Let $r_k$ be the projection of $\mathbf{r}$ on $ \mathbf{e} _k$ 
$$ r_k= \mathbf{r}\cdot\mathbf{e}_k= 2\PP(0|\theta_k)-1 = 1- 2\PP(1|\theta_k)$$
Now, with a maximum of three independent settings, $\theta_{k_1}$, $\theta_{k_2}$ and  $\theta_{k_3}$, if we have
$$
r_{k_1}^2 + r_{k_2}^2 + r_{k_3}^2   > 1
$$
the constraint is that the three vectors $\mathbf{e} _{k_1}$, $\mathbf{e} _{k_2}$ and  $\mathbf{e} _{k_3}$, cannot be orthonormal in $\mathbb{R}^3$ because the norme of $\mathbf{r}$ is less than $1$.
\subsection[Bipartite  systems]{Bipartite  systems}
\label{bipartite}
We have already described Bell-type system within a particular formalism, devoted to totally correlated entities. Now, we will use the general formalism, valid for any degree of entanglement, including separable states.
\subsubsection{Separable states}
Consider the case of two \emph{independent} entities  $\mathfrak{E}_0$  and $\mathfrak{E}_1$. By definition, the entities are not entangled and the joint probability of the pair is defined by the product 
$$\PP (x_0;x_1|u_0;u_1)=\PPr(x_0|u_0)\times\PPr(x_1|u_1)$$
The entangled entropy matrix of separable states is of course zero.

The collapse mechanism involves physically two steps. Nevertheless, we have always found that a one-step collapse is successful. This opens the possibility of  `purification': a third region entangled with each of the two separable regions can be added to the system and a one-step overall collapse is likely to be triggered from this third region.

\subsubsection{General bipartite 2-setting systems}
Let $K=2$.  When we  drop the condition of total correlation,  Table \ref{bell2} is expanded into Table \ref{gene2}a, depending now on 8 parameters, $q_1, q_2, \dots , q_8$. It is possible to define a one-step collapse as follows: Each gauge distribution is defined by a linear system. We have $2 \times 2=4$ gauge distributions. For each gauge distribution, there are $16$ unknowns for $6$ independent equations when accounting for  local consistency. We have then $16$ ignition states but the rank of the linear system is 6 and thus each distribution can be computed with only 6 non zero components. 
It is easy to compute the partial distribution in each region (Tables \ref{gene2}b and \ref{gene2}c). The maximum bipartite entanglement entropy is equal to $1$ bit. Especially, this maximum is obtained in some totally correlated systems like Bell's states but  also in more exotic objects like the PR-Box (Table. \ref{PRbox}c) below. 

\begin{table}[htb]
\centering%
\subtable [%
$\PP(x_0; x_1|u_0; u_1)$
]
{\footnotesize{%
\begin{tabular}{||C||C|C|C|C||}
\hline
\hline
(x_0, x_1) & (\theta_0,\theta_0 )& (\theta_1,\theta_1)&(\theta_0,\theta_1)&  (\theta_1,\theta_0)  \\  
\hline
\hline
(0, 0) & 1-q_1&1-q_2&1-q_3  & 1-q_4 \\
\hline
(0, 1) & q_5      & q_6     &q_3-q_1+q_5& q_4-q_2 +q_6\\
\hline
(1, 0)& q_7      & q_8     & q_3-q_2+q_8&q_4-q_1 +q_7\\
\hline
(1, 1)& q_1-q_5-q_7&q_2-q_6-q_8&q_1+q_2-q_3-q_5-q_8& q_1+q_2-q_4 -q_6-q_7\\
\hline
\hline
\end{tabular}
}}
%

\vspace{3mm}
\subtable
[
{\footnotesize  $\PPr(x_0|u_0)$.}
]
{\footnotesize
\begin{tabular}{||C||C|C||}
\hline
\hline
 x_0 &{ \theta_0 } & { \theta_1 } \\
\hline
\hline
{ 0 } &  1-q_1+q_5 &  1-q_2+q_6\\
\hline
{ 1 } &  q_1-q_5 &  q_2-q_6
\\
\hline
\hline
\end{tabular}
}
%
%
\qquad
\qquad
%
\subtable
[
{\footnotesize  $\PPr(x_1|u_1)$.}%
]
{\footnotesize 
\begin{tabular}{||C||C|C||}
\hline
\hline
 x_1 &{ \theta_0 } & { \theta_1 } \\
\hline
\hline
{ 0 } &  1-q_1+q_7 &  1-q_2+q_8\\
\hline
{ 1 } &  q_1-q_7 &  q_2-q_8
\\
\hline
\hline
\end{tabular}
} 
\caption{
\label{gene2}
{\footnotesize
General \emph{locally consistent} 2-region system with 2 possible settings $\theta_0$ and $\theta_1$ depending on 8 parameters $q_1$ to $q_8$. All entries have to be non negative. 
}
}
\end{table}
\subsubsection{Revisiting Bell-type systems and the singlet state}
Of course, the general case includes the particular case of totally correlated Bell-type systems described in Sec.~\ref{classical}, and provides new gauge distributions in this context. This inflation of  distributions raises no real problem because ignition states and gauge distributions are not physical entities but only gauge parameters. For instance, we can  recover the previous distributions of the EPR pairs (with the convention recalled in footnote~\ref{footnote1}) by using a particular working set of $2K$ ignition states, namely $j=(2^K+1) D_K(r)$ (with $0\le r\le 2K-1$, $D_K(r)$ being the double-plateau function of order $K$ defined in Sec. \ref{discretesettings}). Then the gauge distributions $\g_k$  are duplicate and $\g_{k+K}=\g_k$.

\paragraph{Singlet state}
Specially, consider the singlet state, $$\ket{\psi^-} = (1/\sqrt{2})(\ket{01}-\ket{10}),$$ describing a pair of isotropic opposite spins with the usual convention,  when e.g., $\ket{0}$ stands for spin up and $\ket{1}$ for spin down, irrespective of the region.   
This situation is generally considered as purely quantum and even inconceivable in classical physics. However,  it is easily shown that this belief is baseless in contextual systems. Now, the probability distribution of the singlet state is given in Table \ref{singlet}a for three settings, $X$, $Y$ and $Z$. It is easy to compute a working set of gauge distributions. It is even possible to find a quadruplet of four ignition states, namely $\lambda_7$, $\lambda_{28}$, $\lambda_{42}$, and $\lambda_{49}$ using a single  gauge distribution $\g(j)$ as shown in Table \ref{singlet}b. Therefore, the singlet state is compatible with a set of four `hidden variables' located at the ignition point. This proves that a potential pair of both isotropic and opposite vectors is surprisingly a classical although contextual concept. The total entanglement, Eq.(\ref{Tcorrelation}) is maximum and equal to $1$ bit, while the entropy matrix $\s_2$ is given in Table~\ref{singlet}c. Nevertheless, with the three settings $X$, $Y$ and $Z$, the CHSH average is just equal to $2$, and this entanglement is not detected by this criterion%
\footnote{%
However with four settings, it is well known that the Tsirelson bound $2\sqrt{2}$ is reached.
}
.

\begin{table}
\centering
\subtable
[
Probabilities $P(\mathbf{x}|\mathbf{u})$  with three settings ($i\not= j$). 
]
{
\begin{tabular}{||>{$}c<{$}||>{$}c<{$}|>{$}c<{$}||}
\hline
\hline
\mathbf{x}\backslash \mathbf{u} &(\theta_i, \theta_i )&(\theta_i,\theta_j)\\
\hline
\hline
 0  0 & 0 &  1/4 \\
\hline
 0  1 & 1/2 &  1/4 \\
\hline
 1  0 & 1/2 &  1/4\\
\hline
 1  1 & 0 &  1/4 \\
\hline
\hline
\end{tabular}
}
\qquad
\subtable
[
{\footnotesize%
Gauge distribution. 
}
]
{
\begin{tabular}{||>{$}c<{$}||>{$}c<{$}||}
\hline
\hline
j &  \g (j) \\
\hline
\hline
{ 7 } &  1/4 \\
\hline
{ 28 } &  1/4 \\
\hline
{ 42 } &  1/4 \\
\hline
{ 49 } &  1/4 \\
\hline
\hline
\end{tabular}
} 
\qquad 
\subtable
[
{\footnotesize%
Entropy Matrix. 
}
]
{
\begin{tabular}{||>{$}c<{$}||>{$}c<{$}  |>{$}c<{$}  |>{$}c<{$}    || }
\hline
\hline
\s_2 &  X & Y& Z\\
\hline
\hline
{ X } &  1 & 0 & 0 \\
\hline
{ Y } &  0 & 1 & 0 \\
\hline
{ Z } &  0 & 0 & 1 \\
\hline
\hline
\end{tabular}
} 
\caption
{
\footnotesize{%
\label{singlet} %
Singlet state: Classical simulation of the two region singlet state with three settings $\theta_0=X$, $\theta_1=Y$ and $\theta_2=Z$: The six gauge distributions $\g_{ \gamma } (j) = \g(j)$ are here identical. This means that the singlet state can be described by a quadruplet of equally likely `hidden variables', namely $\lambda_7$, $\lambda_{28}$, $\lambda_{42}$, and $\lambda_{49}$. At each run, each observer submits its configuration labelled $\gamma_i$ to a referee located at the ignition point. When she received the first configuration, the referee draws at random one of the four ignition states, say $\lambda_{j_{\mathrm{igni}}}$, and transmits backwards to the observers the final outcomes $x_i=\Pi(j_{\mathrm{igni}},\gamma_i)$. Such an isotropic pair of opposite spins in inconceivable in non-contextual systems. Finally, the singlet state is maximally entangled with a total entanglement of $1$ bit. 
}
}
\end{table} 
\subsubsection{PR-Box}
Another example of interest is the so-called `PR-Box' proposed by Popescu and Rohrlich~\cite{popescu} and recalled in Table~\ref{PRbox}a. This example is supposed to be hypothetical or `super-quantum' because it exceeds the Tsirelson bound.
Now, we see that the box corresponds to an `ordinary' 2-region-2-setting-system, with $q_1=q_3=q_4=q_6=q_8=1/2$, $q_2=1$ and $q_5=q_7=0$ in Table~\ref{gene2}.  
Define $A = \theta_0$, $A' = \theta_1$, $B = \theta_1$ and $B' = \theta_0$ in Eq.(\ref{tsirelsonbound}.  We have
$$|\s (A, B) + \s (A', B) + \s (A, B') - \s (A', B')|  = 4 > 2\sqrt{2}.$$
Therefore, the system is `super-quantum'. 
 The collapse can be completed in classical physics, using e.g. the gauge distributions given in Table~\ref{PRbox}b. A possible implementation as a classical game using a fair coin is proposed in Table~\ref{PRbox} caption.  The system is clearly maximally entangled and its entanglement entropy is given in Table~\ref{PRbox}c. This is an example of maximally entangled state which is not totally correlated (as defined in Sec. \ref{totalcorrelation}). The box  is known to solve the problem of `communication complexity'~\cite{yao,buhrman} in the sense that  all distributed computations can be performed with a trivial amount of communication, i.e., with one bit~\cite{vandam}. This seems very surprising because the device can be easily implemented, but on the other hand, this classical realization involves a stage of classical communication.  Finally, we have found that the box also describes the second-step collapse of the `super-quantum'-GHZ-states (Sec. \ref{tripartite} below), demonstrating the `super-quantum' behaviour of this state (see Sec. \ref{revisiting}).

\begin{table}[htb] 
\centering
\subtable 
[
$\PP(\mathbf{x}|\mathbf{u}$  
]
{
\begin{tabular}{||>{$}c<{$}||>{$}c<{$}|>{$}c<{$}|>{$}c<{$}|>{$}c<{$}||   }
\hline
\hline
x_ 0 x_ 1 &(\theta_{ 0 }\theta_{ 0 })&(\theta_{ 0 }\theta_{ 1 })&(\theta_{ 1 }\theta_{ 0 })&(\theta_{ 1 }\theta_{ 1 })\\
\hline
\hline
 0  0 & 1/2 &  1/2 &  1/2 &  0\\
\hline
 0  1 & 0 &  0 &  0 &  1/2\\
\hline
 1  0 & 0 &  0 &  0 &  1/2\\
\hline
 1  1 & 1/2 &  1/2 &  1/2 &  0\\
\hline
\hline
\end{tabular}
}
\qquad
\subtable
[
{\footnotesize%
Gauge distributions $\g_\gamma (j)$
}
]
{
\begin{tabular}{||>{$}c<{$}||>{$}c<{$}|>{$}c<{$}|>{$}c<{$}|>{$}c<{$}||   }
\hline
\hline
j &  \g_{ 0 } (j) &  \g_{ 1 } (j) &  \g_{ 2 } (j) &  \g_{ 3 } (j)\\
\hline
\hline
{ 0 } &  1/2 &  0 &  1/2 &  0\\
\hline
{ 6 } &  0 &  1/2 &  0 &  1/2\\
\hline
{ 9 } &  0 &  1/2 &  0 &  1/2\\
\hline
{ 15 } &  1/2 &  0 &  1/2 &  0\\
\hline
\hline
\end{tabular}
} 
\vspace{3mm}
\subtable
[
{\footnotesize%
Entanglement entropy (in bits)
}
]
{
\begin{tabular}{||>{$}c<{$}||>{$}c<{$}|>{$}c<{$}|>{$}c<{$}|>{$}c<{$}||   }
\hline
\hline
(u_0,u_1) & (\theta_0, \theta_0)& (\theta_0, \theta_1) & (\theta_1, \theta_0) & (\theta_1, \theta_1) \\
\hline
\hline
\s_2(u_0,u_1) & 1 & 1 &1&1\\
\hline
\hline
\end{tabular}
} 
\caption
{%
\footnotesize{%
\label{PRbox} %
PR-Box: (a) Probabilities $\PP(\mathbf{x}|\mathbf{u})$ proposed by Popescu and Rohrlich~\cite{popescu}. This system is maximally entangled (c) and `super-quantum'. The process can be implemented in classical physics, using the gauge distributions given in (b) with 4 ignition states (out of 16). This proves that the PR-box, while exceeding the Tsirelson bound Eq.(\ref{tsirelsonbound}), can be implemented in classical physics. We note that each distribution has only two equally likely outcomes. Therefore, it is possible to emulate the system as a simple game between two players $\mathcal{O}_0$ and  $\mathcal{O}_1$ using a fair coin: Each player $\mathcal{O}_i$ chooses her/his setting $k_i =0$ or $1$ and sends to a referee (located at the ignition point)  her/his  configuration, $\gamma_i=k_i+2i$. The referee selects at random between $\gamma_0$ and $\gamma_1$ a gauge configuration,  $ \gamma_\mathrm{igni} \in\llbracket 0,3\rrbracket$, and tosses the coin. The coin is regarded as a gauge distribution $ \g_{\gamma_\mathrm{igni}} (j)$ with an assignment of the relevant ignition states $j$ to `head' and `tail' respectively, e.g.,`head' for $j\in\{ 3,6\}$ and `tail' for $j\in\{ 9,15\}$. Then, the referee draws an ignition state $j$. For example, $j=0$ or $15$ if $ \gamma_\mathrm{igni} =0$. At last, the final outcomes are $x_i=\Pi(\gamma_i,j)$ respectively for $i=0,1$. For example, if $ \gamma_0=0$, $\gamma_1=1$, $ \gamma_\mathrm{igni} =0$ and $j=0$, we have $x_0=x_1=0$. The entanglement entropy  $\s_2 (u_0,u_1) =\mathcal{E}(u_0,u_1) $ is given in (c). We have $\s_2=1=n-1$  where $n=2$ is the number of regions. This proves that the system is maximally entangled while not totally correlated.
}
}
\end{table} 

\subsection[Tripartite entangled systems]{Tripartite entangled systems}
\label{tripartite}
It is well known that there are two irreducible families of tripartite quantum entangled systems, referred to as  GHZ states and W-states. We will first analyse the W-states with only two settings.
\subsubsection{W-type states}
Consider an ensemble of three entities $\{{\mathfrak E}_0, {\mathfrak E}_1, {\mathfrak E}_{2}\} $,  placed respectively in three distant space regions $ {\mathcal R}_0, {\mathcal R}_1$ and ${\mathcal R}_2$ located in a plane  $(X,Y)$. 
The physical system is equivalent to a set of three spin 1/2 particles, initially entangled in the state
$$\ket{\psi}= \frac{1}{\sqrt{3}}(\ket{001}_Z+\ket{010}_Z +\ket{100}_Z)$$
specified along an axis $Z$ perpendicular to the plane $(X,Y)$, where, e.g.,  $\ket{0}_Z$ stands for spin up and $\ket{1}_Z$ for spin down. Later the spins of the three particles are measured independently in the distant regions along either the axis $X$ or $Y$.
In each region, an observer selects freely a setting $u_i$ ($i=0,1$ or $2$). Therefore, we consider the tripartite system with only two settings\footnote{We have nevertheless considered the third setting $Z$ in Sec.~\ref{multipartiteEE} to compute its entanglement entropy (Table~\ref{muatomeWG}).}.  Let $\mathbf{\Theta}=\{ X,Y\}$. 
%
\begin{table}[htbp] 
\begin{center}
\subtable 
[
W-state with two settings: Target probabilities $\PP(\mathbf{x}|\mathbf{u})$  
]
{
{
\begin{tabular}{||C||C|C|C|C||}
\hline
\hline
  & (XXX) & (XXY) & (XYX) & (XYY) \\
\mathbf{x}\backslash \mathbf{u} & (YYY) & (YYX) & (YXY) & (YXX) \\
\hline
\hline
 (0  0  0) & 3/8 &  5/24 &  5/24 &  5/24 \\
\hline
 (0  0  1) & 1/24 &  5/24 &  1/24 &  1/24 \\
\hline
 (0  1  0) & 1/24 &  1/24 &  5/24 &  1/24 \\
\hline
 (0  1  1) & 1/24 &  1/24 &  1/24 &  5/24 \\
\hline
 (1  0  0) & 1/24 &  1/24 &  1/24 &  5/24 \\
\hline
 (1  0  1) & 1/24 &  1/24 &  5/24 &  1/24 \\
\hline
 (1  1  0) & 1/24 &  5/24 &  1/24 &  1/24 \\
\hline
 (1  1  1) & 3/8 &  5/24 &  5/24 &  5/24 \\
\hline
\hline
\end{tabular}
}
}
\\
\vspace{3mm}
\subtable
[
{\footnotesize%
Gauge distributions $\g_\gamma(j)$
}
]
{
{
\begin{tabular} {||C||C|C|C|C|C|C|||C||C|C|C|C|C|C||}
\hline
\hline
j &  \g_{ 0} (j) &  \g_{ 1} (j) &  \g_{ 2} (j) &  \g_{ 3} (j) &  \g_{ 4} (j) &  \g_{ 5} (j) & j &  \g_{ 0} (j) &  \g_{ 1} (j) &  \g_{ 2} (j) &  \g_{ 3} (j) &  \g_{ 4} (j) &  \g_{ 5} (j)\\
\hline
\hline
{ 0} &  1/6 &  1/6 &  1/6 &  1/6 &  1/6 &  1/6 & { 31} &  0 &  0 &  0 &  0 &  1/6 &  0\\
\hline
{ 2} &  0 &  0 &  0 &  0 &  1/24 &  0 & { 33} &  0 &  0 &  1/24 &  1/24 &  0 &  0\\
\hline
{ 5} &  0 &  0 &  0 &  0 &  1/24 &  5/24 & { 34} &  0 &  0 &  5/24 &  1/24 &  0 &  0\\
\hline
{ 9} &  0 &  0 &  0 &  0 &  1/24 &  1/24 & { 36} &  1/24 &  1/24 &  0 &  0 &  0 &  0\\
\hline
{ 10} &  0 &  0 &  0 &  0 &  1/6 &  0 & { 37} &  1/24 &  0 &  1/24 &  0 &  0 &  1/24\\
\hline
{ 14} &  0 &  0 &  0 &  0 &  1/24 &  0 & { 38} &  0 &  1/24 &  1/24 &  0 &  0 &  1/24\\
\hline
{ 17} &  0 &  0 &  1/24 &  5/24 &  0 &  0 & { 40} &  5/24 &  1/24 &  0 &  0 &  0 &  0\\
\hline
{ 18} &  0 &  0 &  1/24 &  1/24 &  0 &  0 & { 41} &  1/24 &  0 &  0 &  1/24 &  0 &  1/24\\
\hline
{ 20} &  1/24 &  5/24 &  0 &  0 &  0 &  0 & { 42} &  0 &  5/24 &  0 &  5/24 &  0 &  5/24\\
\hline
{ 21} &  5/24 &  0 &  5/24 &  0 &  5/24 &  0 & { 47} &  0 &  0 &  0 &  0 &  0 &  1/6\\
\hline
{ 22} &  0 &  1/24 &  1/24 &  0 &  1/24 &  1/24 & { 55} &  0 &  0 &  1/6 &  0 &  0 &  0\\
\hline
{ 24} &  1/24 &  1/24 &  0 &  0 &  0 &  0 & { 59} &  0 &  0 &  0 &  1/6 &  0 &  0\\
\hline
{ 25} &  1/24 &  0 &  0 &  1/24 &  1/24 &  0 & { 61} &  1/6 &  0 &  0 &  0 &  0 &  0\\
\hline
{ 26} &  0 &  1/24 &  0 &  1/24 &  1/24 &  1/24 & { 62} &  0 &  1/6 &  0 &  0 &  0 &  0\\
\hline
\hline
\end{tabular}
} 
} 
\vspace{3mm}
\subtable
[
{\footnotesize%
Entanglement entropy (in bits)
}
]
{
{
\begin{tabular}{||C||C|C|C|C|C|C|C|C||}
\hline
\hline
\mathbf{u} & { (XXX) } & { (XXY) } & { (XYX) } & { XYY }  & { (YXX) } & { (YXY) }  & { (YYX) } & { (YYY) }   \\
\hline
{ \s_3 (\mathbf{u}) } &  0.257 &  0 & 0 & 0 & 0 & 0 & 0 & 0.257\\
\hline
{ \mathcal{E}(\mathbf{u}) } &  0.792 &  0.350 & 0.350 & 0.350 & 0.350 & 0.350 & 0.350 & 0.792\\
\hline
\hline
\end{tabular}
} 
} 
\caption
{%
\footnotesize{%
\label{toutW} %
W-states: 
(a) Conditional probability $\PP(x_0;x_1;x_2|u_0;u_1;u_2)$. 
(b) Gauge distribution working set with 28 ignition states $\lambda_j$ (out of 64) and 6 gauge distributions $\g_\gamma (j)$.
(c) Tripartite entanglement entropy $\s_3 (u_0, u_1, u_2) $ and total entanglement $\mathcal{E}  ( (u_0, u_1, u_2) ) $ (in bits). Since $\mathcal{E}(X,X,X) = \mathcal{E} (Y,Y,Y)= 0.792 < 2$~bits, this state is not maximally entangled.
}
}
\end{center}
\end{table} 
%
%
The conditional probabilities
 $\PP(x_0;x_1;x_2|u_0;u_1;u_2)=\PP(\mathbf{x}|\mathbf{u})$  are given in Table \ref{toutW}a and clearly depends upon the settings in the three regions. 
Complete consistency holds  and, e.g.,  the partial probability $\PPr(x_0|u_0)$ in region $\mathcal{R}_0$ does not depend on the setting in $\mathcal{R}_1$ and $\mathcal{R}_2$. Thus, we have  $ (\forall u_0, \forall u_1, \forall u_2 \in \mathbf{\Theta})$:    

\begin{equation}
\label{localW}
\PPr(x_0|u_0)=\sum_{x_1=0}^{1 }  \sum_{x_2=0}^{1 } \PP(x_0;x_1;x_2|u_0;u_1;u_2) =\frac{1}{2} 
\end{equation}
and similarly for ports $\mathcal{R}_1$ and $\mathcal{R}_2$. 
If we ignore the last port,  the probability $\PPr(x_0;x_1|u_0;u_1)$ in regions $ (\mathcal{R}_0, \mathcal{R}_1), $ does not depend on the setting  in $\mathcal{R}_2$. Therefore, we have $ (\forall u_0, \forall u_1, \forall u_2 \in \mathbf{\Theta})$:  
\begin{equation}
\label{completeW}
\PPr(x_0;x_1|u_0;u_1)= \sum_{x_2=0}^{1 } \PP(x_0;x_1;x_2|u_0;u_1;u_2)  
\end{equation}
The result is given in Table \ref{partialW}a.
%
\begin{table} [htbp]
\begin{center}
\subtable 
[
Partial  probability $\PPr(x_0;x_1|u_0;u_1)$. 
]
{
{
\begin{tabular}{||C||C|C|C|C||}
\hline
\hline
\mathbf{x}\backslash \mathbf{u} & (XX) & (XY) & (YX) & (YY)\\
\hline
\hline
 (0  0) & 5/12 &  1/4 &  1/4 &  5/12\\
\hline
 (0  1) & 1/12 &  1/4 &  1/4 &  1/12\\
\hline
 (1  0) & 1/12 &  1/4 &  1/4 &  1/12\\
\hline
 (1  1) & 5/12 &  1/4 &  1/4 &  5/12\\
\hline
\hline
\end{tabular}
}
}
\qquad
\subtable 
[
Gauge  probability $\PP'(x_0;x_1|u_0;u_1)$. 
]
{
{
\begin{tabular}{||C||C|C|C|C||}
\hline
\hline
\mathbf{x}\backslash \mathbf{u} & (XX) & (XY) & (YX) & (YY)\\
\hline
\hline
 (0  0) & 3/4 &  5/12 &  5/12 &  5/12\\
\hline
 (0  1) & 1/12 &  5/12 &  1/12 &  1/12\\
\hline
 (1  0) & 1/12 &  1/12 &  5/12 &  1/12\\
\hline
 (1  1) & 1/12 &  1/12 &  1/12 &  5/12\\
\hline
\hline
\end{tabular}
}
}
\\
\subtable
[
{\footnotesize%
Partial entropy $\s_2$ (in bits)
}
]
{
{
\begin{tabular}{||C||C|C||}
\hline
\hline
u_0 \backslash u_1 &{ X } & { Y } \\
\hline
\hline
{ X } &  0.35 &  0\\
\hline
{ Y } &  0 &  0.35\\
\hline
\hline
\end{tabular}
} 
} 
\qquad
\subtable
[
{\footnotesize%
Gauge entropy $ \s_2 '$ (in bits)
}
]
{
{
\begin{tabular}{||C||C|C||}
\hline
\hline
u_0 \backslash u_1 &{ X } & { Y } \\
\hline
\hline
{ X } &  0.09 &  0\\
\hline
{ Y } &  0 &  0.35\\
\hline
\hline
\end{tabular}
} 
} 
%
\caption
{%
\label{partialW} %
\footnotesize{%
Partial and gauge bipartite subsystems in regions $\mathcal{R}_0$ and $\mathcal{R}_1$  of the W-state with two settings. 
(a) The partial probability in regions $\mathcal{R}_0$ and $\mathcal{R}_1$ describes an entangled 2-region 2-setting system. 
(b) Gauge probability in regions $\mathcal{R}_0$ and $\mathcal{R}_1$ following a first collapse in region $\mathcal{R}_2$ with the setting $u_2 = X$ and the outcome $x_2=0$. 
(c) The partial entropy computed from (a) shows that the maximum total entanglement is $\s_2(X,X)=\mathcal{E}(X,X) = \s_2(Y,Y)= \mathcal{E}(Y,Y) = 0.35$ bit $ < 1$ bit. The system is not maximally entangled.
(d) The gauge entropy computed from (b) shows that the total entanglement is $\mathcal{E} '(X,X) = 0.09$ bit while $\mathcal{E} '(Y,Y) = 0.35$ bit is unchanged.
}
}
\end{center}
\end{table} 

%
This is the probability distribution of an entangled system obtained from the general 2-region 2-setting system (Table \ref{gene2}a) for $q_1=q_2=7/12$, $q_3=q_4=3/4$ and $q_5=q_6=q_6=q_7=q_8=1/12$. It is easy to compute the bipartite entanglement entropy matrix. The result for one pair of regions is given in Table \ref{partialW}c. As a result the bipartite degree of entanglement is $e_2 = 3$ and the bipartite entanglement entropy $\s_2$, equal to the total entanglement is $ \mathcal{E}(X,X) = \mathcal{E}(Y,Y) = 0.35 < 1$ bit. Therefore, the entanglement is not maximum and, for instance, is not detected by the CHSH inequalities.

\paragraph{Two-step collapse}
It is possible to proceed to a partial collapse, e. g. in region $\mathcal{R}_2$. We have to select a setting, e.g., $u_2 =X$. The partial probability distribution in $\mathcal{R}_2$ is given by  Eq.(\ref{localW}) and the two outcomes are equally likely. Suppose that we draw, e.g., the outcome $x_2 = 0$. In regions $\mathcal{R}_0$ and $\mathcal{R}_1$ the new probability distribution is easily computed and the result is given in Table~\ref{partialW}b. This is a two-region two-setting system with $q_1=1/4$,$q_2=q_3=q_4=7/12$ and $q_5=q_6=q_7=q_8=1/12$. We have computed also the total correlation in Table~\ref{partialW}d. The collapse can be completed in one or two steps. 
  
\paragraph{One-step collapse}
Coming back to the full system, the number of configurations is $nK=3\times 2=6$. We can use the standard projection function, Eq. (\ref {projecgene}), with just $2^{nK} = 2^6=64 $ ignition states and 6 stochastic gauge distributions, $\g_\gamma(\lambda_j)$ with  $\gamma=0,\dots,5$. The result is given in Table~\ref{toutW}b.


\subsubsection{GHZ-type states} 
\label{paraGHZ}
The second family of tripartite quantum entangled system is the GHZ-states, defined by Greenberger et al~\cite{GHZ}. The physical system is a set of three spin 1/2 particles, initially entangled in the state 
\begin{equation}
\label{ketGHZ}
\ket{\psi}= \frac{1}{\sqrt{2}}(\ket{000}_Z+\ket{111}_Z)
\end{equation}
specifies along an axis $Z$ perpendicular to the plane $(X,Y)$. Again, the spins of the three particles are measured along either the axis $X =\theta_0$ or the axis $Y = \theta_1$ and we only consider a 2-setting system, with $\mathbf{\Theta}=\{ X, Y\} $.

The target probabilities $\PP(\mathbf{x}|\mathbf{u})$  are given in Table \ref{toutGHZ}a. In non-contextual systems, one would expect that each local outcome only depends on a local trial. This assumption is easily checked by assuming the existence of  local random functions, defined as $\Xi({\mathcal R}_i|\theta_ k)=x_i$ when the entity $\mathfrak{E}_i$ is measured with the setting $\theta_k$. Let $f(u_0,u_1,u_2)\equiv \Xi({\mathcal R}_0|u_0)+\Xi({\mathcal R}_1|u_1)+\Xi({\mathcal R}_2|u_2) \pmod {2}$. Clearly, $f(X,X,X)\equiv f( Y,Y,X) + f(Y,X ,Y) + f(X,Y,Y)  \pmod {2}$. The assumption is falsified by Table \ref{toutGHZ}a, since $f(X,X,X)\equiv 1 $ with certainty, while $ f( Y,Y,X)\equiv 0 $,  $ f(Y,X,Y)  \equiv 0 $ and $ f(X,Y,Y)  \equiv 0$ with certainty. Therefore, the system is contextual.
%
\begin{table} [htb] 
\begin{center}
\subtable 
[
GHZ state with two settings: Target probabilities $\PP(\mathbf{x}|\mathbf{u})$  
]
{
\begin{tabular}{||C||C|C|C|C|C|C|C|C||}
\hline
\hline
\mathbf{x}\backslash\mathbf{u}& (X X X) &(X X Y) & (X Y X) & (X Y Y) & (Y X X) & (Y X Y) & (Y Y X) & (Y Y Y) \\
\hline
\hline
 (0 0 0)& 1/4 &  1/8 &  1/8 &  0   &  1/8 &  0   &  0   &  1/8\\
\hline
 (0 0 1)& 0   &  1/8 &  1/8 &  1/4 &  1/8 &  1/4 &  1/4 &  1/8\\
\hline
 (0 1 0)& 0   &  1/8 &  1/8 &  1/4 &  1/8 &  1/4 &  1/4 &  1/8\\
\hline
 (0 1 1)& 1/4 &  1/8 &  1/8 &  0   &  1/8 &  0   &  0   &  1/8\\
\hline
 (1 0 0)& 0   &  1/8 &  1/8 &  1/4 &  1/8 &  1/4 &  1/4 &  1/8\\
\hline
 (1 0 1)& 1/4 &  1/8 &  1/8 &  0   &  1/8 &  0   &  0   &  1/8\\
\hline
 (1 1 0)& 1/4 &  1/8 &  1/8 &  0   &  1/8 &  0   &  0   &  1/8\\
\hline
 (1 1 1)& 0   &  1/8 &  1/8 &  1/4 &  1/8 &  1/4 &  1/4 &  1/8\\
\hline
\hline
\end{tabular}
}
\\ 
\vspace{3mm}
\subtable 
[
{\footnotesize%
Gauge distributions $\g_\gamma(j)$
}
]
{
{\footnotesize
\begin{tabular} {||C||C|C|C|C|C|C|||C||C|C|C|C|C|C||}
\hline
\hline
j &  \g_{ 0} (j) &  \g_{ 1} (j) &  \g_{ 2} (j) &  \g_{ 3} (j) &  \g_{ 4} (j) &  \g_{ 5} (j) & j &  \g_{ 0} (j) &  \g_{ 1} (j) &  \g_{ 2} (j) &  \g_{ 3} (j) &  \g_{ 4} (j) &  \g_{ 5} (j)\\
\hline
\hline
{ 2} &    0   &    0   &   1/8  &    0   &   1/8  &    0   & { 34} &    0   &    0   &    0   &   1/8  &    0   &    0  \\
\hline
{ 3} &    0   &   1/8  &    0   &    0   &    0   &   1/8  & { 38} &    0   &   1/8  &    0   &    0   &    0   &   1/8 \\
\hline
{ 5} &    0   &    0   &    0   &    0   &    0   &   1/8  & { 39} &    0   &    0   &   1/8  &    0   &    0   &    0  \\
\hline
{ 7} &   1/8  &    0   &    0   &   1/8  &   1/8  &    0   & { 40} &    0   &   1/8  &    0   &    0   &    0   &    0  \\
\hline
{ 8} &   1/8  &    0   &    0   &    0   &   1/8  &    0   & { 41} &    0   &    0   &    0   &   1/8  &    0   &   1/8 \\
\hline
{ 10} &    0   &    0   &    0   &    0   &    0   &   1/8  & { 45} &   1/8  &    0   &    0   &    0   &    0   &    0  \\
\hline
{ 12} &    0   &    0   &    0   &   1/8  &    0   &    0   & { 47} &    0   &    0   &    0   &    0   &    0   &   1/8 \\
\hline
{ 13} &    0   &   1/8  &   1/8  &    0   &   1/8  &    0   & { 48} &    0   &   1/8  &    0   &   1/8  &    0   &    0  \\
\hline
{ 17} &   1/8  &    0   &    0   &   1/8  &   1/8  &    0   & { 49} &    0   &    0   &   1/8  &    0   &    0   &    0  \\
\hline
{ 20} &    0   &   1/8  &   1/8  &    0   &   1/8  &    0   & { 52} &   1/8  &    0   &    0   &    0   &    0   &    0  \\
\hline
{ 26} &    0   &   1/8  &    0   &   1/8  &    0   &    0   & { 54} &    0   &    0   &   1/8  &    0   &    0   &    0  \\
\hline
{ 27} &    0   &    0   &   1/8  &    0   &   1/8  &    0   & { 57} &   1/8  &    0   &    0   &    0   &    0   &    0  \\
\hline
{ 28} &   1/8  &    0   &    0   &    0   &    0   &   1/8  & { 59} &    0   &    0   &    0   &   1/8  &    0   &    0  \\
\hline
{ 30} &    0   &    0   &    0   &    0   &   1/8  &    0   & { 62} &    0   &   1/8  &    0   &    0   &    0   &    0  \\
\hline
{ 32} &   1/8  &    0   &   1/8  &    0   &    0   &   1/8 &&&&&&&\\
\hline
\hline
\end{tabular}
} 
} 
\\
\vspace{3mm}
\subtable
[
{\footnotesize%
Entropy (in bits) 
}
]
{
{\footnotesize
\begin{tabular}{||C||C|C|C|C|C|C|C|C||}
\hline
\hline
(u_0, u_1, u_2) & { (XXX) } & { (XXY) } & { (XYX) } & { (XYY) }  & { (YXX) } & { (YXY) }  & { (YYX )} & { (YYY) }   \\
\hline
{ \s_3 (u_0, u_1, u_2) } &  -1 &  0         & 0              & -1             & 0             & -1             & -1             & 0              \\
\hline
{ \mathcal{E}(u_0, u_1, u_2) } &  1 &  0 & 0           & 1               & 0             & 1              & 1              & 0              \\
\hline
\hline
\end{tabular}
} 
} 
\caption
{%
\footnotesize{%
\label{toutGHZ} %
GHZ states for two settings $X$ and $Y$ : 
(a) Conditional probability $\PP(x_0;x_1;x_2|u_0;u_1;u_2)$. Local realism is  supposed to be falsified by computing a convenient function, $f(u_0,u_1,u_2)\equiv x_0+x_1+x_2 \pmod {2}$. Whatever the outcomes, we see by inspection that $f(X,X,X)\equiv 0 $ with certainty, while $f( X,Y,Y)\equiv 1 $, $f(Y,X,Y)\equiv 1 $ and $f(Y,Y,X) \equiv 1$ with certainty. 
(b) Gauge distribution working set with 29 ignition states (out of 64).
(c) Tripartite entanglement entropy $\s_3 (u_0, u_1, u_2) $ and total entanglement $\mathcal{E}(u_0, u_1, u_2) $ (in bits). Partial subsystems are separable. Therefore, bipartite entanglement entropies $\s_2 (u_i, u_j)$ are identically zero. 
}
}
\end{center}
\end{table} 

Complete local consistency holds. For instance,  the partial  probability $\PPr(x_0|u_0)$ in region $\mathcal{R}_0$ does not depend on the setting in $\mathcal{R}_1$ and $\mathcal{R}_2$ (Table~\ref{partialGHZ}a). Thus, we have  $ (\forall u_0, \forall u_1, \forall u_2 \in \mathbf{\Theta})$:    
\begin{equation}
\label {localGHZ}
\PPr(x_0|u_0)=\sum_{x_1=0}^{1 }  \sum_{x_2=0}^{1 } \PP(x_0;x_1;x_2|u_0;u_1;u_2) =\frac{1}{2} 
\end{equation}
and similarly for ports $\mathcal{R}_1$ and $\mathcal{R}_2$. 
If we ignore the last port,  the probability $\PPr(x_0;x_1|u_0;u_1)$ in regions $ (\mathcal{R}_0, \mathcal{R}_1), $ does not depend on the setting  in $\mathcal{R}_2$. Therefore, we have $ (\forall u_0, \forall u_1, \forall u_2 \in \mathbf{\Theta})$:
\begin{equation}
\label {completeGHZ}
\PPr(x_0;x_1|u_0;u_1)= \sum_{x_2=0}^{1 } \PP(x_0;x_1;x_2|u_0;u_1;u_2) =\frac{1}{4} 
\end{equation}
This is obtained from the general 2-region 2-setting system (Table \ref{gene2}a) for $q_1=q_2=q_3=q_4=3/4$ and $q_6=q_6=q_7=q_8=1/4$. Thus, the two regions are separable and the probabilities given by Eq(\ref
{completeGHZ}) are simply the product of the probability obtained in Eq.(\ref{localGHZ}). Therefore the entanglement scheme is $(0,1)$. The bipartite entanglement entropy for any pair of region is zero (Table~\ref{partialGHZ}c).  By contrast, depending upon the settings, the signed tripartite entanglement entropy is equal to $0$ or $-1$ bit and the total entanglement to $0$ or $1$ bit(Table~\ref{toutGHZ}c). We have seen that the GHZ-state is maximally entangled with the third setting $(Z,Z,Z)$.

%
\begin{table} [htb]
\begin{center}
\subtable 
[
Partial  probability $\PPr(x_0;x_1|u_0;u_1)$. 
]
{
{
\begin{tabular}{||C||C|C|C|C||}
\hline
\hline
\mathbf{x}\backslash \mathbf{u} & (XX) & (XY) & (YX) & (YY)\\
\hline
\hline
 (0  0) & 1/4  &   1/4  &   1/4  &   1/4 \\
\hline
 (0  1) & 1/4  &   1/4  &   1/4  &   1/4 \\
\hline
 (1  0) & 1/4  &   1/4  &   1/4  &   1/4 \\
\hline
 (1  1) & 1/4  &   1/4  &   1/4  &   1/4 \\
\hline
\hline
\end{tabular}
}
}
\qquad
\subtable 
[
Gauge  probability $\PP'(x_0;x_1|u_0;u_1)$. 
]
{
{
\begin{tabular}{||C||C|C|C|C||}
\hline
\hline
\mathbf{x}\backslash \mathbf{u} & (XX) & (XY) & (YX) & (YY)\\
\hline
\hline
 (0  0) & 1/2    &   1/4  &   1/4  &    0  \\
\hline
 (0  1) & 0      &   1/4  &   1/4  &   1/2 \\
\hline
 (1  0) & 0      &   1/4  &   1/4  &   1/2 \\
\hline
 (1  1) & 1/2    &   1/4  &   1/4  &    0  \\
\hline
\hline
\end{tabular}
}
}
\\
\subtable
[
{\footnotesize%
Partial entropy $\s_2$
}
]
{
{
\begin{tabular}{||C||C|C||}
\hline
\hline
u_0 \backslash u_1 &{ X } & { Y } \\
\hline
\hline
{ \ \ X \ \ } &   \ \  0 \ \   &   \ \  0 \ \  \\
\hline
{  \ \ Y \ \  } &   \ \  0 \ \   &   \ \  0 \ \  \\
\hline
\hline
\end{tabular}
} 
} 
\qquad
\qquad
\qquad
\qquad
\subtable
[
{\footnotesize%
Gauge entropy $ \s_2 '$
}
]
{
{
\begin{tabular}{||C||C|C||}
\hline
\hline
u_0 \backslash u_1 &{ X } & { Y } \\
\hline
\hline
{  \ \ X \ \  } &   \ \  1 \ \   &   \ \  0 \ \  \\
\hline
{  \ \ Y \ \  } &   \ \  0 \ \   &   \ \  1 \ \  \\
\hline
\hline
\end{tabular}
} 
} 
%
\caption
{%
\label{partialGHZ} %
\footnotesize{%
Partial and gauge bipartite subsystems in regions $\mathcal{R}_0$ and $\mathcal{R}_1$  of the GHZ-state with two settings. 
(a) The partial probability in regions $\mathcal{R}_0$ and $\mathcal{R}_1$ describes a separable 2-region 2-setting system. 
(b) Gauge probability in regions $\mathcal{R}_0$ and $\mathcal{R}_1$ following a first collapse in region $\mathcal{R}_2$ with the setting $u_2 = X$ and the outcome $x_2=0$. This is a Bell's state $\phi^+$.
(c) The partial subsystem (a) is separable. Therefore the partial entropy computed from (a) is zero. 
(d) By contrast, the gauge subsystem (b) is entangled. The gauge entropy (in bits) computed from (b) exhibits a maximum entanglement:  $\mathcal{E} '(X,X) = \mathcal{E} '(Y,Y) = 1$ bit. 
}
}
\end{center}
\end{table} 

\paragraph{Two-step collapse}
We can proceed to a 2-step collapse. Whatever the gauge region and the gauge setting, the bipartite gauge system is a Bell-state. Table~\ref{partialGHZ}b,d is computed for  a  collapse in region $\mathcal{R}_2$ with the setting $u_2 = X$ and the outcome $x_2=0$. 

\paragraph{One-step collapse}
The one-step collapse rises no problem. The result is given in Table~\ref{toutGHZ}b.

\subsubsection {Super-quantum GHZ state}
\label{superQGHZ}
We are now going to describe a surprising `super-quantum' GHZ-state%
\footnote{This example stems from a flaw in the previous versions of this paper.}%
. Actually, in the GHZ-state (Table~\ref{toutGHZ}), the settings $X$ and $Y$ are not on an equal footing, even if this is not manifest in the definition, Eq.~(\ref{ketGHZ}). Now, if we demand a symmetrical behaviour between the two settings, we obtain the probability distribution of Table~\ref{superGHZ}a, that we will name \emph{Super-quantum GHZ state}. It is easily shown that this distribution describes a locally consistent system. In addition, the partial probabilities are identical to the partial probability of the conventional GHZ-states and given by Eq.(\ref{localGHZ}, \ref{completeGHZ}). As a result, the entanglement scheme is also $(0,1)$. The interest of this example is threefold. Firstly, we will show that this state is indeed super-quantum by proceeding to a 2-step collapse. Secondly, this example cannot collapses in one step and this can be  easily proved. Thirdly, we may construct a continuous set of similar modes, allowing an adjustment of the total entanglement in the range $[0,1]$ bit and a swiching  between super-quantum, quantum and non contextual behaviours.
%
\begin{table} [htb] 
\begin{center}
\subtable 
[
`Super-quantum' GHZ state with two settings $\theta_0$ and $\theta_1$: Probabilities $\PP(\mathbf {x}|\mathbf{u})$  
]
{
{
\begin{tabular}{||C||C|C|C|C||}
\hline
\hline
\quad \ (u_0, u_1,u_2)\rightarrow &(\theta_{ 0 },\theta_{ 0 },\theta_{ 0 })&(\theta_{ 0 },\theta_{ 0 },\theta_{ 1 })&(\theta_{ 0 },\theta_{ 1 },\theta_{ 0 })&(\theta_{ 0 },\theta_{ 1 },\theta_{ 1 })\\
(x_0, x_1,x_2)&( \theta_{ 1 },\theta_{ 1 },\theta_{ 1 } ) & (\theta_{ 1 },\theta_{ 1 },\theta_{ 0 }) & (\theta_{ 1 },\theta_{ 0 },\theta_{ 1 } )& (\theta_{ 1 },\theta_{ 0 },\theta_{ 0 } )\\
\downarrow & &&&  \\
\hline
\hline
 (0 , 0 , 0) & 0 &  1/4 &  1/4 &  1/4 \\
\hline
 (0 , 0 , 1) & 1/4 &  0&  0&  0\\
\hline
 (0 , 1,  0) & 1/4 &  0&  0&  0\\
\hline
 (0 , 1 , 1) & 0&  1/4 &  1/4 &  1/4 \\
\hline
 (1 , 0 , 0) & 1/4 &  0&  0&  0\\
\hline
 (1 , 0 , 1) & 0&  1/4 &  1/4 &  1/4 \\
\hline
 (1  ,1 , 0) &0&  1/4 &  1/4 &  1/4 \\
\hline
 (1 , 1 , 1) & 1/4 &  0&  0&  0\\
\hline
\hline
\end{tabular}
}
}
\\
\subtable
[
{\footnotesize%
Tripartite entropy $\s_3$ and total entanglement $\mathcal{E}$ (in bits)
}
]
{
{\footnotesize
\begin{tabular}{||C||C|C|C|C|C|C|C|C||}
\hline
\hline
(u_0, u_1, u_2) & { (\theta_0\theta_0\theta_0) } & { (\theta_0\theta_0\theta_1) } & { (\theta_0\theta_1\theta_0) } & { (\theta_0\theta_1\theta_1) }  & { (\theta_1\theta_0\theta_0) } & { (\theta_1\theta_0\theta_1) }  & { (\theta_1\theta_1\theta_0 )} & { (\theta_1\theta_1\theta_1) }   \\
\hline
{ \s_3 (u_0, u_1, u_2) } &  -1 &  -1         & -1              & -1             & -1             & -1             & -1             & -1              \\
\hline
{ \mathcal{E}(u_0, u_1, u_2) } &  1 &  1 & 1           & 1               & 1             & 1              & 1              & 1              \\
\hline
\hline
\end{tabular}
}
}  
\\
\subtable
[
{\footnotesize%
Second-step probability $\PP'$ when $u_2=\theta_0$
}
]
{
{
\begin{tabular}{||C||C|C|C|C||}
\hline
\hline
x_ 0 x_ 1 &(\theta_{ 0 }\theta_{ 0 })&(\theta_{ 0 }\theta_{ 1 })&(\theta_{ 1 }\theta_{ 0 })&(\theta_{ 1 }\theta_{ 1 })\\
\hline
\hline
 0  0 & 0 &  1/2 &  1/2 &  1/2 \\
\hline
 0  1 & 1/2 &  0 &  0 &  0\\
\hline
 1  0 & 1/2 &  0 &  0 &  0\\
\hline
 1  1 & 0 &  1/2 &  1/2 &  1/2 \\
\hline
\hline
\end{tabular}
} 
} 
\subtable
[
{\footnotesize%
Second-step probability $\PP'$ when $u_2=\theta_1$
}
]
{
{
\begin{tabular}{||C||C|C|C|C||}
\hline
\hline
x_ 0 x_ 1 &(\theta_{ 0 }\theta_{ 0 })&(\theta_{ 0 }\theta_{ 1 })&(\theta_{ 1 }\theta_{ 0 })&(\theta_{ 1 }\theta_{ 1 })\\
\hline
\hline
 0  0 & 1/2 &  1/2 &  1/2 &  0 \\
\hline
 0  1 & 0 &  0  &  0  &  1/2 \\
\hline
 1  0 & 0 &  0  &  0  &  1/2  \\
\hline
 1  1 & 1/2  &  1/2 &  1/2 &  0 \\
\hline
\hline
\end{tabular}
} 
} 
\\
\subtable
[
{\footnotesize%
Gauge distributions for (c) 
}
]
{
{
\begin{tabular}{||C||C|C|C|C||}
\hline
\hline
j &  \g_{ 0 } (j) &  \g_{ 1 } (j) &  \g_{ 2 } (j) &  \g_{ 3 } (j)\\
\hline
\hline
{ 1 } &  0 &  1/2 &  1/2 &  0\\
\hline
{ 4 } &  1/2 &  0 &  0 &  1/2\\
\hline
{ 11 } &  1/2 &  0 &  0 &  1/2\\
\hline
{ 14 } &  0 &  1/2 &  1/2 &  0\\
\hline
\hline
\end{tabular}
} 
} 
\qquad
\subtable
[
{\footnotesize%
Gauge distributions for (d) 
}
]
{
{
\begin{tabular}{||C||C|C|C|C||}
\hline
\hline
j &  \g_{ 0 } (j) &  \g_{ 1 } (j) &  \g_{ 2 } (j) &  \g_{ 3 } (j)\\
\hline
\hline
   2 & 1/2 &  0 &  0 &  1/2 \\
\hline
 7 & 0 &  1/2  &  1/2  &  0 \\
\hline
 8 & 0 &  1/2  &  1/2  &  0  \\
\hline
 13 & 1/2  &  0 &  0 &  1/2 \\
\hline
\hline
\end{tabular}
} 
} 
\caption
{%
\label{superGHZ} %
\footnotesize{%
`Super-quantum' GHZ states: 
(a) Conditional probability $\PP(x_0;x_1;x_2|u_0;u_1;u_2)$ describing  a `super-quantum' GHZ  system with two different settings $\theta_0$ and $\theta_1$. It can be seen that this system is locally consistent and behaves exactly like the conventional GHZ state with respect to  `local realism'. 
(b) Tripartite entanglement entropy $\s_3$ and total entanglement $\mathcal{E}$ of the state.
A one-step collapse is impossible. A partial collapse in region $\mathcal{R}_2$ leads to a PR-box. 
(c) Second-step probability $\PP'(x_0;x_1|u_0;u_1)$ in regions $ \mathcal{R}_0, \mathcal{R}_1 $,  following a gauge selection of region $ \mathcal{R}_2 $ and a free choice $u_2=\theta_0$. 
(d) Second-step probability $\PP'(x_0;x_1|u_0;u_1)$ when the free choice is now $u_2=\theta_1$. 
(e) Gauge distribution working set with 4 ignition states (out of 16) for the second-step probability $\PP'(x_0;x_1|u_0;u_1)$ given in (c). This is a relabelling of Table~\ref{PRbox}b. 
(f) Same as (e) for the distribution (d).
}
}
\end{center}
\end{table} 

We first try a one-step collapse. The number of configurations is $3\times 2=6$. In order to classically encode Table \ref{superGHZ}a, we construct an ignition set $\mathbf{\Lambda}=\{ \lambda_j \} $ with $2^6=64$ ignition states, labelled $j$ from $j=0$ to $63$, and 6 stochastic gauge distributions, $\g_\gamma(\lambda_j)$ with  $\gamma=0,\dots,5$. However, none of the six linear systems admits non negative solution.  To circumvent this problem,  we may try a different ignition set. The maximum number  of distinct fine grains is $2^{2^3} = 256$. However, it is easily shown that these $256$ fine grains are all included into at least one target of zero probability. Therefore, whatever the gauge, the  probability of all ignition states will always be zero. In conclusion, within the realm of local realism, the super-quantum GHZ state \emph{cannot collapse in one step}.

We have thus to try a two-step collapse. Let $\mathcal{I}$ be the ignition point at the boundary of the three regions. Each observer selects freely his own setting, $u_i=k_i$ and sends his choice towards $\mathcal{I}$, with a finite velocity. Suppose that the first received configuration is $u_2=k_2$ from region $\mathcal{R}_2$. As a result, we break the equilibrium at $\mathcal{I}$  from region $\mathcal{R}_2$. We draw the outcome $x_2=\xi_2$ from the local probability distribution $\PPr(x_2|k_2)$, Eq.(\ref{localGHZ}). The two possible outcomes $\xi_2\in\{ 0,1\}$ are equally likely. Suppose for instance that $\xi_2=0$. Next, in regions $ \mathcal{R}_0, \mathcal{R}_1 $ we compute the second-step probability distribution $ \PP'(x_0;x_1|u_0;u_1)=\PPr(x_0;x_1|\xi_2;k_2;u_0;u_1) $. The resulting distribution is given in Table \ref{superGHZ}c for $k_2=0$ and $\xi_2=0$. On the other hand, if the observer $\mathcal{O}_2$ had chosen $k_2=1$ instead of $k_2=0$, the distribution $ \PP'(x_0;x_1|u_0;u_1)$, now given in Table \ref{superGHZ}d for $\xi_2=0$, would have been different. These probability distributions are identical to the probabilities of the PR-box, Table~\ref{PRbox}a, (apart from the labelling). Therefore, this `super-quantum' GHZ state is actually `super-quantum'.
To complete the two-step collapse, we have now to construct an ignition set for the two regions $ \mathcal{R}_0$ and $\mathcal{R}_1 $,  accounting for the relevant distribution $\PP'(x_0;x_1|u_0;u_1)$.  
This is easily achieved because the new system is actually a PR-box. Each gauge distribution requires just two ignition states, but we need four  ignition states for the whole, as shown in Table \ref{superGHZ}e,f. This proves that the `super-quantum' GHZ-states can be classically implemented, but only a two-step collapse is feasible.

\begin{table}[htbp]
$$
\begin{array}{||c||c|c|c|c||}
\hline
\hline
\quad \ (u_0, u_1,u_2)\rightarrow &(\theta_{ 0 },\theta_{ 0 },\theta_{ 0 })&(\theta_{ 0 },\theta_{ 0 },\theta_{ 1 })&(\theta_{ 0 },\theta_{ 1 },\theta_{ 0 })&(\theta_{ 0 },\theta_{ 1 },\theta_{ 1 })\\
(x_0, x_1,x_2)&( \theta_{ 1 },\theta_{ 1 },\theta_{ 1 } ) & (\theta_{ 1 },\theta_{ 1 },\theta_{ 0 }) & (\theta_{ 1 },\theta_{ 0 },\theta_{ 1 } )& (\theta_{ 1 },\theta_{ 0 },\theta_{ 0 } )\\
\downarrow & &&&  \\
\hline
\hline
 (0 , 0 , 0) & \epsilon &  1/4 -\epsilon &  1/4-\epsilon  &  1/4-\epsilon  \\
\hline
 (0 , 0 , 1) & 1/4-\epsilon  &  \epsilon &  \epsilon &  \epsilon \\
\hline
 (0 , 1,  0) & 1/4-\epsilon  &  \epsilon &  \epsilon &  \epsilon \\
\hline
 (0 , 1 , 1) & \epsilon &  1/4-\epsilon  &  1/4 -\epsilon &  1/4-\epsilon  \\
\hline
 (1 , 0 , 0) & 1/4-\epsilon  &  \epsilon &  \epsilon &  \epsilon \\
\hline
 (1 , 0 , 1) & \epsilon &  1/4 -\epsilon &  1/4-\epsilon  &  1/4-\epsilon  \\
\hline
 (1  ,1 , 0) &\epsilon &  1/4-\epsilon  &  1/4-\epsilon  &  1/4-\epsilon  \\
\hline
 (1 , 1 , 1) & 1/4-\epsilon  &  \epsilon &  \epsilon &  \epsilon \\
\hline
\hline
\end{array}
$$
\caption{\label {pseudoGHZ} {\footnotesize `Quasi-super-quantum' GHZ states: Conditional probability $\PP(x_0;x_1;x_2|u_0;u_1;u_2)$ of a quasi-GHZ  system.  Target probabilities are derived from the super-quantum GHZ state by replacing 0 by $\epsilon$  in Table~\ref{superGHZ}a. A one-step collapse is feasible for  $1/16<\epsilon < 3/16$. Partial probabilities Eq.(\ref{localGHZ} and \ref{completeGHZ}) are unchanged.  }}
\end{table}

\paragraph{A continuous set of similar modes}
We have checked whether a  \emph{quasi-super} state can collapse in one step. Indeed, in the super-quantum GHZ state,  a number of targets have zero probability (Table \ref{toutGHZ}a). It is possible to replace 0 by a small probability $\epsilon$  as shown in Table~\ref{pseudoGHZ}. In spite of this change, the partial probabilities, Eq. (\ref{localGHZ}) and (\ref{completeGHZ}) are strictly conserved and therefore, any pair of regions remains separable. As a result, the entanglement scheme is still $(0,1)$, except when the three regions become independent. Thus, the new system may actually be viewed as a \emph{quasi-super GHZ} state, provided that $\epsilon$ remains small. By computer simulation, we have found that a one-step collapse is feasible only when $\epsilon$ is in the range $]1/16,3/16[$. For $\epsilon = 2/16$, the three regions become independent. 

\begin{center}
\begin{tabular}{CCc}
\epsilon & \mathcal{E} \ \mathrm{(bits)}             & Comment                                           \\
\hline
0.1250           &                0.0                          &  separable classical object                  \\
                     &                                               &   quantum-like,   1-step collapse          \\
0.0625           &                0.2                          &   limit of 1-step collapse                      \\
                     &                                               &   quantum-like, 2-step collapse            \\
0.0366           &                0.4                          &   Tsirelson bound                 \\
                     &                                               &   super-quantum system            \\
0.0000           &                1.0                          &  super-quantum GHZ-state (Table~\ref{superGHZ}) \\
\hline
\\
\end{tabular}

Behaviour of the 'quasi-super-quantum' object versus $\epsilon$
\end{center}

For $\epsilon = 1/16=0.0625$, we have  checked whether the system passes the GHZ test of `local realism' (in fact, non-contextuality). With the same notations as above (Table~\ref{toutGHZ}), the probability that $f(\theta_ 1,\theta_ 1,\theta_ 1)\equiv  1 \pmod {2}$ is now $0.75$ (instead of $1$), while the probability that $ f( \theta_0,\theta_ 0,\theta_ 1)\equiv 0 $, $f(\theta_ 0,\theta_ 1 ,\theta_ 0)\equiv 0 $ and $f(\theta_ 1,\theta_ 0,\theta_ 0) \equiv 0$ is $0.56$ (instead of $1$). We can conclude that the behaviour of a quasi-GHZ state is similar to a strict mode, but there is no dramatic all-versus-nothing outcome as in the strict state.

Finally, we have  checked by simulation whether the system passes the Tsirelson test. We have found that a local collapse in one region gives a `super-quantum' bipartite system when $\epsilon < 0.0366$. When $\epsilon = 0.0366$, the CHSH average (Eq.~\ref{tsirelsonbound}) is about $2\sqrt{2}$. The total tripartite entanglement is then $\mathcal{E} = 0.4$~bit, irrespective of the settings. We can conclude that the system behaviour is `super-quantum' at least for $0\le\epsilon\le 0.0366$. Note that this example indicates that a partial collapse can give an issue about the super-quantum behaviour of a tripartite system while direct 3-region witnesses are unable to conclude.

\paragraph{}
More generally, the current model should open the possibility to \emph{classically} synthesize any locally consistent system, whether quantum or not, whatever its degree of entanglement. One has to define a number of regions, $n$, a number of settings, $K$,  $n$ local measurement entropies and  $ \binom{n}{m} K^m $ coefficients of entanglement for $m=2$ to $n$. This is similar to the analysis of the so-called `nonsignaling polytope' in quantum information. This discussion is beyond the scope of the present paper.

\section{Conclusion}

Contextuality is not the privilege of the quantum world. Quite the reverse, we have shown in a previous paper~\cite{mf1} that a strictly classical dice game can be contextually dependent and even nonsignaling in the quantum sense. Therefore, the difficult task to reconcile quantum mechanics with non-contextual logic is  unnecessary. In the current paper we have constructed a model named `stochastic gauge system' to compute contextually dependent classical systems. We have shown that the theory described quantum randomness as well and can be used to simulate quantum states and even non-local boxes in classical physics, e.g., EPR pairs, GHZ states or PR-boxes. Quantum systems are compared with classical extended objects in equilibrium and quantum collapses are therefore identified with classical breaks of equilibrium. Furthermore, the model gives a straightforward  classical interpretation of entanglement and entanglement entropy and provides simple tools for characterizing multipartite entanglements. 

In our opinion, the present theory opens the way for a number of important questionings,  regarding quantum theory, quantum information and even the general conception of physics, obviously beyond the scope of this paper.

In quantum mechanics, we have shown that classical local realism is compatible with the conventional formulation of the theory, contrary to a general belief. This result is likely to explain a number of difficulties and opens up also new topics. For instance, in the relativistic domain, the conventional interpretation fails to define a consistent  probability distribution. In our opinion, the concept of gauge probability should be a key ingredient to settle this problem. The notion of quantum collapse should be also revisited and compared with a classical break of equilibrium. New issues,  like collapse kinetics should probably be investigated. Incidentally, while a verification of  Bell's inequality violation is useless, a meaningful test to check the present theory would be an experimental probe of the collapse kinetics between entangled regions, in terms, e.g., of the distance between particles.

The model proposes also an interpretation of quantum parallelism. A quantum system can be viewed as a complex entangled object in equilibrium.  When the equilibrium is broken, a number of new situations become potentially possible. Quantum parallelism describes the set of all these potential situations. In quantum information theory, a final measurement will draw just one outcome. The theory uses entanglement as a resource, but again, this is not a quantum privilege. In our opinion, contextuality should be a resource in classical information as well. Indeed, `quantum  randomness' can be viewed as strictly classical, provided that one includes an additional concept in the conventional stochastic theory, namely, the notion of gauge distribution. Classical simulation of quantum algorithms should open up new paths in classical computation, beside deterministic and probabilistic algorithms.  More generally, a number of  notions like quantum cryptography,  quantum complexity or communication complexity (see above Sec.~\ref{bipartite}) should probably deserve to be revisited.

Nowadays, the principle of gauge invariance plays a key role in modern physics. This is not very surprising because every physical measurement is performed within a particular framework while physical reality has to be independent of any referential and any system of units. Therefore,  mathematical tools have been devised: Dimensional analysis, covariance principle, and more generally, gauge theories. The gauge invariance principle has proved to be extraordinary powerful in all branches of physics.

Surprisingly, the measurement in quantum theory remains nevertheless thought as \emph{absolute}. The seed of this conception lies probably in the very origin of the theory, i.e., the Copenhagen interpretation of quantum mechanics,  according to which the squared amplitude of the wave function is an \emph{absolute probability}. In our opinion, this belief is inconsistent, as highlighted by the EPR paradox or the problematic generalization of this interpretation in special relativity.

The view of an absolute probability space is also the origin of a controversy between `Bayesians' and `Frequentists' in the probability domain. In our opinion, the `Frequentist'  position is simply not sustainable. For instance, we have emphasized in the present paper that a number of fundamental concepts like partial measurements are subjective \emph{by definition}. 

Actually, physics aims to understand the entirety of the universe, in which a number of long range interactions are recorded. By contrast, each observer is embedded into \emph{his} own limited \emph{causal bubble} or \emph{causal diamond}, bounded by light sheets,  black hole and cosmological horizons. Therefore, he can only describe a finite part of the world and, apart from long range interactions, he has to account for the rest of the universe by \emph{his} best evaluation of \emph{his} irreducible ignorance, technically in form of  entropy located on \emph{his} causal horizons. 

This vision of the universe seems very new in physics: Unpredictability is no more in contradiction with determinism but, quite the reverse, an inevitable consequence.  This implies a break between the general Laplace concept of \emph{determinism} and the technical notion of \emph{causality}. In this respect, the dispute concerning the completeness of quantum mechanics appears totally unrealistic.

\bibliography{biblio}

\end{document}